\documentclass[10pt]{revtex4}
\usepackage{graphicx}

\begin{document}



\newcount\driver
\newcount\bozza

\font\ottorm=cmr8\font\ottoi=cmmi8\font\ottosy=cmsy8%
\font\ottobf=cmbx8\font\ottott=cmtt8%
\font\ottocss=cmcsc8%
\font\ottosl=cmsl8\font\ottoit=cmti8%
\font\sixrm=cmr6\font\sixbf=cmbx6\font\sixi=cmmi6\font\sixsy=cmsy6%
\font\fiverm=cmr5\font\fivesy=cmsy5
\font\fivei=cmmi5
\font\fivebf=cmbx5%
\font\tenmib=cmmib10
\font\sevenmib=cmmib10 scaled 800
\font\titolo=cmbx12
\font\titolone=cmbx10 scaled\magstep 2
\font \titolino=cmbx10
\font\cs=cmcsc10
\font\sc=cmcsc10
\font\css=cmcsc8
\font\ss=cmss10
\font\sss=cmss8
\font\crs=cmbx8
\font\ninerm=cmr9
\font\tenrm=cmr10
\font\elevenrm=cmr11
\font\twelverm=cmr12
\font\ottorm=cmr8
\textfont5=\tenmib
\scriptfont5=\sevenmib
\scriptscriptfont5=\fivei
\font\msxtw=msbm9 scaled\magstep1
\font\euftw=eufm9 scaled\magstep1
\font\euftww=eufm7 scaled\magstep1
\font\euftwww=eufm5 scaled\magstep1
\font\msytw=msbm9 scaled\magstep1
\font\msytww=msbm7 scaled\magstep1
\font\msytwww=msbm5 scaled\magstep1
\font\indbf=cmbx10 scaled\magstep2
\font\type=cmtt10
\def\st{\scriptstyle}

\font\ottorm=cmr8\font\ottoi=cmmi8\font\ottosy=cmsy8%
\font\ottobf=cmbx8\font\ottott=cmtt8%
\font\ottocss=cmcsc8%
\font\ottosl=cmsl8\font\ottoit=cmti8%
\font\sixrm=cmr6\font\sixbf=cmbx6\font\sixi=cmmi6\font\sixsy=cmsy6%
\font\fiverm=cmr5\font\fivesy=cmsy5
\font\fivei=cmmi5
\font\fivebf=cmbx5%

\def\ottopunti{\def\rm{\fam0\ottorm}%
\textfont0=\ottorm\scriptfont0=\sixrm\scriptscriptfont0=\fiverm%
\textfont1=\ottoi\scriptfont1=\sixi\scriptscriptfont1=\fivei%
\textfont2=\ottosy\scriptfont2=\sixsy\scriptscriptfont2=\fivesy%
\textfont4=\ottocss\scriptfont4=\sc\scriptscriptfont4=\sc%
\scriptfont4=\ottocss\scriptscriptfont4=\ottocss%
\textfont5=\tenmib\scriptfont5=\sevenmib\scriptscriptfont5=\fivei
\setbox\strutbox=\hbox{\vrule height7pt depth2pt width0pt}%
\normalbaselineskip=9pt\let\sc=\sixrm\normalbaselines\rm}
\let\nota=\ottopunti%

\mathchardef\BDpr = "0540  
\mathchardef\Bg   = "050D  

\def\sf{\textfont1=\amit}

{\count255=\time\divide\count255 by 60 \xdef\hourmin{\number\count255}
        \multiply\count255 by-60\advance\count255 by\time
   \xdef\hourmin{\hourmin:\ifnum\count255<10 0\fi\the\count255}}

\def\openone{\leavevmode\hbox{\elevenrm 1\kern-3.63pt\twelverm1}}%
\def\*{\vglue0.5truecm}


\let\a=\alpha \let\b=\beta  \let\g=\gamma  \let\d=\delta \let\e=\varepsilon
\let\z=\zeta  \let\h=\eta   \let\th=\theta \let\k=\kappa \let\l=\lambda
\let\m=\mu    \let\n=\nu    \let\x=\xi     \let\p=\pi    \let\r=\rho
\let\s=\sigma \let\t=\tau   \let\f=\varphi \let\ph=\varphi\let\c=\chi
\let\ps=\psi  \let\y=\upsilon \let\o=\omega\let\si=\varsigma
\let\G=\Gamma \let\D=\Delta  \let\Th=\Theta\let\L=\Lambda \let\X=\Xi
\let\P=\Pi    \let\Si=\Sigma \let\F=\Phi    \let\Ps=\Psi
\let\O=\Omega \let\Y=\Upsilon

\def\\{\hfill\break} \let\==\equiv
\let\txt=\textstyle\let\dis=\displaystyle

\let\io=\infty \def\Dpr{\V\dpr\,}
\def\aps{{\it a posteriori\ }}\def\ap{{\it a priori\ }}
\let\0=\noindent\def\pagina{{\vfill\eject}}
\def\bra#1{{\langle#1|}}\def\ket#1{{|#1\rangle}}
\def\media#1{{\langle#1\rangle}}
\def\ie{{i.e. }}\def\eg{{e.g. }}
\let\dpr=\partial \def\der{{\rm d}} \let\circa=\cong
\def\arccot{{\rm arccot}}

\def\tende#1{\,\vtop{\ialign{##\crcr\rightarrowfill\crcr
 \noalign{\kern-1pt\nointerlineskip} \hskip3.pt${\scriptstyle
 #1}$\hskip3.pt\crcr}}\,}
\def\circage{\lower2pt\hbox{$\,\buildrel > \over {\scriptstyle \sim}\,$}}
\def\otto{\,{\kern-1.truept\leftarrow\kern-5.truept\to\kern-1.truept}\,}
\def\fra#1#2{{#1\over#2}}

\def\PPP{{\cal P}}\def\EE{{\cal E}}\def\MM{{\cal M}}
\def\CC{{\cal C}}\def\FF{{\cal F}} \def\HHH{{\cal H}}\def\WW{{\cal W}}
\def\TT{{\cal T}}\def\NN{{\cal N}} \def\BBB{{\cal B}}\def\III{{\cal I}}
\def\RR{{\cal R}}\def\LL{{\cal L}} \def\JJ{{\cal J}} \def\OO{{\cal O}}
\def\DD{{\cal D}}\def\AAA{{\cal A}}\def\GG{{\cal G}} \def\SS{{\cal S}}
\def\KK{{\cal K}}\def\UU{{\cal U}} \def\QQ{{\cal Q}} \def\XXX{{\cal X}}

\def\T#1{{#1_{\kern-3pt\lower7pt\hbox{$\widetilde{}$}}\kern3pt}}
\def\VVV#1{{\VV #1}_{\kern-3pt
\lower7pt\hbox{$\widetilde{}$}}\kern3pt\,}
\def\W#1{#1_{\kern-3pt\lower7.5pt\hbox{$\widetilde{}$}}\kern2pt\,}
\def\Re{{\rm Re}\,}\def\Im{{\rm Im}\,}
\def\lis{\overline}\def\tto{\Rightarrow}
\def\etc{{\it etc}} \def\acapo{\hfill\break}
\def\mod{{\rm mod}\,} \def\per{{\rm per}\,} \def\sign{{\rm sign}\,}
\def\indica{\leaders \hbox to 0.5cm{\hss.\hss}\hfill}
\def\guida{\leaders\hbox to 1em{\hss.\hss}\hfill}

\def\hh{{\bf h}} \def\HH{{\bf H}} \def\AA{{\bf A}} \def\qq{{\bf q}}
\def\VV{{\cal V}}
\def\BB{{\bf B}} \def\XX{{\bf X}} \def\PP{{\bf P}} \def\pp{{\bf p}}
\def\vv{{\bf v}} \def\xx{{\bf x}} \def\yy{{\bf y}} \def\zz{{\bf z}}
\def\aaa{{\bf a}}\def\bbb{{\bf b}}\def\hhh{{\bf h}}\def\II{{\bf I}}
\def\ii{{\bf i}}\def\jj{{\bf j}}\def\kk{{\bf k}}\def\bS{{\bf S}}
\def\Vm{{\bf m}}\def\Vn{{\bf n}}

\def\VV#1{{\,\underline#1\,}}
\def\ul{\underline}
\def\olu{{\overline{u}}}
\def\defin{{\buildrel def\over=}}
\def\wt{\widetilde}
\def\wh{\widehat}
\def\Dpr{\BDpr\,}


\mathchardef\dd   = "050E
\mathchardef\aa   = "050B
\mathchardef\bb   = "050C
\mathchardef\ggg  = "050D
\mathchardef\xxx  = "0518
\mathchardef\zzzzz= "0510
\mathchardef\oo   = "0521
\mathchardef\lll  = "0515
\mathchardef\mm   = "0516
\mathchardef\Dp   = "0540
\mathchardef\H    = "0548
\mathchardef\FFF  = "0546
\mathchardef\ppp  = "0570
\mathchardef\Bn   = "0517
\mathchardef\pps  = "0520
\mathchardef\fff  = "0527
\mathchardef\FFF  = "0508
\mathchardef\nnnnn= "056E

\def\ol#1{{\overline #1}}

\def\to{\rightarrow}
\def\la{\left\langle}
\def\ra{\right\rangle}

\def\Overline#1{{\bar#1}}
\let\ciao=\bye
\def\qed{\hfill\raise1pt\hbox{\vrule height5pt width5pt depth0pt}}
\def\hf#1{{\hat \f_{#1}}}
\def\barf#1{{\tilde \f_{#1}}} \def\tg#1{{\tilde g_{#1}}}
\def\bq{{\bar q}}
\def\Val{{\rm Val}}
\def\indic{\hbox{\raise-2pt \hbox{\indbf 1}}}

\def\RRR{\hbox{\msytw R}} \def\rrrr{\hbox{\msytww R}}
\def\rrr{\hbox{\msytwww R}} \def\CCC{\hbox{\msytw C}}
\def\cccc{\hbox{\msytww C}} \def\ccc{\hbox{\msytwww C}}
\def\NNN{\hbox{\msytw N}} \def\nnnn{\hbox{\msytww N}}
\def\nnn{\hbox{\msytwww N}} \def\ZZZ{\hbox{\msytw Z}}
\def\zzzz{\hbox{\msytww Z}} \def\zzz{\hbox{\msytwww Z}}
\def\SSS{\hbox{\msytw S}}
\def\ssss{\hbox{\msytww S}} \def\sss{\hbox{\msytwww S}}
\def\TTT{\hbox{\msytw T}} \def\tttt{\hbox{\msytww T}}
\def\ttt{\hbox{\msytwww T}}\def\MMM{\hbox{\euftw M}}
\def\QQQ{\hbox{\msytw Q}} \def\qqqq{\hbox{\msytww Q}}
\def\qqq{\hbox{\msytwww Q}}

\def\vvv{\hbox{\euftw v}}    \def\vvvv{\hbox{\euftww v}}
\def\vvvvv{\hbox{\euftwww v}}\def\www{\hbox{\euftw w}}
\def\wwww{\hbox{\euftww w}}  \def\wwwww{\hbox{\euftwww w}}
\def\vvr{\hbox{\euftw r}}    \def\vvvr{\hbox{\euftww r}}

\def\ul#1{{\underline#1}}
\def\Sqrt#1{{\sqrt{#1}}}
\def\V0{{\bf 0}}
\def\defi{\,{\buildrel def\over=}\,}
\def\lhs{{\it l.h.s.}\ }
\def\rhs{{\it r.h.s.}\ }

\font\tenmib=cmmib10 \font\eightmib=cmmib8
\font\sevenmib=cmmib7\font\fivemib=cmmib5
\font\ottoit=cmti8
\font\fiveit=cmti5\font\sixit=cmti6
\font\fivei=cmmi5\font\sixi=cmmi6\font\ottoi=cmmi8
\font\ottorm=cmr8\font\fiverm=cmr5\font\sixrm=cmr6
\font\ottosy=cmsy8\font\sixsy=cmsy6\font\fivesy=cmsy5
\font\ottobf=cmbx8\font\sixbf=cmbx6\font\fivebf=cmbx5%
\font\ottott=cmtt8%
\font\ottocss=cmcsc8%
\font\ottosl=cmsl8%

\textfont5=\tenmib\scriptfont5=\sevenmib\scriptscriptfont5=\fivemib
\mathchardef\Ba   = "050B  
\mathchardef\Bb   = "050C  
\mathchardef\Bg   = "050D  
\mathchardef\Bd   = "050E  
\mathchardef\Be   = "0522  
\mathchardef\Bee  = "050F  
\mathchardef\Bz   = "0510  
\mathchardef\Bh   = "0511  
\mathchardef\Bthh = "0512  
\mathchardef\Bth  = "0523  
\mathchardef\Bi   = "0513  
\mathchardef\Bk   = "0514  
\mathchardef\Bl   = "0515  
\mathchardef\Bm   = "0516  
\mathchardef\Bn   = "0517  
\mathchardef\Bx   = "0518  
\mathchardef\Bom  = "0530  
\mathchardef\Bp   = "0519  
\mathchardef\Br   = "0525  
\mathchardef\Bro  = "051A  
\mathchardef\Bs   = "051B  
\mathchardef\Bsi  = "0526  
\mathchardef\Bt   = "051C  
\mathchardef\Bu   = "051D  
\mathchardef\Bf   = "0527  
\mathchardef\Bff  = "051E  
\mathchardef\Bch  = "051F  
\mathchardef\Bps  = "0520  
\mathchardef\Bo   = "0521  
\mathchardef\Bome = "0524  
\mathchardef\BG   = "0500  
\mathchardef\BD   = "0501  
\mathchardef\BTh  = "0502  
\mathchardef\BL   = "0503  
\mathchardef\BX   = "0504  
\mathchardef\BP   = "0505  
\mathchardef\BS   = "0506  
\mathchardef\BU   = "0507  
\mathchardef\BF   = "0508  
\mathchardef\BPs  = "0509  
\mathchardef\BO   = "050A  
\mathchardef\BDpr = "0540  
\mathchardef\Bstl = "053F  
\def\BK{{\bf K}}
\def\bT{{\bf T}}

\def\V#1{{\bf#1}}
\let\aa=\Ba\let\fff=\Bf\let\defin=\defi\def\HHH{{\cal H}}
\def\VV{{\cal V}}

\let\wt=\widetilde\def\AAA{{\cal A}}\let\oo=\Bo\let\nn=\Bn
\let\aaa=\Ba\let\pps=\Bps\def\hhh={\V h}\def\bbb{{\V b}}
\let\bb=\Bb\def\ss{\ul{\s}}

\def\RRR{\hbox{\msytw R}} \def\rrrr{\hbox{\msytww R}}
\def\rrr{\hbox{\msytwww R}} \def\CCC{\hbox{\msytw C}}
\def\cccc{\hbox{\msytww C}} \def\ccc{\hbox{\msytwww C}}
\def\NNN{\hbox{\msytw N}} \def\nnnn{\hbox{\msytww N}}
\def\nnn{\hbox{\msytwww N}} \def\ZZZ{\hbox{\msytw Z}}
\def\zzzz{\hbox{\msytww Z}} \def\zzz{\hbox{\msytwww Z}}
\def\TTT{\hbox{\msytw T}} \def\tttt{\hbox{\msytww T}}
\def\ttt{\hbox{\msytwww T}}
\def\QQQ{\hbox{\msytw Q}} \def\qqqq{\hbox{\msytww Q}}
\def\qqq{\hbox{\msytwww Q}}
\def\SSS{\hbox{\msytw S}} \def\BBBB{\hbox{\msytw B}}

\let\ul=\underline\def\hh{{\V h}}
\def\cfr{{cf. }}\let\ig=\int
\def\Tr{{\rm Tr}}
\def\dist{{\rm dist}}

%
%
%
\def\ins#1#2#3{\vbox to0pt{\kern-#2 \hbox{\kern#1 #3}\vss}\nointerlineskip}
%
%
%
\newdimen\xshift \newdimen\xwidth \newdimen\yshift
\newcount\griglia

\def\insertplot#1#2#3#4#5#6{%
\begin{figure}[h]
\begin{center}
\vspace{#2pt}
\begin{minipage}{#1pt}
#3
\ifnum\driver=1
\griglia=#6
\ifnum\griglia=1
\openout13=griglia.ps
\write13{gsave .2 setlinewidth}
\write13{0 10 #1 {dup 0 moveto #2 lineto } for}
\write13{0 10 #2 {dup 0 exch moveto #1 exch lineto } for}
\write13{stroke}
\write13{.5 setlinewidth}
\write13{0 50 #1 {dup 0 moveto #2 lineto } for}
\write13{0 50 #2 {dup 0 exch moveto #1 exch lineto } for}
\write13{stroke grestore}
\closeout13
\includegraphics{griglia.ps}\fi
\includegraphics{#4.ps}\fi
\ifnum\driver=2
\fi
\end{minipage}
\end{center}
\caption{#5}
\end{figure}
}

\def\gtopl{\hbox{\msxtw \char63}}
\def\ltopg{\hbox{\msxtw \char55}}

\newdimen\shift \shift=-1truecm
\def\lb#1{%
\ifnum\bozza=1
\label{#1}\rlap{\kern\shift{$\scriptstyle#1$}}
\else\label{#1}
\fi}

\def\be{\begin{equation}}
\def\ee{\end{equation}}
\def\bea{\begin{eqnarray}}\def\eea{\end{eqnarray}}
\def\bean{\begin{eqnarray*}}\def\eean{\end{eqnarray*}}
\def\bfr{\begin{flushright}}\def\efr{\end{flushright}}
\def\bc{\begin{center}}\def\ec{\end{center}}
\def\ba#1{\begin{array}{#1}} \def\ea{\end{array}}
\def\bd{\begin{description}}\def\ed{\end{description}}
\def\bv{\begin{verbatim}}\def\ev{\end{verbatim}}
\def\nn{\nonumber}
\def\Halmos{\hfill\vrule height10pt width4pt depth2pt \par\hbox to \hsize{}}
\def\pref#1{(\ref{#1})}
\def\Dim{{\bf Dim. -\ \ }} \def\Sol{{\bf Soluzione -\ \ }}
\def\virg{\quad,\quad}
\def\bsl{$\backslash$}

\newtheorem{lemma}{Lemma}[section]
\newtheorem{theorem}{Theorem}[section]
\renewcommand{\theequation}{\arabic{section}.\arabic{equation}}

\newdimen\xshift \newdimen\xwidth \newdimen\yshift \newdimen\ywidth

\def\ins#1#2#3{\vbox to0pt{\kern-#2\hbox{\kern#1 #3}\vss}\nointerlineskip}

\def\eqfig#1#2#3#4#5{
\par\xwidth=#1 \xshift=\hsize \advance\xshift
by-\xwidth \divide\xshift by 2
\yshift=#2 \divide\yshift by 2
\line{\hglue\xshift \vbox to #2{\vfil
#3 \includegraphics{#4.ps}
}\hfill\raise\yshift\hbox{#5}}}

\def\8{\write12}

\def\figini#1{
\catcode`\%=12\catcode`\{=12\catcode`\}=12
\catcode`\<=1\catcode`\>=2
\openout12=#1.ps}

\def\figfin{
\closeout12
\catcode`\%=14\catcode`\{=1%
\catcode`\}=2\catcode`\<=12\catcode`\>=12}


\title{The 2D Hubbard model on the honeycomb lattice}
\*

\author{Alessandro Giuliani}
\affiliation{Dipartimento di Matematica,
Universit\`a di Roma Tre, L.go S. Leonardo Murialdo 1, 00146 Roma Italy}
\author{Vieri Mastropietro}
\affiliation{Dipartimento di Matematica, Universit\`a di Roma Tor
Vergata, V.le della Ricerca Scientifica, 00133 Roma Italy}
\begin{abstract} We consider the 2D Hubbard model on the honeycomb lattice,
as a model for a single layer graphene sheet in the presence of
screened Coulomb interactions. At half filling and weak enough coupling,
we compute the free energy, the ground state energy and we
construct the correlation functions up to zero temperature in terms of 
convergent series; analiticity is proved by making use of 
constructive fermionic renormalization group methods. 
We show that the interaction produces
a modification of the Fermi velocity and of the wave function renormalization
without changing the asymptotic infrared properties of the model
with respect to the unperturbed non-interacting case; this rules
out the possibility of superconducting or magnetic
instabilities in the ground state. We also prove that the correlations verify
a Ward Identity similar to the one for massless Dirac fermions, up
to asymptotically negligible corrections and a renormalization
of the charge velocity.
\end{abstract}

\maketitle

\section{Introduction}

The recent experimental realization of a monocrystalline graphitic
film, known as {\it graphene} \cite{N}, revived the interest in
the low temperature physics of two--dimensional electron systems
on the honeycomb lattice, which is the typical underlying
structure displayed by single--layer graphene sheets. Graphene is
quite different from most conventional quasi--two dimensional
electron gases, because of the peculiar quasi--particles
dispersion relation, which closely resembles the one of massless
Dirac fermions in $2+1$ dimensions. This was already pointed out
in \cite{W} and further exploited in \cite{S}, where the analogy
between graphene and $2+1$-dimensional quantum electrodynamics (QED)
was made explicit, and used to predict a condensed-matter analogue of
the axial anomaly in QED. From this point of view, graphene can be
considered as a sort of testing bench to investigate the
properties of infrared QED in $2+1$ dimensions. Recently, the
experimental observation of graphene greatly enhanced the study of
the anomalous effects induced by the pseudo-relativistic
dispersion relation of its quasi particles, see \cite{C} for an
up-to-date description of the state of art. Among the most unusual
and exciting phenomena displayed by graphene, and already
experimentally observed, let us mention the anomalous integer
quantum Hall effect and the insensitivity to localization effects
generated by disorder. It is reasonable to guess that the unique
properties of graphene will have in the next few years several
important applications in condensed matter and in
nano-technologies.

The main reason behind these anomalous effects lies in the
geometry of the Fermi surface, which at half filling is not given by a curve, 
as in usual 2D Fermi systems, but is completely degenerate: it
consists of {\it two isolated points}, as in one dimensional Fermi
systems. From a theoretical point of view, this fact completely
changes the infrared scaling properties of the propagator. It has been 
pointed out, see for instance
\cite{GGV1} and references therein, that, in the case of
short-range electron-electron interactions, all the operators with
four or more fermionic fields are irrelevant in a Renormalization
Group (RG) sense; this suggests that the interaction should not
affect too much the asymptotic behavior of the model, at least at
small coupling. It should be remarked however that such RG
analyses were performed only at a perturbative level, without any
control on the convergence of the expansion, and directly in the
relativistic approximation, consisting in replacing the 
actual dispersion relation by its linear approximation around the singularity;
such approximation implies in particular the validity of 
a continuous Lorentz $U(1)$ symmetry that is not present 
in the original model.

Aim of this paper is to present the first rigorous
construction of the low temperature and ground state properties 
of the 2D Hubbard
model on the honeycomb lattice with weak local interactions; this
is achieved by rewriting the correlation functions in terms of  
resummed series, convergent uniformly in the temperature
up to zero temperature, as we prove by making use of 
constructive fermionic renormalization group. We
show that indeed the interaction does not change the
asymptotic infrared properties of the model with respect to the
unperturbed non-interacting case, but it produces a
renormalization of the Fermi velocity and of the wave function
(note that no renormalization of the Fermi surface would be
present in the relativistic approximation). Our result rules out the
presence of superconducting or magnetic instabilities at weak
coupling; this is in striking contrast with the Hubbard model on
the square lattice, where quantum instabilities (corresponding to
the magnetic or superconducting long range order that are
presumably present in the ground state) prevent the convergence of
the perturbative expansion in $U$ for low enough temperatures. We
also prove that indeed the 2D Hubbard model on a honeycomb
lattice is asymptotically described by a QED$_{2+1}$ in the
presence of an ultraviolet cutoff, massive ``photons'' and
massless electrons; however the bare parameters of the QED theory must be
carefully chosen to include lattice effects.

\section{The Model and the Main Results}

\subsection{The model}

The grandcanonical
Hamiltonian of the 2D Hubbard model on the honeycomb lattice at half
filling in second quantized form is given by:
\bea
H_\L&=&-\sum_{\vec x\in \L\atop i=1,2,3}\sum_{\s=\uparrow\downarrow} \Big(
a^+_{\vec x,\s} b_{\vec x+\vec \d_i,\s}^-+
b_{\vec x+\vec \d_i,\s}^+ a^-_{\vec x,\s} \Big)
+\label{1.1}\\
&+&\frac{U}3 \sum_{\vec x\in \L\atop i=1,2,3}\Big[
\big(a^+_{\vec x,\uparrow} a^-_{\vec x,\uparrow} -\frac12\big)\big(
a^+_{\vec x,\downarrow}
a^-_{\vec x,\downarrow}-\frac12\big)+ \big(
b_{\vec x+\vec \d_i,\uparrow}^+b_{\vec x+\vec \d_i,
\uparrow}^--\frac12\big)\big(b_{\vec x+\vec \d_i,\downarrow}^+
b_{\vec x+\vec \d_i,\downarrow}^--\frac12\big)\Big]\nn\eea
where:
\begin{enumerate}
\item $\L$ is a periodic triangular lattice, defined as
$\L=\BBBB/L\BBBB$, where $L\in\NNN$ and $\BBBB$ is the triangular
lattice with basis
${\vec a_1}={1\over 2}(3,\sqrt{3})$, ${\vec a_2}={1\over
2}(3,-\sqrt{3})$.
\item The vectors $\vec \d_i$ are defined as
\be {\vec \d_1}=(1,0)\;,\quad {\vec \d_2}={1\over 2}
(-1,\sqrt{3})\;,\quad{\vec \d_3}={1\over
2}(-1,-\sqrt{3})\;.\label{1.2}\ee
\item
$a_{\vec x,\s}^\pm$ are creation or annihilation
fermionic operators with spin index $\s=\uparrow\downarrow$ and
site index $\vec x\in\L$, satisfying periodic boundary conditions
in $\vec x$
\item $b_{\vec x+\vec\d_i,\s}^\pm$ are creation or
annihilation
fermionic operators with spin index $\s=\uparrow\downarrow$ and
site index $\vec x+\vec\d_i\in\L+\vec\d_1$,
satisfying periodic boundary conditions in $\vec x$.
\item $U$ is the strength of the
on--site density--density interaction;
it can be either positive or negative.
\end{enumerate}

Note that the Hamiltonian (\ref{1.1}) is hole-particle symmetric,
i.e., it is invariant under the exchange $a^\pm_{\vec x,\s}\otto
a^\mp_{\vec x,\s}$, $b^\pm_{\vec x+\vec\d_1,\s}\otto-b^\mp_{\vec
x+ \vec\d_1,\s}$. This invariance implies in particular that, if
we define the average density of the system to be
$\r=(2|\L|)^{-1}\media{N}_{\b,\L}$, with $N=\sum_{\vec
x,\s}(a^+_{\vec x,\s} a^-_{\vec x,\s}+b^+_{\vec x+\vec\d_1,\s}
b^-_{\vec x+\vec\d_1,\s})$ the total particle number operator and
$\media{\cdot}_{\b,\L} =\Tr\{e^{-\b H_\L}\cdot\}/\Tr\{e^{-\b
H_\L}\}$ the average with respect to the (grandcanonical) Gibbs
measure at inverse temperature $\b$, one has $\r\=1$, for any
$|\L|$ and any $\b$. We also recall that a
theorem \cite{L} guarantees
that at half filling the ground state of (\ref{1.1}) is unique and
its total spin is equal to zero.
\\

Our goal is to characterize the low and zero temperature properties of the
system described by (\ref{1.1}), by computing thermodynamic functions
(e.g., specific free energy and specific ground state energy) and a complete
set of correlations at low or zero temperatures. To this purpose it is
convenient to introduce the notions of specific free energy
\be f_\b(U)=-\frac1\b\lim_{|\L|\to\io}|\L|^{-1}\log\Tr\{e^{-\b H_\L}\}\;,\ee
of specific ground state energy $e(U)=\lim_{\b\to\io}f_\b(U)$, and of
Schwinger functions, defined as follows.

Let us introduce the two component fermionic operators $\Psi^\pm_{\vec x,\s}=
\big(a^\pm_{\vec x,\s}, b^\pm_{\vec x+\vec\d_1,
\s}\big)$ and let us write $\Psi^\pm_{\vec x,\s,1}=a^\pm_{\vec x,\s}$ and
$\Psi^\pm_{\vec x,\s,2}=b^\pm_{\vec x+\vec\d_1,\s}$.
We shall also consider the operators $\Psi^\pm_{\xx,\s}= e^{H
x_0}\Psi^\pm_{\vec x,\s} e^{-H x_0}$ with $\xx=(x_0,\vec x)$ and
$x_0\in[0,\b]$, for some $\b>0$; we shall call $x_0$ the
time variable. We shall write $\Psi^\pm_{\xx,\s,1}=a^\pm_{\xx,\s}$
and $\Psi^\pm_{\xx,\s,2}=b^\pm_{\xx+\dd_1,\s}$, with
$\dd_1=(0,\vec\d_1)$. We define
\be S_n^{\b,\L}(\xx_1,\e_1,\s_1,\r_1;\ldots;\xx_n,\e_n,\s_n,\r_n)
=\media{\bT\{
\Psi^{\e_1}_{\xx_1,\s_1,\r_1}\cdots \Psi^{\e_n}_{\xx_n,\s_n,\r_n}\}}_{\b,\L}
\label{1.3}\ee
where: $\xx_i\in[0,\b]\times\L$, $\s_i=\uparrow\downarrow$,
$\e_i=\pm$, $\r_i=1,2$ and
$\bT$ is the operator of fermionic time ordering, acting on a
product of fermionic fields as:
\be \bT(
\Psi^{\e_1}_{\xx_1,\s_1,\r_1}\cdots \Psi^{\e_n}_{\xx_n,\s_n,\r_n})= (-1)^\p
\Psi^{\e_{\p(1)}}_{\xx_{\p(1)},\s_{\p(1)},\r_{\p(1)}}\cdots
\Psi^{\e_{\p(n)}}_{
\xx_{\p(n)},\s_{\p(n)},\r_{\p(n)}}\label{1.4}\ee
where $\p$ is a permutation of $\{1,\ldots,n\}$, chosen in such a
way that $x_{\p(1)0}\ge\cdots\ge x_{\p(n)0}$, and $(-1)^\p$ is its
sign. [If some of the time coordinates are
equal each other, the arbitrariness of the definition is solved by
ordering each set of operators with the same time coordinate so
that creation operators precede the annihilation operators.]

Taking the limit $\L\to\io$ in (\ref{1.3}) we get the finite
temperature $n$-point Schwinger functions, denoted by
$S^\b_n(\xx_1,\e_1,\s_1,\r_1; \ldots;\xx_n,\e_n,\s_n,\r_n) $,
which describe the properties of the infinite volume system at
finite temperature. Taking the $\b\to\io$ limit of the finite
temperature Schwinger functions, we get the zero temperature
Schwinger functions, simply denoted by
$S_n(\xx_1,\e_1,\s_1,\r_1;\ldots; \xx_n,\e_n,\s_n,\r_n)$, which
describe the properties of the ground state of (\ref{1.1}) in the
thermodynamic limit (note that in this case,  by the uniqueness of
the ground state proved in \cite{Lieb}, the infinite volume and zero
temperature limits commute).

\subsection{The non interacting case}

In the non--interacting case $U=0$ the Schwinger functions of any
order $n$ can be exactly computed as linear combinations of
products of two--point Schwinger functions, via the well--known
{\it Wick rule}. The two--point Schwinger function itself, also
called the {\it free propagator}, for $\xx\neq\yy$ and $\xx-\yy\neq
(\pm\b,\vec 0)$,
is equal to (see Appendix \ref{A} for details):
\bea S_0^{\b,\L}
(\xx-\yy)_{\r,\r'} &\=& S_{2}^{\b,\L}(\xx,\s,-,\r;\yy,\s,+,\r')\Big|_{U=0}
=\nn\\
&=&\lim_{M\to\io} {1\over \b|\L|}
\sum_{\kk\in\DD_{\b,L}}\frac{e^{-i\kk\cdot(\xx-\yy)}}{k_0^2+|v(\vec
k)|^2} \pmatrix{i k_0 & -v^*(\vec k) \cr -v(\vec k) &
ik_0}_{\r,\r'}\label{1.5}\eea
where:
\begin{enumerate}
\item $M\in\NNN$, $\kk=(k_0,\vec k)$ and $\DD_{\b,L}=\DD_\b\times\DD_L$; \item
$\DD_\b=\{k_0={2\pi\over \b}(n_0+{1\over 2})\;:\;n_0=-M,\ldots,M-1
\}$ and $\DD_L=\{\vec k={n_1\over L}{\vec b_1}+ {n_2\over L}{\vec
b_2}\ :\ 0\le n_1,n_2\le L-1\}$, where $\vec b_1= {2\pi\over
3}(1,\sqrt{3})$, ${\vec b_2}={2\pi\over 3}(1,-\sqrt{3})$ are a
basis of the dual lattice $\L^*$; \item $v(\vec
k)=\sum_{i=1}^3e^{i\vec k(\vec\d_i-\vec\d_1)}= 1+2 e^{-i 3/2
k_1}\cos{\sqrt{3}\over 2}k_2$; its modulus $|v(\vec k)|$ is the
{\it dispersion relation}, given by
\be |v_{\vec k}|=\sqrt{\big(1+2\cos(3k_1/2)\cos(\sqrt{3} k_2/2)\big)^2+
4\sin^2(3 k_1/2)\cos^2(\sqrt{3} k_2/2)}\;.\label{1.6}\ee

\end{enumerate}

At $\xx=\yy$ or $\xx-\yy=(\pm\b,\vec 0)$, the free propagator has a jump
discontinuity, see discussion at the end of Appendix \ref{A}.
Note that $S_0^{\b,\L}(\xx)$ is antiperiodic in $x_0$,
i.e. $S_0^{\b,\L}(x_0+\b,\vec x)=-S_0^{\b,\L}(x_0,\vec x)$,
and that its Fourier transform $\hat S_0^{\b,\L}(\kk)$ is
well--defined for any $\kk\in\DD_{\b,L}$, even in the
thermodynamic limit $L\to\io$, since $|k_0|\ge {\pi\over \b}$. We
shall refer to this last property by saying  that the inverse
temperature $\b$ acts as an infrared cutoff for our theory.

If we take $\b,L\to\io$, the limiting propagator $\hat S_0(\kk)$ becomes
singular at $\{k_0=0\}\times \{\vec k=\vec p_F^{\pm}\}$,
where
\be \vec p_{F}^{\ \pm}=({2\pi\over 3},\pm{2\pi\over 3\sqrt{3}})\label{1.7}\ee
are the {\it Fermi points} (also called {\it Dirac points}, for an analogy with
massive QED$_{2+1}$ that will become clearer below). Note that the
asymptotic behavior of $v(\vec k)$ close to the Fermi points is given by
$v(\vec p_F^\pm+\vec k')\simeq \frac32(i k_1'\pm k'_2)$. In particular,
if $\o=\pm$, the Fourier transform of the 2-point Schwinger function
close to the Fermi point $\vec p_F^\o$
can be rewritten in the form:
\be\hat S_{0}(k_0,\vec p_F^\o+\vec k')={1\over Z_0}\;
\pmatrix{-i k_0 & -v_F^{(0)}(-ik_1'+\o k_2') +r_\o(\vec k')\cr -v_F^{(0)}
(ik_1'+\o k_2') +
r_\o^*(\vec k')& -ik_0}^{-1}\;,\label{1.8}\ee
where $Z_0=1$ is the {\it
free wave function renormalization} and $v_F^{(0)}=3/2$
is the {\it free Fermi velocity}. Moreover,
$|r_\o(\vec k')|\le C \big|\vec k'|^2$, for small values of $\vec k'$
and for some positive constant $C$.

\subsection{The interacting case}

We are now interested in what happens by adding a local interaction.
In the case $U\neq 0$, the Schwinger functions are not
exactly computable anymore. It is well--known that they can be
written as formal power series in $U$, constructed
in terms of {\it Feynmann diagrams}, using as free propagator the
function $S_0(\xx)$ in (\ref{1.5}). Our main result consists in a proof
of convergence of this formal expansion for $U$ small enough, after
the implementation of suitable resummations of the original power series.
Our main result can be
described as follows. \\

{\bf Theorem 1.} {\it Let us consider the 2D Hubbard model on the
honeycomb lattice at half filling, defined by (\ref{1.1}). There
exist a constant $U_0>0$ such that, if $|U|\le U_0$, the specific
free energy $f_\b(U)$ and the finite temperature Schwinger
functions are analytic functions of $U$, uniformly in $\b$ as
$\b\to\io$, and so are the specific ground state energy $e(U)$ and
the zero temperature Schwinger functions. The Fourier transform of
the zero temperature two point Schwinger function $S(\xx)_{\r,\r'}\defin
S_2(\xx,\s,-,\r;\V0,\s,+,\r')$, denoted by $\hat S(\kk)$, is
singular only
at the Fermi points $\kk=\pp_F^{\pm}=(0,\vec p_F^{\,\pm})$, see
(\ref{1.7}), and, close to the
singularities, if $\o=\pm$, it can be written as
\be \hat S(k_0,\vec p_F^{\;\o}+\vec k')= \frac1{Z}
\pmatrix{-i k_0 & -v_F(-ik_1'+\o k_2')\cr -v_F
(ik_1'+\o k_2') & -ik_0}^{-1} \Big(\openone + R(\kk')\Big)\;,\label{1.9}\ee
with $\kk'=(k_0,\vec k')$, and with $Z$ and $v_F$ two real constants such that
\be Z=1+ a U^2+O(U^3)\;,\qquad\qquad v_F=\frac32 +
b U^2+O(U^3)\label{1.10}\ee
where $a$ and $b$ are non-vanishing constants. Moreover the matrix
$R(\kk')$ satisfies $||R(\kk')||\le C |\kk'|^\vartheta$ for some
constants $C, \vartheta>0$ and for $|\kk'|$ small enough.}
\\

{\bf Remarks.}\\
1) Theorem 1 says that the location of the singularity does not
change in the presence of interaction; on the contrary, the wave
function renormalization and Fermi velocity are modified by the
interaction. Note also that, in the presence of the interaction, 
the Fermi velocity
remains the same in the two coordinate direction even though the
model does not display $90^o$ discrete rotational symmetry, but
rather a $120^o$ rotational symmetry.
\\
2) The resulting theory is not quasi-free: the Wick rule is not valid
anymore in the presence of interactions. However, the long distance asymptotics
of the higher order Schwinger functions can be estimated by the same methods
used to prove Theorem 1, and it is the same suggested by the Wick rule.\\
3) The fact that the interacting correlations decay as in the non-interacting
case implies in particular the absence of
long range order at zero temperature,
e.g., the absence of N\'eel order in the ground state at weak coupling.
In fact, as a corollary of our construction, we find:
\be \Big|\lim_{\b,|\L|\to\io}
\media{\vec S_{\vec x}\cdot \vec S_{\vec y}}_{\b,\L}\Big|\le C
\frac1{|\vec x-\vec y|^4}\;,\label{1.11}\ee
where, if $\vec x\in\L$, the spin operator $\vec S_{\vec x}$ is
defined as: $\vec S_{\vec x} =a^+_{\vec x,\cdot} \vec \s\,
a^-_{\vec x,\cdot}$, with $\s_i$, $i=1,2,3$, the Pauli matrices;
similarly, if $\vec x\in\L+\vec \d_1$, $\vec S_{\vec x} =\sum_{\s}
b^+_{\vec x,\cdot} \vec \s b^-_{\vec x,\cdot}$. Note that it is
well known that the ground state has zero total spin \cite{Lieb},
however existence of N\'eel order was neither proven nor ruled out
by the results in \cite{Lieb}.\\ 
4) Similarly to what remarked in
the previous item, one can exclude the existence of
superconducting long range order: the Cooper pairs correlations
decay to zero at infinity at least as fast
as the spin-spin correlations in (\ref{1.11}).\\
5) Our analysis can be extended in a straightforward way to the
case of exponentially decaying interactions (instead of local
interactions). However, if the decay is
slower, the result may change. In particular, in the presence of
3D Coulomb interactions, the electron-electron interaction becomes
marginal (instead of irrelevant), in a renormalization group
sense \cite{GGV2}.
\\
6) Previous analyses of the Hubbard model on the honeycomb lattice
were performed only at a perturbative level, without any control
on the convergence of the weak coupling expansion, and directly in
the Quantum Field Theory approximation, consisting in the replacement of $\hat
S_0(\kk)$ by its linear approximation around the Fermi points,
see for instance \cite{GGV1} and references therein.
\\
7) In Appendix \ref{C} we prove that indeed the 2D Hubbard
model on a honeycomb lattice is asymptotically described by a
QED$_{2+1}$ model in the presence of an ultraviolet cutoff, massive
``photons'' and massless electrons, provided that the bare
parameters of the QED$_{2+1}$ are carefully chosen to include
lattice effects. As a result, the correlations asymptotically
verify a modified Ward Identity (WI) 
related to an approximate local $U(1)$ Lorentz 
symmetry: note, however, that the renormalized charge velocity appearing in the
modified WI for graphene explicitly breaks rotational invariance, contrary
to what happens to the Fermi velocity, or to the charge velocity of 
a pure relativistic QED model.\\

The proof of the Theorem is based on constructive fermionic
Renormalization Group (RG) methods, see
\cite{BG,M2,Salmhofer} for extensive reviews.
It is worth remarking that the result summarized in
Theorem 1 is one of the few rigorous construction of the ground state
properties (including correlations)
of a weak coupling 2D Hubbard model. The only other example we are aware
of is the Fermi liquid construction in \cite{FKT}, applicable to cases
of weakly interacting 2D Fermi systems with a highly asymmetric
interacting Fermi surface. Related results include
the construction of the state
at temperatures larger than a BCS-like critical temperature
\cite{DR,BGM},
or the computation of the first contribution to the ground state
energy in a weak coupling limit \cite{LSS,G,SY}.

The rest of the paper will be devoted to the proof of Theorem 1. In Sec.
\ref{IIIa} we review the Grassmann integral representation for
the free energy and the Schwinger functions. In Sec.\ref{IIIb} we start
to describe the integration procedure leading to the computation of the
free energy, and in particular we describe how to integrate out the
ultraviolet degrees of freedom. In Sec.\ref{IIIc} we complete the proof of
convergence of the series for the free energy and the ground state energy.
In Sec.\ref{IIId} we describe the proof of convergence for the series for the
Schwinger functions, with particular emphasis to the case of the two-point
Schwinger function. In the Appendices we provide further details concerning the
non-interacting theory, the ultraviolet integration and the
equivalence (as far as the long distance behavior is concerned) between
the Hubbard model and a massive QED theory in 2+1 dimensions.

\section{Renormalization Group Analysis}

\subsection{Grassmann Integration}\label{IIIa}
It is well--known that the usual formal power series in $U$ for
the partition function and for the Schwinger functions of model
(\ref{1.1}) can be equivalently rewritten in terms of Grassmann
functional integrals, defined as follows.

We consider the
Grassmann algebra generated by the Grassmannian variables
$\{\hat\Psi^\pm_{\kk,\s,\r}\}_{ \kk \in
\DD_{\b,L}}^{\s=\uparrow\downarrow,\ \r=1,2}$ and a Grassmann
integration $\int
\big[\prod_{\kk\in\DD_{\b,L}}\prod_{\s=\uparrow\downarrow}^{\r=1,2}
d\hat\Psi_{\kk,\s,\r}^+ d\hat\Psi_{\kk,\s,\r}^-\big]$ defined as
the linear operator on the Grassmann algebra such that, given a
monomial $Q( \hat\Psi^-, \hat\Psi^+)$ in the variables
$\hat\Psi^\pm_{\kk,\s,\r}$, its action on $Q(\hat\Psi^-,
\hat\Psi^+)$ is $0$ except in the case $Q(\hat\Psi^-,
\hat\Psi^+)=\prod_{\kk\in\DD_{\b,L}}
\prod_{\s=\uparrow\downarrow}^{\r=1,2} \hat\Psi^-_{\kk,\s,\r}
\hat\Psi^+_{\kk,\s,\r}$, up to a permutation of the variables. In
this case the value of the integral is determined, by using the
anticommuting properties of the variables, by the condition
\be \int \Big[\prod_{\kk\in\DD_{\b,L}}\prod_{\s=\uparrow\downarrow}^{\r=1,2}
d\hat\Psi_{\kk,\s,\r}^+
d\hat\Psi_{\kk,\s,\r}^-\Big]\prod_{\kk\in\DD_{\b,L}}
\prod_{\s=\uparrow\downarrow}^{\r=1,2}
\hat\Psi^-_{\kk,\s,\r} \hat\Psi^+_{\kk,\s,\r}=1\label{2.1}\ee
Defining the free propagator matrix $\hat g_\kk$ as
\be \hat g_\kk=\pmatrix{-i k_0 & -v^*(\vec k) \cr -v(\vec k) & -i k_0}^{-1}
\label{2.2}\ee
and the ``Gaussian integration'' $P(d\Psi)$ as
\bea P(d\Psi) &=& \Big[\prod_{\kk\in\DD_{\b,L}}
^{\s=\uparrow\downarrow}\frac{-\b^2|\L|^2}{k_0^2+|v(\vec k)|^2}
d\hat\Psi_{\kk,\s,1}^+
d\hat\Psi_{\kk,\s,1}^-d\hat\Psi_{\kk,\s,2}^+
d\hat\Psi_{\kk,\s,2}^-\Big]\cdot\nn\\&&\hskip3.truecm
\cdot\;\exp \Big\{-(\b|\L|)^{-1}
\sum_{\kk\in\DD_{\b,L}}^{\s=\uparrow\downarrow}
\hat\Psi^{+}_{\kk,\s,\cdot}\,{\hat g}_\kk^{-1}\hat\Psi^{-}_{\kk,\s,\cdot}
\Big\}\;,
\label{2.3}\eea
it turns out that
\be\int P(d \Psi) \hat \Psi^-_{\kk_1,\s_1,\r_1}\hat \Psi^+_{\kk_2,\s_2,\r_2} =
\b|\L|\d_{\s_1,\s_2}\d_{\kk_1,\kk_2} \big[\hat g_{\kk_1}\big]_{\r_1,\r_2}
\;,\label{2.4}\ee
so that, if $\xx-\yy\not\in\b\ZZZ\times\{\vec 0\}$,
\be\lim_{M\to\io} \fra{1}{\b|\L|}\sum_{\kk\in\DD_{\b,L}}
e^{-i\kk(\xx-\yy)}\hat g_{\kk}= \lim_{M\to\io} \int
P(d\Psi)\Psi^-_{\xx,\s}\Psi^+_{\yy,\s}= S_0(\xx-\yy)\;,\label{2.5}\ee
where $S_0(\xx-\yy)$ was defined in (\ref{1.4}) and the Grassmann
fields $\Psi_{\xx,\s}^\pm$ are defined by
\be\Psi_{\xx,\s,\r}^{\pm}=\fra{1}{\b|\L|}\sum_{\kk\in\DD_{\b,L}}
e^{\pm i\kk\xx}\hat\Psi^\pm_{\kk,\s,\r}\;,\qquad \xx\in\L_{\b,M}\times\L\;,
\label{2.6}\ee
with $\L_{\b,M}=\{m\b/M : m=-M,\ldots,M-1\}$. Let us now consider
the function on the Grassmann algebra
\bea V(\Psi)
&=&U\sum_{\r=1,2}\int d\xx \, \Psi^+_{\xx,\uparrow,\r}
\Psi^-_{\xx,\uparrow,\r}\Psi^+_{\xx,\downarrow,\r}
\Psi^-_{\xx,\downarrow,\r}=\nn\\
&=& \frac{U}{(\b|\L|)^3}\sum_{\r=1,2}\;\sum_{\kk,\kk',\pp}\;
\hat \Psi^+_{\kk-\pp,\uparrow,\r}
\hat \Psi^-_{\kk,\uparrow,\r}\hat \Psi^+_{\kk'+\pp,\downarrow,\r}
\hat \Psi^-_{\kk',\downarrow,\r}\;,
\label{2.6xyz}
\eea
where, in the first line, the symbol $\int d\xx$ must be interpreted as
\be \int d\xx=\frac{\b}{2M}\sum_{x_0\in\L_{\b,M}}\sum_{\vec x\in\L}\;,
\label{2.7c}\ee
and, in the second line, the sums over $\kk,\kk'$ run over the set
$\DD_{\b,L}$, while the sums over $\pp$ run over the set $2\p\b^{-1}
\ZZZ\times
\DD_L$ ($\pp$ is the {\it transferred momentum}).
Note that the integral $\int P(d\Psi)e^{-V(\Psi)}$ is well
defined for any $U$; it is indeed a polynomial in $U$, of degree
depending on $M$ and $L$. Standard arguments
show that, if there exists the limit of $\int P(d\Psi)e^{-V(\Psi)}$ as
$M\to\io$, then the normalized partition function can be written
as
\be e^{-\b|\L| F_{\b,L}}\defin\fra{\Tr[e^{-\b H_\L}]}{\Tr[e^{-\b
H_0}]}= \lim_{M\to\io} \int P(d\Psi)e^{-V(\Psi)}\label{2.6a}\ee
where $H_0$ is equal to (\ref{1.1}) with $U=0$. A possible way to
prove (\ref{2.6a}) is to compare the perturbation theory obtained
by expanding in powers of $U$ via Trotter's product formula the
trace $\Tr\{e^{-\b H_\L}\}/\Tr\{e^{-\b H_0}\}$ with the one
obtained by expanding in $U$ the Grassmann functional integral,
and then show that they are the same, order by order, see
\cite{BG-jsp}. This proof also shows that the correct choice of
the interaction (\ref{2.6xyz}) expressed in Grassmann variables
does not include terms bilinear in the fields, contrary to the
interaction in second quantized form, see (\ref{1.1}): in fact,
with this choice, in both perturbative expansions the "tadpoles"
are exactly vanishing, as required by the condition that the
system is at half filling.

Similarly, the Schwinger functions {\it at distinct space-time
points}, defined in (\ref{1.3}), can be computed as
\be S(\xx_1,\s_1,\e_1,\r_1;\ldots;\xx_n,\s_n,\e_n,\r_n)=
\lim_{M\to\io} { \int P(d\Psi) e^{-V(\Psi)}
\Psi^{\e_1}_{\xx_1,\s_1,\r_1}\cdots\Psi^{\e_n}_{\xx_n,\s_n,\r_n}
\over \int P(d\Psi) e^{-V(\Psi)}}\;.\label{2.8}\ee
Note that the limit $\xx_i-\xx_j\to{\bf 0}$ and the limit $M\to\io$
do not commute in general.

In the following we shall study the functional integrals by
introducing suitable expansions where the value of $M$ plays no essential role
and we shall indeed be able to control such expansions {\it
uniformly in $M$, if $U$ is small enough} and that 
the limit $M\to\io$ can be taken in the resulting expressions
for the free energy and the Schwinger functions. 
For this reason, from now on we shall not stress anymore the dependence on $M$,
unless for the cases where the presence of a finite $M$ is relevant, e.g., 
for the analysis of the ultraviolet integration described in Appendix \ref{A}.

It is important to note that
both the Gaussian integration $P(d\Psi)$ and the
interaction $V(\Psi)$ are invariant under the action of a number
of remarkable symmetry transformations,
which will be preserved by the subsequent
iterative integration procedure and will guarantee
the vanishing of some running coupling constants (see below for details).
Let us collect in the following lemma
all the symmetry properties we will need in the following.\\

{\bf Lemma 1.} {\it For any choice of $M,\b,\L$, both the
quadratic
Grassmann measure $P(d\Psi)$ defined in (\ref{2.3}) and the quartic
Grassmann interaction $V(\Psi)$ defined in (\ref{2.6xyz})
are invariant under the following transformations:\\
\noindent(1) \ul{spin exchange}: $\hat \Psi^\e_{\kk,\s,\r}\otto
\hat\Psi^\e_{\kk,-\s,\r}$;\\
\noindent(2) \ul{global
$U(1)$}: $\hat\Psi^\e_{\kk,\s,\r}\to e^{i\e\a_\s}\hat\Psi^\e_{\kk,\s,\r}$, with
$\a_\s\in \RRR$ independent of $\kk$;\\
\noindent(3)  \ul{ spin $SO(2)$}:
$\pmatrix{\hat\Psi^\e_{\kk,\uparrow,\r}\cr
\hat\Psi^\e_{\kk,\downarrow,\r}\cr}\to R_\th\pmatrix{\hat\Psi^\e_{
\kk,\uparrow,\r}\cr
\hat\Psi^\e_{\kk,\downarrow,\r}\cr}$, with
$R_\th=\pmatrix{\cos\th & \sin\th \cr
-\sin\th & \cos\th \cr}$ and $\th\in\TTT$ independent of
$\kk$;\\
\noindent(4)
 \ul{discrete spatial rotations}: $\hat \Psi_{(k_0,\vec k),\s,\r}^\pm\to
e^{\mp i\vec k(\vec \d_3
-\vec \d_1)(\r-1)}\hat \Psi_{(k_0,T_1\vec k),\s,\r}^\pm$,
with $T_1\vec x\defin R_{2\p/3}\vec x$; note that in real space this
transformation simply reads $a^\pm_{(x_0,\vec x),\s}\to
a^\pm_{(x_0,T_1\vec x),\s}$ and
$b^\pm_{(x_0,\vec x),\s}\to
b^\pm_{(x_0,T_1\vec x),\s}$;\\
\noindent(5) \ul{complex conjugation}: $\hat \Psi^{\pm}_{\kk,\s,\r}
\to \hat\Psi^\pm_{-\kk,\s,\r}$, $c\to c^*$, where $c$ is a generic
constant appearing in $P(d\Psi)$ and/or in $V(\Psi)$;\\
\noindent(6.a) \ul{horizontal reflections}: $\hat\Psi^{\pm}_{(k_0,k_1,k_2),
\s,1}
\otto \hat\Psi^\pm_{(k_0,-k_1,k_2),\s,2}$;\\
\noindent(6.b) \ul{vertical reflections}: $\hat\Psi^{\pm}_{(k_0,k_1,k_2),\s,\r}
\to \hat\Psi^\pm_{(k_0,k_1,-k_2),\s,\r}$;\\
\noindent(7) \ul{particle-hole}: $\hat\Psi^{\pm}_{(k_0,\vec k),\s,\r}\to
i\hat\Psi^{\mp}_{(k_0,-\vec k),\s,\r}$.\\
\noindent(8) \ul{inversion}: $\hat\Psi^{\pm}_{(k_0,\vec k),\s,\r}\to
i(-1)^\r\hat\Psi^{\pm}_{(-k_0,\vec k),\s,\r}$.}
\\

{\cs Proof.} A moment's thought shows that the invariance of $V(\Psi)$
under the above symmetries is obvious, and so is the invariance of $P(d\Psi)$
under (1)-(2)-(3). Let us then prove the invariance of $P(d\Psi)$ under
(4)-(5)-(6.a)-(6.b)-(7)-(8). More precisely, let us consider the term
\bea&&
\sum_\kk
\hat\Psi^{+}_{\kk,\s,\cdot}\,{\hat g}_\kk^{-1}\hat\Psi^{-}_{\kk,\s,\cdot}=
\label{2.8a}\\
&&-i\sum_\kk\hat\Psi^{+}_{\kk,\s,1}k_0\hat\Psi^{-}_{\kk,\s,1}-
\sum_\kk\hat\Psi^{+}_{\kk,\s,1}v^*(\vec k)\hat\Psi^{-}_{\kk,\s,2}
-\sum_\kk\hat\Psi^{+}_{\kk,\s,2}v(\vec k)\hat\Psi^{-}_{\kk,\s,1}
-i\sum_\kk\hat\Psi^{+}_{\kk,\s,2}k_0\hat\Psi^{-}_{\kk,\s,2}\nn\eea
in (\ref{2.3}), and let us prove its invariance under the transformations
(4)-(5)-(6.a)-(6.b)-(7)-(8).

Under the transformation (4), the first and fourth term
in the second line of (\ref{2.8a}) are obviously invariant,
while the sum of the second and third is changed into
\bea&& -
\sum_\kk\Big[
\hat\Psi^{+}_{(k_0,T_1\vec k),\s,1}v^*(\vec k)e^{+i\vec k(\vec\d_3-
\vec\d_1)}
\hat\Psi^{-}_{(k_0,T_1\vec k),\s,2}
+\hat\Psi^{+}_{(k_0,T_1\vec k),\s,2}
e^{-i\vec k(\vec\d_3-
\vec\d_1)}
v(\vec k)\hat\Psi^{-}_{(k_0,T_1\vec k),\s,1}\Big]=\nn\\
&&=-\sum_\kk\Big[\hat\Psi^{+}_{\kk,\s,1}v^*(T_1^{-1}\vec k)e^{+i\vec k
(\vec\d_1-
\vec\d_2)}
\hat\Psi^{-}_{\kk,\s,2}+\hat\Psi^{+}_{\kk,\s,2}
e^{-i\vec k(\vec\d_1-
\vec\d_2)}
v(T_1^{-1}\vec k)\hat\Psi^{-}_{\kk,\s,1}\Big]\;.\label{2.8b}
\eea
Using that $v(T_1^{-1}\vec k)=e^{i\vec k(\vec\d_1-\vec\d_2)}v(\vec k)$, as
it follows by the definition $v(\vec k)=\sum_{i=1,2,3}e^{i\vec k(\vec\d_i-
\vec\d_1)}$, we find that the last line of (\ref{2.8b}) is equal to
the sum of the second and third term in (\ref{2.8a}), as desired.

The invariance of (\ref{2.8a}) under the transformation (5) is very simple,
if one notes that $v(-\vec k)=v^*(\vec k)$, as it follows by the definition of
$v(\vec k)$.

Under the transformation (6.a), the sum of the first and fourth term
in the second line
of (\ref{2.8a}) is obviously invariant, while the sum of the second and third
is changed into
\bea && -
\sum_\kk\hat\Psi^{+}_{(k_0,-k_1,k_2),\s,2}v^*(\vec k)\hat\Psi^{-}_{
(k_0,-k_1,k_2),\s,1}
-\sum_\kk\hat\Psi^{+}_{(k_0,-k_1,k_2),\s,1}v(\vec k)\hat\Psi^{-}_{
(k_0,-k_1,k_2),\s,2}=\nn\\
&&=-
\sum_\kk\hat\Psi^{+}_{\kk,\s,2}v^*((-k_1,k_2))\hat\Psi^{-}_{\kk,\s,1}
-\sum_\kk\hat\Psi^{+}_{\kk,\s,1}v((-k_1,k_2))\hat\Psi^{-}_{\kk,\s,2}
\;.\eea
Noting that $v((-k_1,k_2))=v^*(\kk)$, one sees that this is the same as
the sum of the second and third term in (\ref{2.8a}), as desired.

Similarly, noting that $v((k_1,-k_2))=v(\kk)$, one finds that (\ref{2.8a})
is invariant under the transformation (6.b).

Under the transformation (7), the sum of the first and fourth
term in (\ref{2.8a}) is obviously invariant,
while the sum of the second and third term is changed into
\bea && +
\sum_\kk\hat\Psi^{-}_{(k_0,-\vec k),\s,1}v^*(\vec k)\hat\Psi^{+}_{
(k_0,-\vec k),\s,2}+
\sum_\kk\hat\Psi^{-}_{(k_0,-\vec k),\s,2}v(\vec k)\hat\Psi^{+}_{
(k_0,-\vec k),\s,1}=\nn\\
&&=-
\sum_\kk\hat\Psi^{+}_{\kk,\s,2}v^*(-\vec k)\hat\Psi^{-}_{\kk,\s,1}
-\sum_\kk\hat\Psi^{+}_{\kk,\s,1}v(-\vec k)\hat\Psi^{-}_{\kk,\s,2}
\;.\eea
Using, again, that $v(-\vec k)=v^*(\vec k)$, we see that the latter sum is
the same as the sum of the second and third term in (\ref{2.8a}), as desired.

Finally, under the transformation (8), all the terms in the right hand side of
(\ref{2.8a}) are separately invariant, and the proof of Lemma 1 is concluded.
\qed

\subsection{Free energy: The ultraviolet integration}\label{IIIb}
We start by studying the partition function
\be \Xi_{\b,L}=e^{-\b|\L|F_{\b,L}}=\int P(d\Psi)e^{-V(\Psi)}\;.\label{2.9}\ee
Note that our lattice model has an intrinsic ultraviolet cut-off in the
$\vec k$ variables, while the $k_0$ variable is unbounded. A preliminary
step to our infrared analysis is the integration of the
ultraviolet degrees of freedom corresponding to the large values
of $k_0$. We proceed in the following way. We decompose the free
propagator $\hat g_\kk$ into a sum of two propagators supported in
the regions of $k_0$ ``large'' and ``small'', respectively. The
regions of $k_0$ large and small are defined in terms of a smooth
support function $\chi_0(t)$ which is $1$ for $t\le a_0$ and
$0$ for $t\ge a_0 \g$, $\g>1$; $a_0$ is chosen so that the support
of $\chi_0\Big(\sqrt{k_0^2+|\vec k-\vec p_F^+|^2}\; \Big)$ and
$\chi_0\Big(
\sqrt{k_0^2+|\vec k-\vec p_F^-|^2}\;\Big)$ are disjoint (here $|\cdot|$
is the euclidean norm over $\RRR^2/\L^*$). In order for this condition
to be satisfied, it is enough that $2a_0\g<4\p/(3\sqrt{3})$; in the following,
for reasons that will become clearer later, we shall assume the slightly
more restrictive condition $2a_0\g<4\p/3-4\p/(3\sqrt3)$. We define
\be f_{u.v.}(\kk)=1-\chi_0\Big(\sqrt{k_0^2+|\vec k-\vec p_F^+|^2}\; \Big)
-\chi_0\Big(
\sqrt{k_0^2+|\vec k-\vec p_F^-|^2}\;\Big)\label{2.10}\ee
and $f_{i.r.}(\kk)=1-f_{u.v.}(\kk)$, so that we can rewrite $\hat g_\kk$ as:
\be \hat g_\kk=f_{u.v.}(\kk)\hat g_\kk+ f_{i.r.}(\kk)\hat g_\kk\defin
\hat g^{(u.v.)}(\kk)+\hat g^{(i.r.)}(\kk)\;.\label{2.11}\ee
We now introduce two independent set of Grassmann fields $\{
\Psi^{(u.v.)\pm}_{\kk,\s,\r}\}$ and $\{
\Psi^{(i.r.)\pm}_{\kk,\s,\r}\}$, with $\kk\in\DD_{\b,L}$,
$\s=\uparrow\downarrow$, $\r=1,2$,
and the Gaussian integrations $P(d\Psi^{(u.v.)})$ and
$P(d\Psi^{(i.r.)})$ defined by
\bea &&
\int P(d \Psi^{(u.v.)}) \hat \Psi^{(u.v.)-}_{\kk_1,\s_1,\r_1}\hat
\Psi^{(u.v.)+}_{ \kk_2,\s_2,\r_2} = \b|\L|\d_{\s_1,\s_2}\d_{\kk_1,\kk_2}
\hat g^{(u.v.)}(\kk_1)_{\r_1,\r_2}\;,\nn\\
&&\int P(d \Psi^{(i.r.)}) \hat \Psi^{(u.v.)-}_{\kk_1,\s_1,\r_1}\hat
\Psi^{(i.r.)+}_{ \kk_2,\s_2,\r_2} = \b|\L|\d_{\s_1,\s_2}\d_{\kk_1,\kk_2}
\hat g^{(i.r.)}(\kk_1)_{\r_1,\r_2}\;.\label{2.12}\eea
Similarly to $P(d\Psi)$, the Gaussian integrations $P(d
\Psi^{(u.v.)})$, $P(d \Psi^{(i.r.)})$ also admit an explicit
representation analogous to (\ref{2.2}), with $\hat g_\kk$ replaced
by $\hat g^{(u.v.)}(\kk)$ or $\hat g^{(i.r.)}(\kk)$ and the sum
over $\kk$ restricted to the values in the support of $f_{u.v.}(\kk)$ or
$f_{i.r.}(\kk)$, respectively. It easy to verify that the
ultraviolet propagator $g^{(u.v.)}(\xx-\yy)=(\b|\L|)^{-1}\sum_{\kk
\in\DD_{\b,L}}e^{-i\kk(\xx-\yy)}\hat g^{(u.v.)}(\kk)$ satisfies
\be |g^{(u.v.)}(\xx-\yy)|\le {C_N\over 1+|\xx-\yy|^N}\;.\label{2.12a}\ee
The definition of Grassmann integration
implies the following identity (``addition principle''):
\be \int P(d\Psi)e^{-V(\Psi)}=\int P(d\Psi^{(i.r.)})\int P(d\Psi^{(u.v.)})
e^{-V(\Psi^{(i.r.)}+\Psi^{(u.v.)})}\label{2.13}\ee
so that we can rewrite the partition function as
\bea \Xi_{\b,L}&=& e^{-\b|\L| F_{L,\b}}=
\int P(d\Psi^{(i.r.)})\exp\,\big\{\, \sum_{n\ge
1}\fra{1}{n!}\EE^T_{u.v.}(-V(\Psi^{(i.r.)}+\cdot);n)\big\}\=\nn\\
&\=&
e^{-\b |\L| F_0}\int P(d\Psi^{(i.r.)}) e^{-{\cal V}(\Psi^{(i.r.)})}\;,
\label{2.14}\eea
where the {\it truncated expectation} $\EE^T_{u.v.}$ is defined, given
any polynomial $V_1(\Psi^{(u.v.)})$ with coefficients depending on
$\Psi^{(i.r.)}$, as
\be \EE^T_{u.v.}(V_1(\cdot);n)=\fra{\dpr^n}{\dpr\l^n}
\log\int P(d\Psi^{(u.v.)})e^{\l
V_1(\Psi^{(u.v.)})}\Big|_{\l=0}\label{2.15}\ee
and ${\cal V}$ is fixed by the condition ${\cal V}(0)=0$.
It can be shown (see discussion after (\ref{2.17}) and Appendix \ref{B})
that ${\cal V}$ can be written as
\bea && {\cal V}(\Psi)=\sum_{n=1}^\io (\b|\L|)^{-2n}
\sum_{\s_1,\ldots,\s_n=\uparrow
\downarrow}\sum_{\r_1,\ldots,\r_{2n}=1,2}\sum_{\kk_1,\ldots,\kk_{2n}}
\Big[\prod_{j=1}^n\hat \Psi^{(i.r.)+}_{\kk_{2j-1},\s_j,\r_{2j-1}}
\hat \Psi^{(i.r.)-}_{\kk_{2j},\s_j,\r_{2j}}\Big]\;\cdot\nn\\
&&\hskip5.truecm\cdot\hat W_{2n,\ul\r}(\kk_1,\ldots,
\kk_{2n-1})\;\d(\sum_{j=1}^n(\kk_{2j-1}-\kk_{2j}))\;,\label{2.16}\eea
where $\ul\r=(\r_1,\ldots,\r_{2n})$ and we used the notation
\be \d(\kk)=\d(\vec k)\d(k_0)\;,\qquad \d(\vec k)=|\L|\sum_{n_1,n_2\in\zzz}
\d_{\vec k,n_1\vec b_1+n_2\vec b_2}\;,\qquad \d(k_0)=\b\d_{k_0,0}\;,
\label{2.16a}\ee
with $\vec b_1,\vec b_2$ a basis of $\L^*$.
The possibility of representing ${\cal V}$ in the form
(\ref{2.16}), with the {\it kernels} $\hat W_{2n,\ul\r}$ independent of the
spin indices $\s_i$, follows from the symmetries listed in Lemma 1
and from the remark that $P(d\Psi^{(u.v.)})$ and
$P(d\Psi^{(i.r.)})$ are separately invariant under the same symmetries.

The constant $F_0$ in (\ref{2.14}) and the kernels $\hat W_{2n,\ul\r}$ in
(\ref{2.16}) are
given by power series in $U$, convergent
under the condition $|U|\le U_0$, for $U_0$ small enough; after Fourier
transform, the $\xx$-space counterparts of the kernels $\hat W_{2n,\ul\r}$
satisfy the following bounds:
\be \int d\xx_1\cdots d\xx_{2n}\Big[\prod_{1\le i<j\le 2n}
|\xx_i-\xx_j|^{m_{i,j}}\Big]\big|W_{2n,\ul\r}(\xx_1,\ldots,\xx_{2n})\big|
\le \b|\L|C_{m}^n |U|^{\max\{1,n-1\}}\;,\label{2.17}\ee
for some constant $C_m>0$, where $m=\sum_{1\le i<j\le 2n}m_{i,j}$.

The proof of convergence of the power series defining $F_0$ and
$W_{2n,\ul\r}$, as well as the proof of the bounds (\ref{2.17}),
uses the decay property (\ref{2.12a}) combined with standard
fermionic cluster expansion methods. The proof is much simpler
than the infrared integration that we shall study below, and is
based on similar ideas; see Appendix \ref{B} for a proof. Note
that the decay (\ref{2.12a}) suggests the possibility of a single
scale integration of the ultraviolet degrees of freedom. However,
the discontinuity of the propagator at $x_0-y_0=0$ implies that
$\hat g_\kk$ does not admit a Gram representation \cite{PS}
and this fact prevents the direct implementation of a single step fermionic
cluster expansion (see next section for the notion of
Gram determinant and for a description of the use of Gram
determinants in the infrared multiscale integration). A possible
way out of this problem is to decompose the ultraviolet propagator
as a sum of propagators, each admitting a Gram representation, and
to perform a simple multiscale analysis of the ultraviolet
problem. This strategy was described many times before in the
literature, see for instance \cite{GK,L,BGPS94,BM1,BGM}; for
completeness, it will be presented in a self-contained form in
Appendix \ref{B}. Recently, a different proof based on a single
scale integration step and using improved bounds on determinants
associated to ``chronological products''
was proposed \cite{PS}.\\

It is important for the incoming discussion to note that the symmetries
listed in Lemma 1 also imply some non trivial invariance
properties of the kernels. We will be particularly interested in the invariance
properties of the quadratic part $\hat W_{2,(\r_1,\r_2)}(\kk)$, which will
be used below to show that the structure of the
quadratic part of the new effective interaction has the same symmetries
as the free integration. The crucial properties that we will need are the
following.
\\

{\bf Lemma 2.} {\it Let $\hat W_{aa}(\kk)\= \hat
W_{2,(1,1)}(\kk)$, $\hat W_{b\,b}(\kk)= \hat W_{2,(2,2)}(\kk)$,
$\hat W_{ab}(\kk)=\hat W_{2,(1,2)}(\kk)$
and $\hat W_{ba}(\kk)=\hat W_{2,(2,1)}(\kk)$. Then the following properties
are valid:\\
(i) $W_{aa}(\kk)=W_{bb}(\kk)$ and $W_{ab}(\kk)=W^*_{ba}(\kk)$;\\
(ii) as $\b\to\io$, for $\o=\pm$, $W_{aa}(0,\vec p_F^\o)=W_{ab}(0,\vec p_F^\o)
=0$;\\
(iii) as $\b,|\L|\to\io$, for $\o=\pm$,
\bea && \partial_{\vec k} \hat W_{aa}(0,\vec
p_F^{\;\o})=\vec 0\;,\qquad {\rm Re}\big\{
\partial_{k_0}\hat W_{aa}(0,\vec p_F^{\;\o})\big\}=0\;,
\qquad \partial_{k_0} \hat
W_{ab}(0,\vec p_F^{\;\o})=0\;,
\label{2.21}\\
&& {\rm
Re}\big\{\partial_{k_1}\hat W_{ab}(0,\vec p_F^{\;\o})\big\}= {\rm
Im}\big\{\partial_{k_2}\hat W_{ab}(0,\vec p_F^{\;\o})\big\}=0\;,
\qquad i\partial_{k_1} \hat W_{ab}(0,\vec p_F^{\;\o})=\o\partial_{k_2}
\hat W_{ab}(0,\vec p_F^{\;\o})\;. \nn\eea}

{\bf Remarks.}\\
1) For simplicity, the properties (ii) and (iii)
are spelled out only in the zero temperature limit and in
the thermodynamic limit; however, as it will be clear from the proof,
those properties all have a finite
temperature/volume counterpart.\\
2) Lemma 2 implies that in the vicinity of the Fermi points the kernel
$W_{2,(\r,\r')}(\kk)$ can be rewritten in the form
\be W_{2,(\r,\r')}(k_0,\vec p_F^\o+\vec k')\simeq
\pmatrix{ -iz_0k_0 & \d_0(ik_1'-\o k_2') \cr
\d_0(-ik_1'-\o k_2') & -iz_0k_0}_{\r,\r'}\;,\ee
for some real constants $z_0,\d_0$,
modulo higher order terms in $(k_0,\vec k')$. Therefore, it is apparent that
its structure is the same as the one of $\hat S_0(\kk)$, modulo
higher order terms in $(k_0,\vec k')$.\\

{\cs Proof.} As remarked after (\ref{2.16a}),
$P(d \Psi^{(u.v.)})$ and $P(d \Psi^{(i.r.)})$ are separately invariant under
the symmetry properties listed in Lemma 1. Therefore ${\cal V}(\Psi)$
is also invariant under the same symmetries, and so is the quadratic
part of ${\cal V}(\Psi)$, that is
\bea &&(\b|\L|)^{-2}
\sum_{\s}\sum_{\kk,\pp}\d(\pp)
\Big[\hat \Psi^{(i.r.)+}_{\kk,\s,1}
\hat \Psi^{(i.r.)-}_{\kk+\pp,\s,1}W_{aa}(\kk)
+\hat \Psi^{(i.r.)+}_{\kk,\s,1}
\hat \Psi^{(i.r.)-}_{\kk+\pp,\s,2}W_{ab}(\kk)
+\nn\\
&&\hskip2.truecm+\hat \Psi^{(i.r.)+}_{\kk,\s,2}
\hat \Psi^{(i.r.)-}_{\kk+\pp,\s,1}W_{ba}(\kk)+
\hat \Psi^{(i.r.)+}_{\kk,\s,2}
\hat \Psi^{(i.r.)-}_{\kk+\pp,\s,2}W_{bb}(\kk)\Big]\;.\label{2.17a}
\eea
Recall that, as assumed in the lines preceding (\ref{2.10}),
the support of $\hat \Psi^{(i.r.)}$ consists of two disjoint
regions around $\vec p_F^+$ and $\vec p_F^-$, respectively; in particular,
we assumed that $2a_0\g<4\p/3-4\p/(3\sqrt3)$. Under this condition,
it is easy to realize that if both $\kk$ and $\pp+\kk$ belong
to the support of $\hat\Psi^{(i.r.)}$, then $|\pp|<4\p/3$.
As a consequence, in (\ref{2.17}), the only non zero
contributions correspond to the terms with $\pp={\bf 0}$ (in fact,
if $\pp$ is $\neq{\bf 0}$
and belongs to the support of $\d(\pp)$, then $|\pp|\ge 4\p/3$, which means
that either $\kk$ or $\kk+\pp$ is outside the support of $\hat\Psi^{(i.r.)}$,
and the corresponding term in the sum is identically zero). This means that
the sum
\bea &&
\sum_{\s,\kk}
\Big[\hat \Psi^{(i.r.)+}_{\kk,\s,1}
\hat \Psi^{(i.r.)-}_{\kk,\s,1}W_{aa}(\kk)
+\hat \Psi^{(i.r.)+}_{\kk,\s,1}
\hat \Psi^{(i.r.)-}_{\kk,\s,2}W_{ab}(\kk)
+\nn\\
&&\hskip2.truecm+\hat \Psi^{(i.r.)+}_{\kk,\s,2}
\hat \Psi^{(i.r.)-}_{\kk,\s,1}W_{ba}(\kk)+
\hat \Psi^{(i.r.)+}_{\kk,\s,2}
\hat \Psi^{(i.r.)-}_{\kk,\s,2}W_{bb}(\kk)\Big]\;.\label{2.17b}
\eea
is invariant under the symmetries (1)--(7) listed in Lemma 1.

Invariance under symmetry (4) implies that:
\bea && W_{aa}(k_0,\vec k)=W_{aa}(k_0,T_1^{-1}\vec k)\;,\qquad
W_{bb}(k_0,\vec k)=W_{bb}(k_0,T_1^{-1}\vec k)\;,\label{2.17c}\\
&& W_{ab}(k_0,\vec k)=
e^{i\vec k(\vec\d_1-\vec\d_2)}W_{ab}(k_0,T_1^{-1}\vec k)
\;,\qquad
W_{ba}(k_0,\vec k)=e^{-i\vec k(\vec\d_1-\vec\d_2)}W_{ab}
(k_0,T_1^{-1}\vec k)\;;
\nn\eea

invariance under (5) implies that:
\bea && W_{aa}(\kk)=W_{aa}(-\kk)^*\;,\qquad
W_{bb}(\kk)=W_{bb}(-\kk)^*\;,\label{2.17d}\\
&&W_{ab}(\kk)=W_{ab}(-\kk)^*\;,\qquad
W_{ba}(\kk)=W_{ba}(-\kk)^*\;;\nn\eea

invariance under (6.a) implies that:
\bea && W_{aa}(k_0,k_1,k_2)=W_{bb}(k_0,-k_1,k_2)\;,\qquad
W_{ab}(k_0,k_1,k_2)=
W_{ba}(k_0,-k_1,k_2)\;;
\label{2.17e}\eea

invariance under (6.b) implies that:
\bea && W_{aa}(k_0,k_1,k_2)=W_{aa}(k_0,k_1,-k_2)\;,\qquad
W_{bb}(k_0,k_1,k_2)=W_{bb}(k_0,k_1,-k_2)\;,\label{2.17f}\\
&& W_{ab}(k_0,k_1,k_2)=W_{ab}(k_0,k_1,-k_2)\;,\qquad
W_{ba}(k_0,k_1,k_2)=W_{ba}(k_0,k_1,-k_2)
\;;
\nn\eea

invariance under (7) implies that:
\bea && W_{aa}(k_0,\vec k)=W_{aa}(k_0,-\vec k)\;,\qquad
W_{bb}(k_0,\vec k)=W_{bb}(k_0,-\vec k)\;,\label{2.17g}\\
&&W_{ab}(k_0,\vec k)=
W_{ba}(k_0,-\vec k)\;;\nn\eea

Finally, invariance under (8) implies that:
\bea && W_{aa}(k_0,\vec k)=-W_{aa}(-k_0,\vec k)\;,\qquad
W_{bb}(k_0,\vec k)=-W_{bb}(-k_0,\vec k)\;,\label{2.17h}\\
&&W_{ab}(k_0,\vec k)=W_{ab}(-k_0,\vec k)\;,\qquad
W_{ba}(k_0,\vec k)=W_{ba}(-k_0,\vec k)\;;\nn\eea

Now, combining the first of (\ref{2.17e}), the second of (\ref{2.17f})
and the second of (\ref{2.17g}), we find that $W_{aa}(\kk)=W_{bb}(\kk)$.
Combining the third of (\ref{2.17d}), the third of (\ref{2.17g}) and the
last of (\ref{2.17h}), we find that $W_{ab}(\kk)=W_{ba}(\kk)^*$. This concludes
the proof of item (i).

The first of (\ref{2.17h}) implies that, as $\b\to\io$,
$W_{aa}(0,\vec k)=0$, and this proves, in particular, that
$W_{aa}(0,\vec p_F^\o)=0$ and that, in the limit $|\L|\to\io$,
$\partial_{\vec k}W_{aa}(0,\vec p_F^\o)=\vec 0$.

Using that $\vec p_F^\o$ is invariant under the action of $T_1$, we see that
the third of (\ref{2.17c})
implies that $(1-e^{i\vec p_F^\o(\vec\d_1-\vec\d_2)})
W_{ab}(k_0,\vec p_F^\o)=0$. Since $e^{i\vec p_F^\o(\vec\d_1-\vec\d_2)}=
-e^{i\o\p/3}\neq1$, this identity proves, in particular,
that $W_{ab}(0,\vec p_F^\o)=0$, and $\partial_{k_0}W_{ab}(0,\vec p_F^\o)=0$.
This concludes the proof of item (ii).

Now, combining the first of (\ref{2.17d}) with the first of (\ref{2.17g}),
we find that $W_{aa}(k_0,\vec k)=W_{aa}(-k_0,\vec k)^*$, which implies,
in particular, that ${\rm
Re}\big\{\partial_{k_0}\hat W_{aa}(0,\vec p_F^{\;\o})\big\}=0$.

Finally, let $W_{ab}(0,\vec p_F^\o+\vec k')\simeq \a_1^\o k'_1+\a_2^\o k'_2$,
modulo higher order terms in $\vec k'$. Using that
$T_1^{-1}=\pmatrix{-1/2 & \sqrt3/2 \cr -\sqrt3/2 & -1/2}$ in the third
of (\ref{2.17c}), we find that
\be \a_1^\o k'_1+\a_2^\o k'_2=e^{-i\o\p/3}\Big[\a_1^\o(k'_1/2-\sqrt3 k'_2/2)
+\a_2^\o(\sqrt3 k'_1/2+k'_2/2)\Big]\;,\ee
which implies $\a_1^\o=-i\o\a_2^\o$. Moreover, using the third of
(\ref{2.17d}) we find that $\a_i^\o=-(\a_i^{-\o})^*$, and using the third
of (\ref{2.17f}) we find that $\a_2^\o=-\a_2^{-\o}$. Therefore,
$\a_2^\o=-\a_2^{-\o}=-(\a_2^{-\o})^*$, and we see that $\a_2^\o$ is real and
odd in $\o$, that is $\a_2^\o=\o a$, for some real constant $a$. Therefore,
$\a_1^\o=-i\o\a_2^\o=-i a$, and this concludes the proof of item (iii).
\qed

\subsection{Free energy: The infrared integration}\label{IIIc}
\noindent {\it Multiscale analysis.} In order to compute
(\ref{2.14}) we shall proceed in an iterative fashion, using
standard functional Renormalization Group methods \cite{BG,GM,M2}.
As a starting point, it is convenient to decompose the infrared
propagator as:
\be g^{(i.r.)}(\xx,\yy)=\sum_{\o=\pm} e^{-i \vec p_F^{\;\o}(\vec x-\vec
y)}g_{\o}^{(\le 0)}(\xx,\yy)\;,\label{2.22}\ee
where, if $\kk'=(k_0,\vec k')$,
\be g_{\o}^{(\le 0)}(\xx,\yy)=\frac1{\b|\L|}\sum_{\kk'\in\DD^\o_{\b,L}}
\chi_0(|\kk'|)e^{-i\kk'(\xx-\yy)}\pmatrix{-i k_0 & -v^*(\vec k'+
\vec p_F^{\;\o})
\cr -v(\vec k'+\vec p_F^{\;\o}) & -i k_0}^{-1}\label{2.23}\ee
and $\DD^\o_{\b,L}=\DD_\b\times\DD^\o_L$, with $\DD^\o_L=\{\frac{n_1}L\vec b_1
+\frac{n_2}L\vec b_2-\frac{(L\,|\, {\mod} 3)}
L\vec p_F^{\;\o}\;, \big[\frac{-L}{2}\big]+1
\le n_1,n_2\le \big[\frac{L}2\big]\}$.

Correspondingly, we rewrite $\Psi^{(i.r.)}$ as a sum of two independent
Grassmann fields:
\be \Psi^{(i.r.)\pm}_{\xx,\s,\r}=\sum_{\o=\pm}e^{i\vec p_F^{\;\o}\vec x}
\Psi^{(\le 0)\pm}_{\xx,\s,\r,\o}\label{2.24}\ee
and we rewrite (\ref{2.14}) in the form:
\be \Xi_{\b,L}=e^{-\b |\L| F_0}\int P_{\c_0,A_0}(d\Psi^{(\le 0)})
e^{-{\cal V}^{(0)}(\Psi^{(\le 0)})}
\;,\label{2.25}\ee
where ${\cal V}^{(0)}(\Psi^{(\le 0)})$ is equal to
${\cal V}(\Psi^{(i.r.)})$, once $\Psi^{(i.r.)}$ is rewritten as in
(\ref{2.24}), i.e.,
\bea && {\cal V}^{(0)}(\Psi^{(\le 0)})=\label{V0}\\
&&=\sum_{n=1}^\io (\b|\L|)^{-2n}\sum_{\s_1,\ldots,\s_n=\uparrow
\downarrow}\;\sum_{\r_1,\ldots,\r_{2n}=1,2}^{\o_1,\ldots,\o_{2n}=\pm}\;
\sum_{\kk_1',\ldots,\kk_{2n}'}
\Big[\prod_{j=1}^n\hat \Psi^{(\le 0)+}_{\kk_{2j-1}',\s_j,\r_{2j-1},\o_{2j-1}}
\hat \Psi^{(\le 0)-}_{\kk_{2j}',\s_j,\r_{2j},\o_{2j}}\Big]\;\cdot\nn\\
&&\hskip3.8truecm\cdot\hat W_{2n,\ul\r,\ul\o}^{(0)}(\kk_1',\ldots,
\kk_{2n-1}')\;\d\big(\sum_{j=1}^{2n}(-1)^j(\pp_F^{\o_{j}}+\kk_{j}')\big)=\nn\\
&&=\sum_{n=1}^\io \sum_{\ul\s,\ul\r,\ul\o}
\int\,d\xx_1\cdots d\xx_{2n}
\Big[\prod_{j=1}^n \Psi^{(\le 0)+}_{\xx_{2j-1},\s_j,\r_{2j-1},\o_{2j-1}}
\Psi^{(\le 0)-}_{\xx_{2j},\s_j,\r_{2j},\o_{2j}}\Big]
W_{2n,\ul\r,\ul\o}^{(0)}(\xx_1,\ldots,
\xx_{2n})\;,\nn\eea
with:\\
1) $\ul\o=(\o_1,\ldots,\o_{2n})$, $\ul\s=(\s_1,\ldots,\s_{n})$ and
$\pp_F^{\o}=(0,\vec p_F^{\;\o})$;\\
2) $\hat W_{2n,\ul\r,\ul\o}^{(0)}
(\kk_1',\ldots,\kk_{2n-1}')=\hat W_{2n,\ul\r}(\kk_1'+\pp_F^{\o_j},\ldots,
\kk_{2n-1}'+\pp_F^{\o_{2n-1}})$, see (\ref{2.16});\\
3) the kernels $W_{2n,\ul\r,\ul\o}^{(0)}(\xx_1,\ldots,
\xx_{2n})$ are defined as:
\bea &&W_{2n,\ul\r,\ul\o}^{(0)}(\xx_1,\ldots,
\xx_{2n})=\label{xspace}\\
&&=(\b|\L|)^{-2n}\sum_{\kk'_1,\ldots,\kk_{2n}'}
e^{i\sum_{j=1}^{2n}(-1)^j\kk_{j}\xx_{j}}
\hat W_{2n,\ul\r,\ul\o}^{(0)}(\kk_1',\ldots,
\kk_{2n-1}')\;\d\big(\sum_{j=1}^{2n}(-1)^j(\pp_F^{\o_{j}}+\kk_{j}')\big)
\;.\nn\eea
Moreover,
$P_{\c_0,A_0}(d\Psi^{(\le 0)})$ is defined as
\bea &&P_{\c_0,A_0}(d\Psi^{(\le 0)})={{\cal N}_0}^{-1}\Biggl[\;
\prod_{\kk'\in\DD_{\b,L}^\o}^{\c_0(|\kk'|)>0}
\;\prod_{\s,\o,\r}
d\hat\Psi^{(\le 0)+}_{\kk',\s,\r,\o}d\hat\Psi^{(\le 0)-}_{\kk',\s,\r,\o}\Biggr]
\cdot\label{2.26}\\
&&\hskip2.truecm\cdot
\exp \Big\{-(\b|\L|)^{-1}\sum_{\o=\pm,\s=\uparrow\downarrow}
\sum_{\kk'\in\DD_{\b,L}^\o}^{\c_0(|\kk'|)>0}\c_0^{-1}(|\kk'|)
\hat\Psi^{(\le 0)+}_{\kk',\s,\cdot,\o}A_{0,\o}(\kk')\hat\Psi^{(\le 0)-}_{
\kk',\s,\cdot,\o}\}\;,\nn\eea
where:
\bea &&
A_{0,\o}(\kk')=\pmatrix{-i k_0 & -v^*(\vec k'+\vec p_F^{\;\o})
\cr -v(\vec k'+\vec p_F^{\;\o}) & -i k_0}=\nn\\
&&\hskip1.2truecm
=\pmatrix{-i \z_0 k_0 +s_{0}(\kk') & c_0(ik_1'-\o k_2')+t_{0,\o}(\kk')
\cr c_0(-ik_1'-\o k_2') +t_{0,\o}^*(\kk')& -i \z_0 k_0+s_{0}(\kk') }
\;,\nn\eea
${\cal N}_0$ is chosen in such a way that $\int P_{\c_0,A_0}
(d\Psi^{(\le 0)})=1$, $\z_0=1$, $c_0=3/2$,
$s_{0}\=0$ and $|t_{0,\o}(\kk')|\le C|\kk'|^2$.

It is apparent that the $\Psi^{(\le 0)}$ field
has zero mass (i.e., its propagator
decays polynomially at large distances in $\xx$-space). Therefore,
its integration requires an infrared multiscale analysis. We consider the
scaling parameter $\g>1$ introduced above, see the lines preceding
(\ref{2.10}), and we define a sequence of geometrically
decreasing momentum scales $\g^h$, $h=0,-1,-2,\ldots$ Correspondingly
we introduce compact support functions $f_h(\kk')=\c_0(\g^{-h}|\kk'|)
-\c_0(\g^{-h+1}|\kk'|)$ and we rewrite
\be \c_0(|\kk'|)=\sum_{h=-\io}^0 f_h(\kk')\;.\label{2.27}\ee
The purpose is to perform the integration of (\ref{2.25}) in an iterative way.
We step by step decompose the
propagator into a sum of two propagators, the first supported on
momenta $\sim \g^{h}$, $h\le 0$, the second supported
on momenta smaller than $\g^h$. Correspondingly we rewrite the
Grassmann field as a sum of two independent fields: $\Psi^{(\le
h)}=\Psi^{(h)}+ \Psi^{(\le h-1)}$ and we integrate the field
$\Psi^{(h)}$. In this way
we inductively prove that, for any $h\le 0$, \pref{2.25} can be
rewritten as
\be \Xi_{\b,L}=e^{-\b|\L|F_h}\int P_{\c_h,A_h}(d\Psi^{(\le h)})
e^{-{\cal V}^{(h)}(\Psi^{(\le h)})}\;,\label{2.28}\ee
where $F_h, A_h, {\cal V}^{(h)}$ will be defined recursively,
$\c_h(|\kk'|)=
\sum_{k=-\io}^h f_k(\kk')$ and $P_{\c_h,A_h}(d\Psi^{(\le h)})$ is defined
in the same way as $P_{\c_0,A_0}(d\Psi^{(\le 0)})$ with
$\Psi^{(\le 0)}, \c_0, A_{0,\o}, \z_0, c_0, s_{0}, t_{0,\o}$ replaced
by $\Psi^{(\le h)}, \c_h, A_{h,\o}, \z_h, c_h, s_h, t_{h,\o}$,
respectively.
Moreover ${\cal V}^{(h)}(0)=0$ and
\bea && {\cal V}^{(h)}(\Psi)=\sum_{n=1}^\io (\b|\L|)^{-2n}
\sum_{\ul\s,\ul\r,\ul\o}\;
\sum_{\kk_1',\ldots,\kk_{2n}'}
\Big[\prod_{j=1}^n\hat\Psi^{(\le h)+}_{\kk_{2j-1}',\s_j,\r_{2j-1},\o_{2j-1}}
\hat \Psi^{(\le h)-}_{\kk_{2j}',\s_j,\r_{2j},\o_{2j}}\Big]\;\cdot\nn\\
&&\hskip3.truecm\cdot\hat W_{2n,\ul\r,\ul\o}^{(h)}(\kk_1',\ldots,
\kk_{2n-1}')\;\d(\sum_{j=1}^{2n}(-1)^j(\pp_F^{\o_{j}}+\kk_{j}'))=\label{2.29}\\
&&=\sum_{n=1}^\io \sum_{\ul\s,\ul\r,\ul\o}
\int\,d\xx_1\cdots d\xx_{2n}
\Big[\prod_{j=1}^n \Psi^{(\le h)+}_{\xx_{2j-1},\s_j,\r_{2j-1},\o_{2j-1}}
\Psi^{(\le h)-}_{\xx_{2j},\s_j,\r_{2j},\o_{2j}}\Big]
W_{2n,\ul\r,\ul\o}^{(h)}(\xx_1,\ldots,
\xx_{2n})\;.\nn\eea

Note that the field $\Psi^{(\le h)}_{\kk',\s,\cdot,\o}$,
whose propagator is given by $\c_h(|\kk'|)
[A^{(h)}_\o(\kk')]^{-1}$, has the same support as $\c_h$, that is on a
neighborood of size $\g^h$ around the singularity $\kk'={\bf 0}$ (that, in the
original variables, corresponds to the Dirac point $\kk=\pp_F^\o$). It is
important for the following to think $\hat W^{(h)}_{2n,\ul\r,\ul\o}$, $h\le 0$,
as functions of the variables $\{\z_k,c_k\}_{h<k\le 0}$. The iterative
construction below will inductively imply that the dependence on these
variables is well defined.

The iteration will continue up to the scale $h_\b$, where $h_\b$
is the largest scale such that
\be a_0 \g^{h_\b-1}<\frac{\p}{\b}\z_{h_\b}\label{2.30}\ee
where $a_0$ is the constant appearing in the definition of $\c_0(|\kk'|)$.
By the properties of $\z_h$ that will be described and proved below, it will
turn out that $h_\b$ is finite and larger than $\log_\g\frac\p{2 a_0\b}$.
The result of the last iteration will be $\Xi_{\b,L}$, i.e.,
the value of the partition function.
\\
\\
{\it Localization and renormalization.}
In order to inductively prove \pref{2.28}
we write
\be {\cal V}^{(h)} =\LL{\cal V}^{(h)}+\RR{\cal V}^{(h)}\label{2.loc}
\ee
where
\be
\LL{\cal V}^{(h)}=
{1\over \b|\L|}\sum_{\s_1,\s_2=\uparrow
\downarrow}\;\sum_{\r_1,\r_{2}=1,2\atop \o=\pm}\;
\sum_{\kk'}^{\c_h(|\kk'|)>0} \hat\Psi^{(\le h)+}_{\kk',\s_1,\r_{1},\o}
\hat\Psi^{(\le h)-}_{\kk',\s_2,\r_{2},\o}
\hat W_{2,\ul\r,(\o,\o)}^{(h)}(\kk')
\;,\label{2.30a}\ee
and $\RR{\cal V}^{(h)}$ is given by \pref{2.29} with
$\sum_{n=1}^\io$ replaced by $\sum_{n=2}^\io$, that is it contains
only the monomials with more than four fields.

Note that in (\ref{2.30a}) the $\o$-index of the $\Psi$ fields is the same;
this follows from the fact that in the terms with different $\o$'s the momenta
verify $\kk'_1-\kk'_2+\pp_F^{\o}-\pp_F^{-\o}=n_1 \vec b_1+n_2 \vec b_2$,
for some choice of $n_1,n_2$, and such a condition cannot
be verified if $\kk'_1,\kk'_2$ are in the support of the $\Psi^{(\le h)}$
fields, because $\pp_F^{\o}-\pp_F^{-\o}\not\in \L^*$ and $2a_0\g$
is smaller than $4\p/3-4\p/(3\sqrt3)$, see the lines preceding
(\ref{2.10}) and the discussion after (\ref{2.17a}).

\*
{\bf Remark.} The fact that
the quadratic terms with different $\o$'s, i.e., the one particle
{\it umklapp processes}, do not
contribute to the infrared effective potential is a crucial fact,
which reduces the number of {\it relevant running coupling constants} and,
in particular, tells us that the interaction does not generate
{\it mass terms}. Note, in fact, that the presence of one particle umklapp
terms with a non zero contribution at the Fermi points could produce
an exponential decay of the interacting correlations.
\*

The symmetries of the action,
listed in Lemma 1, which are preserved by the iterative integration
procedure, imply that, in the zero temperature and thermodynamic limit,
$\hat W^{(h)}_{2,\ul\r,(\o,\o)}({\bf 0})=0$ and
\be \kk'\partial_{\kk'}
\hat W^{(h)}_{2,(\r_1,\r_2),(\o,\o)}({\bf 0})=
\pmatrix{ -iz_hk_0 & \d_h(ik_1'-\o k_2') \cr
\d_h(-ik_1'-\o k_2') & -iz_hk_0}_{\r_1,\r_2}\;, \label{2.31}\ee
for suitable real constants $z_h,\d_h$. The proof of (\ref{2.31}) is completely
analogous to the proof of Lemma 2 and will not be repeated here.

Once that the above definitions are given, we can describe our iterative
integration procedure for $h\le 0$. We start from (\ref{2.28})
and we rewrite it as
\be \int P_{\c_h,A_h}(d\Psi^{(\le h)}) \, e^{-\LL{\cal V}^{(h)}
(\Psi^{(\le h)})-\RR{\cal V}^{(h)}
(\Psi^{(\le h)}) -\b|\L| F_h}\;,\label{2.32}\ee
with
\bea &&\LL{\cal V}^{(h)}(\Psi^{(\le h)})=(\b|\L|)^{-1}\sum_{\o,\s}
\sum_{\kk'}^{\c_h(|\kk'|)>0}\cdot\label{2.33}\\
&&\hskip2.truecm\cdot
\hat\Psi^{(\le h)+}_{\kk',\s,\cdot,\o}\pmatrix{
-iz_hk_0 +\s_h(\kk')& \d_h(ik_1'-\o k_2')+\t_{h,\o}(\kk') \cr
\d_h(-ik_1'-\o k_2) +\t_{h,\o}^*(\kk')& -iz_hk_0+\s_h(\kk')}
\hat\Psi^{(\le h)-}_{\kk',\s,\cdot,\o}\;.\nn\eea
Then we include $\LL{\cal V}^{(h)}$
in the fermionic integration, so obtaining
\be
\int P_{\c_h,\lis A_{h-1}}(d\Psi^{(\le h)}) \, e^{-
\RR{\cal V}^{(h)}
(\Psi^{(\le h)}) -\b|\L| (F_h+e_h)}\;,\label{2.34}\ee
where $e_h$ is a constant that takes into account the change in the
normalization factor of the measure and
\be \lis A_{h-1,\o}(\kk')=\pmatrix{-i \lis \z_{h-1} k_0 +\lis s_{h-1}(\kk') &
\lis c_{h-1}(ik_1'-\o k_2')+\lis t_{h-1,\o}(\kk')
\cr \lis c_{h-1}(-ik_1'-\o k_2') +\lis t_{h-1,\o}^*(\kk')& -i \lis \z_{h-1} k_0
+\lis s_{h-1}(\kk')}\label{2.34aa}\ee
with:
\bea&& \lis \z_{h-1}(\kk')= \z_h +z_h \c_h(\kk')\;,\hskip1.7truecm
\lis c_{h-1}(\kk')= c_h +\d_h \c_h(\kk')\;,\nn\\
&& \lis s_{h-1}(\kk')=s_{h}(\kk')+\s_h(\kk')\c_h(\kk')\;,\hskip.5truecm
\lis t_{h-1,\o}(\kk')=t_{h,\o}(\kk')+\t_{h,\o}(\kk')\c_h(\kk')\;.
\label{2.34ab}\eea
Now we can perform the integration of the $\Psi^{(h)}$ field.
We rewrite the Grassmann field $\Psi^{(\le h)}$ as a sum of two independent
Grassmann fields $\Psi^{(\le h-1)}+\Psi^{(h)}$ and correspondingly we rewrite
(\ref{2.34}) as
\be e^{-\b|\L|(F_h+e_h)}\int
P_{\c_{h-1},A_{h-1}}(d\Psi^{(\le h-1)}) \,
\int P_{f^{-1}_h,\lis A_{h-1}}
(d\Psi^{(h)})\, e^{-\RR{\cal V}^{(h)}(\Psi^{(\le h-1)}+\Psi^{(h)})}\;,
\label{2.35}\ee
where
\be A_{h-1,\o}(\kk')=\pmatrix{-i\z_{h-1} k_0 +s_{h-1}(\kk') &
c_{h-1}(ik_1'-\o k_2')+t_{h-1,\o}(\kk')
\cr c_{h-1}(-ik_1'-\o k_2') +t_{h-1,\o}^*(\kk')& -i \z_{h-1} k_0
+s_{h-1}(\kk')}\label{2.34ac}\ee
with:
\bea&& \z_{h-1}= \z_h +z_h\;,\hskip1.7truecm
c_{h-1}= c_h +\d_h \;,\nn\\
&& s_{h-1}(\kk')=s_{h}(\kk')+\s_h(\kk')\;,\hskip.5truecm
 t_{h-1,\o}(\kk')=t_{h,\o}(\kk')+\t_{h,\o}(\kk')\;.
\label{2.34ad}\eea

The single scale propagator is
\be \int P_{f^{-1}_h,\lis A_{h-1}}(d \Psi^{(h)})
\Psi^{(h)-}_{\xx_1,\s_1,\r_1,\o_1}
\Psi^{(h)+}_{ \xx_2,\s_2,\r_2,\o_2} = \d_{\s_1,\s_2}\d_{\o_1,\o_2}
\big[g^{(h)}_\o(\xx_1,\xx_2)\big]_{\r_1,\r_2}\;,\label{2.36}\ee
where
\be g^{(h)}_\o(\xx_1,\xx_2)=\frac1{\b|\L|}\sum_{\kk'\in\DD_{\b,L}^\o}
e^{-i\kk'(\xx_1-\xx_2)}f_h(\kk')\Big[\lis A_{h-1,\o}(\kk')\Big]^{-1}
\;.\label{2.37}\ee
After the integration of the field on scale $h$ we are left with an
integral involving the fields
$\Psi^{(\le h-1)}$ and the new effective interaction
${\cal V}^{(h-1)}$, defined as
\be e^{-{\cal V}^{(h-1)}(\Psi^{(\le h-1)})-\lis e_h \b|\L|}=
\int P_{f^{-1}_h,\lis A_{h-1}}
(d\Psi^{(h)})\, e^{-\RR{\cal V}^{(h)}(\Psi^{(\le h-1)}+\Psi^{(h)})}
\;.\label{2.39}\ee
It is easy to see that ${\cal V}^{(h-1)}$ is of the form (\ref{2.29}) and that
$F_{h-1}=F_h+e_h+\lis e_h$. It is sufficient to use the well known identity
\be \lis e_h+{\cal V}^{(h-1)}(\Psi^{(\le h-1)})=
\sum_{n\ge 1}{1\over n!}(-1)^{n+1}\EE^T_h(\RR{\cal V}^{(h)}\big(
\Psi^{(\le h-1)}
+\Psi^{(h)}\big);n)
\;,\label{2.40}\ee
where $\EE^T_h(X(\Psi^{(h)});n)$
is the truncated expectation of order $n$ w.r.t. the
propagator $g^{(h)}_\o$, which is the analogue of (\ref{2.15})
with $\Psi^{(u.v.)}$ replaced by $\Psi^{(h)}$ and with
$P(d\Psi^{(u.v.)})$ replaced by $P_{f^{-1}_h,\lis A_{h-1}}(d\Psi^{(h)})$.

Note that the above procedure allows us to write the
{\it effective renormalizations} $\vec v_{h}=(\z_{h},c_{h})$, $h\le 0$,
in terms of $\vec v_k$, $h< k\le 0$, namely
$\vec v_{h-1}=\b_h(\vec v_h,\ldots,\vec v_0)$,
where $\b_h$ is the so--called {\it Beta function}.
\*
{\it Tree expansion for the effective potentials.}
An iterative implementation of (\ref{2.40}) leads to a representation of
${\cal V}^{(h)}(\Psi^{(\le h)})$ in terms of a tree expansion, defined as
follows.


\begin{figure}[ht]
\hspace{-5.5truecm}
\includegraphics[height=8.truecm]{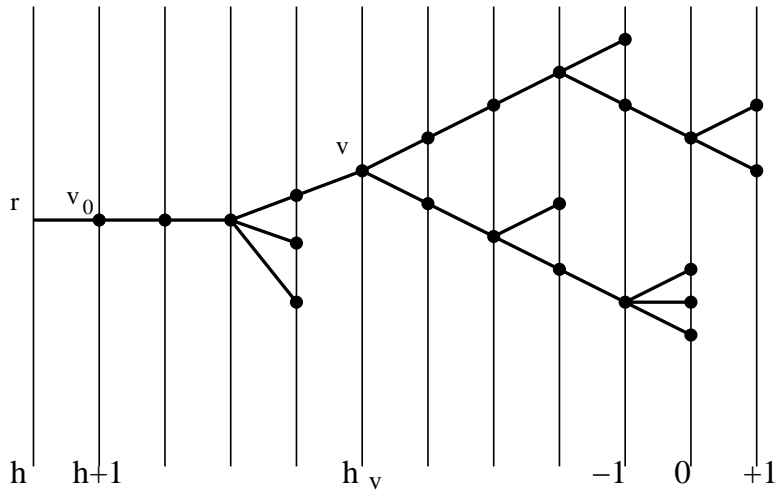}
\vspace{-0.7truecm}
\caption{A tree $\t\in\TT_{h,n}$ with its scale labels.}
\end{figure}


\0 1) Let us consider the family of all trees which can be constructed
by joining a point $r$, the {\it root}, with an ordered set of $n\ge 1$
points, the {\it endpoints} of the {\it unlabeled tree},
so that $r$ is not a branching point. $n$ will be called the
{\it order} of the unlabeled tree and the branching points will be called
the {\it non trivial vertices}.
The unlabeled trees are partially ordered from the root
to the endpoints in the natural way; we shall use the symbol $<$
to denote the partial order.
Two unlabeled trees are identified if they can be superposed by a suitable
continuous deformation, so that the endpoints with the same index coincide.
It is then easy to see that the number of unlabeled trees with $n$ end-points
is bounded by $4^n$.
We shall also consider the {\it labelled trees} (to be called
simply trees in the following); they are defined by associating
some labels with the unlabelled trees, as explained in the
following items.

\0 2) We associate a label $h\le -1$ with the root and we denote
$\TT_{h,n}$ the corresponding set of labeled trees with $n$
endpoints. Moreover, we introduce a family of vertical lines,
labeled by an integer taking values in $[h,1]$, and we represent
any tree $\t\in\TT_{h,n}$ so that, if $v$ is an endpoint or a non
trivial vertex, it is contained in a vertical line with index
$h_v>h$, to be called the {\it scale} of $v$, while the root $r$ is on
the line with index $h$.
In general, the tree will intersect the vertical lines in set of
points different from the root, the endpoints and the branching
points; these points will be called {\it trivial vertices}.
The set of the {\it
vertices} will be the union of the endpoints, of the trivial
vertices and of the non trivial vertices; note that the root is not a vertex.
Every vertex $v$ of a
tree will be associated to its scale label $h_v$, defined, as
above, as the label of the vertical line whom $v$ belongs to. Note
that, if $v_1$ and $v_2$ are two vertices and $v_1<v_2$, then
$h_{v_1}<h_{v_2}$.

\0 3) There is only one vertex immediately following
the root, which will be denoted $v_0$ and cannot be an endpoint;
its scale is $h+1$.

\0 4) Given a vertex $v$ of $\t\in\TT_{h,n}$ that is not an endpoint,
we can consider the subtrees of $\t$ with root $v$, which correspond to the
connected components of the restriction of
$\t$ to the vertices $w\ge v$. If a subtree with root $v$ contains only
$v$ and an endpoint on scale $h_v+1$,
it will be called a {\it trivial subtree}.

\0 5) With each endpoint $v$ we associate one of the monomials
with four or more Grassmann fields contributing to $\RR {\cal
V}^{(0)}(\Psi^{(\le h_v-1)})$, corresponding to the terms with
$n\ge 2$ in the r.h.s. of (\ref{V0}) (with $\Psi^{(\le 0)}$
replaced by $\Psi^{(\le h_v-1)}$) and a set $\xx_v$ of space-time
points (the corresponding integration variables in the $\xx$-space
representation).

\0 6) We introduce a {\it field label} $f$ to distinguish the field variables
appearing in the terms associated with the endpoints as in item 3);
the set of field labels associated with the endpoint $v$ will be called $I_v$;
note that $|I_v|$ is the order of the monomial contributing
to ${\cal V}^{(0)}(\Psi^{(\le h_v-1)})$ and associated to $v$.
Analogously, if $v$ is not an endpoint, we shall
call $I_v$ the set of field labels associated with the endpoints following
the vertex $v$; $\xx(f)$, $\e(f)$, $\s(f)$, $\r(f)$ and $\o(f)$ will denote the
space-time point, the $\e$ index, the $\s$ index, the $\r$ index and the
$\o$ index, respectively, of the Grassmann field variable with label $f$.

In terms of these trees, the effective potential ${\cal V}^{(h)}$, $h\le -1$,
can be written as
\be {\cal V}^{(h)}(\Psi^{(\le h)}) + \b|\L| \lis e_{k+1}=
\sum_{n=1}^\io\sum_{\t\in\TT_{h,n}}
{\cal V}^{(h)}(\t,\Psi^{(\le h)})\;,\label{2.41}\ee
where, if $v_0$ is the first vertex of $\t$
and $\t_1,\ldots,\t_s$ ($s=s_{v_0}$)
are the subtrees of $\t$ with root $v_0$,
${\cal V}^{(h)}(\t,\Psi^{(\le h)})$
is defined inductively as follows:\\
i) if $s>1$, then
\be {\cal V}^{(h)}(\t,\Psi^{(\le h)})={(-1)^{s+1}\over s!} \EE^T_{h+1}
\big[\bar{\cal V}^{(h+1)}(\t_1,\Psi^{(\le h+1)});\ldots; \bar{\cal V}^{(h+1)}
(\t_{s},\Psi^{(\le h+1)})\big]\;,\label{2.42}\ee
where $\bar{\cal V}^{(h+1)}(\t_i,\Psi^{(\le h+1)})$ is equal to $\RR{\cal
V}^{(h+1)}(\t_i,\Psi^{(\le h+1)})$ if the subtree $\t_i$ contains
more than one end-point, or if it contains one end-point but it is
not a trivial subtree;
it is equal to $\RR{\cal V}^{(0)}(\t_i,\Psi^{(\le h+1)})$
if $\t_i$ is a trivial subtree;\\
ii) if $s=1$, then ${\cal V}^{(h)}(\t,\Psi^{(\le h)})$ is equal to
$\EE^T_{h+1}\big[\RR{\cal V}^{(h+1)}(\t_1,\Psi^{(\le h+1)})\big]$
if $\t_1$ is not a trivial
subtree; it is equal to $\EE^T_{h+1}\big[\RR{\cal
V}^{(0)}(\Psi^{(\le h+1)})- \RR{\cal V}^{(0)}(\Psi^{(\le h)})\big]$
if $\t_1$ is a trivial subtree.

Using its inductive definition, the right hand side of (\ref{2.41}) can be
further expanded, and in order to describe the resulting expansion we need some
more definitions.

We associate with any vertex $v$ of the tree a subset $P_v$ of $I_v$,
the {\it external fields} of $v$. These subsets must satisfy various
constraints. First of all, if $v$ is not an endpoint and $v_1,\ldots,v_{s_v}$
are the $s_v\ge 1$ vertices immediately following it, then
$P_v \subseteq \cup_i
P_{v_i}$; if $v$ is an endpoint, $P_v=I_v$.
If $v$ is not an endpoint, we shall denote by $Q_{v_i}$ the
intersection of $P_v$ and $P_{v_i}$; this definition implies that $P_v=\cup_i
Q_{v_i}$. The union ${\cal I}_v$ of the subsets $P_{v_i}\setminus Q_{v_i}$
is, by definition, the set of the {\it internal fields} of $v$,
and is non empty if $s_v>1$.
Given $\t\in\TT_{h,n}$, there are many possible choices of the
subsets $P_v$, $v\in\t$, compatible with all the constraints. We
shall denote ${\cal P}_\t$ the family of all these choices and ${\bf P}$
the elements of ${\cal P}_\t$.

With these definitions, we can rewrite
${\cal V}^{(h)}(\t,\Psi^{(\le h)})$
in the r.h.s. of (\ref{2.41}) as:
\bea &&{\cal V}^{(h)}(\t,\Psi^{(\le
h)})=\sum_{{\bf P}\in{\cal P}_\t}
{\cal V}^{(h)}(\t,\PP)\;,\nn\\
&&{\cal V}^{(h)}(\t,\PP)=\int d\xx_{v_0}
\widetilde\Psi^{(\le h)}(P_{v_0})
K_{\t,\PP}^{(h+1)}(\xx_{v_0})\;,\label{2.43}\eea
where
\be \widetilde\Psi^{(\le h)}
(P_{v})=\prod_{f\in P_v}\Psi^{ (\le
h)\e(f)}_{\xx(f),\s(f),\r(f),\o(f)}\label{2.44}\ee
and $K_{\t,\PP}^{(h+1)}(\xx_{v_0})$ is defined inductively by
the equation, valid for any $v\in\t$ which is not an endpoint,
\be K_{\t,\PP}^{(h_v)}(\xx_v)={1\over s_v !}
\prod_{i=1}^{s_v} [K^{(h_v+1)}_{v_i}(\xx_{v_i})]\; \;\EE^T_{h_v}[
\widetilde\Psi^{(h_v)}(P_{v_1}\setminus Q_{v_1}),\ldots,
\widetilde\Psi^{(h_v)}(P_{v_{s_v}}\setminus
Q_{v_{s_v}})]\;,\label{2.45}\ee
where $\widetilde\Psi^{(h_v)}(P_{v_i}\setminus Q_{v_i})$ has a definition
similar to (\ref{2.44}). Moreover, if $v_i$ is an endpoint
$K^{(h_v+1)}_{v_i}(\xx_{v_i})$ is equal to one
of the kernels of the monomials contributing to
$\RR{\cal V}^{(0)}(\Psi^{(\le h_v)})$, corresponding to the
terms with $n\ge 2$ in the r.h.s. of (\ref{V0})
(with $\Psi^{(\le 0)}$ replaced by $\Psi^{(\le h_v)}$); if $v_i$ is not an
endpoint, $K_{v_i}^{(h_v+1)}=K_{\t_i,\PP_i}^{(h_v+1)}$, 
where ${\bf P}_i=\{P_w, w\in\t_i\}$.

(\ref{2.41})--(\ref{2.45}) is not the final form of our expansion;
we further decompose ${\cal V}^{(h)}(\t,\PP)$, by using the
following representation of the truncated expectation in the r.h.s. of
(\ref{2.45}). Let us put $s=s_v$, $P_i\=P_{v_i}\setminus Q_{v_i}$;
moreover we order in an arbitrary way the sets $P_i^\pm\=\{f\in
P_i,\e(f)=\pm\}$, we call $f_{ij}^\pm$ their elements and we
define $\xx^{(i)}=\cup_{f\in P_i^-}\xx(f)$, $\yy^{(i)}=\cup_{f\in
P_i^+}\xx(f)$, $\xx_{ij}=\xx(f^-_{ij})$,
$\yy_{ij}=\xx(f^+_{ij})$. Note that $\sum_{i=1}^s
|P_i^-|=\sum_{i=1}^s |P_i^+|\=n$, otherwise the truncated
expectation vanishes. A couple
$l\=(f^-_{ij},f^+_{i'j'})\=(f^-_l,f^+_l)$ will be called a line
joining the fields with labels $f^-_{ij},f^+_{i'j'}$, sector
indices $\o_l^-=\o(f^-_l)$, $\o_l^+=\o(f^+_l)$,
$\r$-indices $\r^-_l=\r(f^-_l)$, $\r_l^+=\r(f^+_l)$,
and spin indices
$\s^-_l=\s(f^-_l)$, $\s^+_l=\s(f^+_l)$, connecting the points
$\xx_l\=\xx_{ij}$ and $\yy_l\=\yy_{i'j'}$, the {\it endpoints} of
$l$. Moreover, if $\o^-_l=\o^+_l$, we shall put
$\o_l\=\o^-_l=\o^+_l$.
Then, we use the {\it Brydges-Battle-Federbush} formula (e.g., see
\cite{GM,M2}) saying that, up to a sign, if $s>1$,
\be \EE^T_{h}(\widetilde\Psi^{(h)}(P_1),\ldots,
\widetilde\Psi^{(h)}(P_s))=\sum_{T}\prod_{l\in T}
\d_{\o^-_l,\o^+_l} \d_{\s^-_l,\s^+_l}\,
\big[g^{(h)}_{\o_l}(\xx_l-\yy_l)\big]_{\r^-_l,\r^+_l}\,
\int dP_{T}({\bf t})\; {\rm det}\, G^{h,T}({\bf t})\;,\label{2.46}\ee
where $T$ is a set of lines forming an {\it anchored tree graph} between
the clusters of points $\xx^{(i)}\cup\yy^{(i)}$, that is $T$ is a
set of lines, which becomes a tree graph if one identifies all the
points in the same cluster. Moreover ${\bf t}=\{t_{ii'}\in [0,1],
1\le i,i' \le s\}$, $dP_{T}({\bf t})$ is a probability measure with
support on a set of ${\bf t}$ such that $t_{ii'}={\bf u}_i\cdot{\bf u}_{i'}$
for some family of vectors ${\bf u}_i\in \RRR^s$ of unit norm. Finally
$G^{h,T}({\bf t})$ is a $(n-s+1)\times (n-s+1)$ matrix, whose elements
are given by
\be G^{h,T}_{ij,i'j'}=t_{ii'}
\d_{\o^-_l,\o^+_l} \d_{\s^-_l,\s^+_l}\,
\big[g^{(h)}_{\o_l}(\xx_{ij}-\yy_{i'j'})\big]_{\r^-_l,\r^+_l}\;,
\label{2.48}\ee
with $(f^-_{ij}, f^+_{i'j'})$ not belonging to $T$.
In the following we shall use (\ref{2.44}) even for $s=1$, when $T$
is empty, by interpreting the r.h.s. as equal to $1$, if
$|P_1|=0$, otherwise as equal to ${\rm det}\,G^{h}=
\EE^T_{h}(\widetilde\Psi^{(h)}(P_1))$. \\

{\bf Remark.} It is crucial to note that $G^{h,T}$ is a Gram matrix, i.e.,
defining ${\bf e}_+={\bf e}_\uparrow=(1,0)$ and
${\bf e}_-={\bf e}_\downarrow=(0,1)$,
the matrix elements in (\ref{2.48}) can be written in terms of scalar products:
\bea&& t_{ii'}
\d_{\o^-_l,\o^+_l} \d_{\s^-_l,\s^+_l}\,
\big[g^{(h)}_{\o_l}(\xx_{ij}-\yy_{i'j'})\big]_{\r^-_l,\r^+_l}
=\label{2.48a}\\
&&\hskip.3truecm
=\Big({\bf u}_i\otimes {\bf e}_{\o^-_l}\otimes{\bf e}_{\s^-_l}\otimes
A(\xx_{ij}-\cdot)\;,\ {\bf u}_{i'}\otimes {\bf e}_{\o^+_l}\otimes
{\bf e}_{\s^+_l}\otimes B(\xx_{i'j'}-\cdot)\Big)\=({\bf f}_\a,{\bf g}_\b)
\nn
\;,\eea
where
\bea &&A(\xx)=\frac1{\b|\L|}\sum_{\kk'\in\DD_{\b,L}^\o}
e^{-i\kk'\xx}\sqrt{f_h(\kk')}\;\openone\;,\nn\\
&&B(\xx)=\frac1{\b|\L|}\sum_{\kk'\in\DD_{\b,L}^\o}
e^{-i\kk'\xx}\sqrt{f_h(\kk')}\Big[\lis A_{h-1,\o}(\kk')\Big]^{-1}\;.
\label{2.48b}\eea
The symbol $(\cdot,\cdot)$ denotes the inner product, i.e.,
\bea&&
\big({\bf u}_i\otimes{\bf e}_\o\otimes{\bf e}_\s\otimes A(\xx-\cdot),
{\bf u}_{i'}\otimes{\bf e}_{\o'}\otimes{\bf e}_{\s'}\otimes B(\xx'-\cdot)\big)
=\nn\\
&&=({\bf u}_i\cdot{\bf u}_{i'})\, ({\bf e}_\o\cdot {\bf e}_{\o'})\,
({\bf e}_\s\cdot {\bf e}_{\s'})\cdot\int d\zz A^*(\xx-\zz)B(\xx'-\zz)\;,
\label{2.48c}\eea
and the vectors ${\bf f}_\a,{\bf g}_\b$ with $\a,\b=1,\ldots,n-s+1$ are
implicitely defined by (\ref{2.48a}). The usefulness of the representation
(\ref{2.48a}) is that, by the Gram-Hadamard inequality (see, e.g., \cite{GM}),
$|\det({\bf f}_\a,{\bf g}_\b)|\le \prod_{\a}||f_\a||\,||g_\a||$.
In our case, $||{\bf f}_\a||\le C\g^{3h/2}$ and $||{\bf g}_\a||\le C\g^{h/2}$.
Therefore, $||f_\a||\, ||g_\a||\le C \g^{2h}$, uniformly in $\a$,
so that the Gram determinant
can be bounded by $C^{n-s+1}\g^{2h(n-s+1)}$.
\\

If we apply the expansion (\ref{2.46}) in each vertex of
$\t$ different from the endpoints, we get an expression of the form
\be {\cal V}^{(h)}(\t,\PP) = \sum_{T\in {\bf T}} \int d\xx_{v_0}
\widetilde\Psi^{(\le h)}(P_{v_0}) W_{\t,\PP,T}^{(h)}(\xx_{v_0})
\= \sum_{T\in {\bf T}}
{\cal V}^{(h)}(\t,\PP,T)\;,\label{2.49}\ee
where ${\bf T}$ is a special family of graphs on the set of points
$\xx_{v_0}$, obtained by putting together an anchored tree graph
$T_v$ for each non trivial vertex $v$. Note that any graph $T\in
{\bf T}$ becomes a tree graph on $\xx_{v_0}$, if one identifies
all the points in the sets $\xx_v$, with $v$ an endpoint.
Given $\t\in\TT_{h,n}$ and the labels $\PP,T$,
calling $v_i^*,\ldots,v_n^*$ the endpoints of $\t$ and putting
$h_i=h_{v_i^*}$, the explicit representation of $W_{\t,\PP,T}^{(h)}
(\xx_{v_0})$ in (\ref{2.49}) is
\bea &&
W_{\t,\PP, T}(\xx_{v_0}) =\left[\prod_{i=1}^n
K_{v_i^*}^{(h_i)} (\xx_{v_i^*})\right] \;\cdot\label{2.50}\\ &&\cdot\;
\Bigg\{\prod_{v\,\atop\hbox{\ottorm not e.p.}}{1\over s_v!} \int
dP_{T_v}({\bf t}_v)\;{\rm det}\, G^{h_v,T_v}({\bf t}_v)\Biggl[
\prod_{l\in T_v} \d_{\o^-_l,\o^+_l} \d_{\s^-_l,\s^+_l}\,
\big[g^{(h)}_{\o_l}(\xx_l-\yy_l)\big]_{\r^-_l,\r^+_l}\,\Biggr]
\Bigg\}\;,\nn\eea
{\it Analyticity of the effective potentials.} The tree expansion
described above allows us to express the effective potential
${\cal V}^{(h)}$ in terms of the {\it running coupling constants}
$\z_h, c_h$ and of the {\it renormalization functions}
$\s_k(\kk),t_{k,\o}(\kk)$.

The next goal will be the proof of the following result.\\
\\
{\cs Theorem 2.} {\it There exists a constants $U_0>0$ such that,
if $|U|\le U_0$, then the kernels
$W^{(h)}_{2l,\ul\r,\ul\o}
(\xx_1,\ldots,\xx_{2l})$ in (\ref{2.29}), $h\le -1$,
are analytic functions of $U$,
satisfying, for any $0\le \th<1$ and a suitable constant $C>0$, the following
estimates:
\be \frac1{\b|\L|}\int d\xx_1\cdots d\xx_{2l}|W^{(h)}_{2l,\ul\r,\ul\o}
(\xx_1,\ldots,\xx_{2l})|\le
\g^{h (3-2l+\th)} \,(C\,|U|)^{max(1,l-1)}\;.\label{2.52}\ee
Moreover, the constants $e_{h}$ and $\lis e_h$ defined by (\ref{2.35})
and (\ref{2.39}) are also analytic functions of $U$ in the domain $|U|\le U_0$,
and there they satisfy the estimate
$|\lis e_{h}|+|e_{h}|\le C U_0 \g^{h(3+\th)}$.}
\\
\\
{\bf Remark.}
The above result immediately implies the
analyticity of the specific free energy
$f_\b(U)$ and of its zero temperature limit $e(U)$, i.e., of the
specific ground state energy. In fact, by construction,
$f_\b(U)=F_0+\sum_{h=h_\b}^0
(e_h+\lis e_h)$, with $F_0$ an analytic function of $U$, see the discussion
after (\ref{2.16a}) and in Appendix \ref{B}. Therefore, Theorem 2 implies the
part of the statement of Theorem 1 concerning the free energy and the ground
state energy. For the proof of analyticity of the Schwinger functions, see next
Section.
\*
{\cs Proof of Theorem 2.}
Let us preliminarily assume that, for $h\le -1$,
and for suitable constants $c, c_n$, the corrections $z_h,\d_h,\s_h(\kk')$
and $\t_h(\kk')$ defined in \pref{2.31} and \pref{2.33}, satisfy the following
estimates:
\bea &&\hskip1.25truecm\max\ \{|z_h|, |\d_h|\}\le c |U|\g^{\th h}
\;,\label{2.51}\\
&&\sup_{|\kk'| s.t.
\c_h(\kk')\not=0}\{||\partial_{\kk'}^n
\s_h(\kk')||,||\partial_{\kk'}^n \t_{h,\o}(\kk')||\}\le c_n
|U|\g^{(1+\th-n)h}\;.\nn\eea
Using \pref{2.51} we inductively see that the running coupling functions
$\z_h,c_h,s_h(\kk')$ and $t_h(\kk')$ satisfy similar estimates:
\bea &&\hskip1.25truecm\max\ \{|\z_h-1|,|c_h-3/2|\}\le c |U|\;,\label{2.51a}\\
&&\sup_{|\kk'| s.t.
\c_h(\kk')\not=0}\{||\partial_{\kk'}^n
s_h(\kk')||,||\partial_{\kk'}^n t_{h,\o}(\kk')||\}\le c_n
|U|\g^{(1+\th-n)h}\;.\nn\eea
Now, using the definition of $g_\o^{(h)}$, see (\ref{2.37}) and
(\ref{2.34aa}), we get, after integration by parts, for any $N\ge 0$,
\be \big|\!\big|\big[\dpr^{n_0}_{x_0}\dpr^{n_1}_{x_1}\dpr^{n_2}_{x_2}
g^{(h)}_{\o}(\xx_1,\xx_2)\big]_{\r,\r'}\big|\!\big|
\le C_{N,n}{\g^{(2+n)h}\over 1+(\g^{h}|\xx_1-\xx_2|)^N}\;,\label{2.38}\ee
where $n=n_0+n_1+n_2\ge 0$ and $C_{N,n}$ is a suitable constant.

Using the tree expansion described above
and, in particular, Eqs.(\ref{2.41}), (\ref{2.43}), (\ref{2.49}) and
(\ref{2.50}), we find that the l.h.s. of (\ref{2.52}) can be bounded from above
by
\bea && \sum_{n\ge 1}\sum_{\t\in {\cal T}_{h,n}}
\sum_{\PP\in{\cal P}_\t\atop |P_{v_0}|=2l}\sum_{T\in{\bf T}}
\int\prod_{l\in T^*}
d(\xx_l-\yy_l) \left[\prod_{i=1}^n|K_{v_i^*}^{(h_i)}(\xx_{v_i^*})|\right]\cdot
\label{2.53}\\
&&\hskip3.truecm\cdot\Bigg[\prod_{v\ {\rm not}\ {\rm e.p.}}{1\over s_v!}
\max_{{\bf t}_v}\big|{\rm det}\, G^{h_v,T_v}({\bf t}_v)\big|
\prod_{l\in T_v}
\big|\big|g^{(h)}_{\o_l}(\xx_l-\yy_l)\big|\big|\Bigg]\nn\eea
where $||\cdot||$ is the spectral norm and where $T^*$ is a tree graph
obtained from $T=\cup_vT_v$, by adding
in a suitable (obvious) way, for each endpoint $v_i^*$,
$i=1,\ldots,n$, one or more lines connecting the space-time points
belonging to $\xx_{v_i^*}$.

A standard application of Gram--Hadamard inequality, combined with
the dimensional bound on $g^{(h)}_\o(\xx)$ given by (\ref{2.38}),
see the remark after (\ref{2.48}), implies that
\be |{\rm det} G^{h_v,T_v}({\bf t}_v)| \le
c^{\sum_{i=1}^{s_v}|P_{v_i}|-|P_v|-2(s_v-1)}\cdot\;
\g^{{h_v}
\left(\sum_{i=1}^{s_v}|P_{v_i}|-|P_v|-2(s_v-1)\right)}\;.\label{2.54}\ee
By the decay properties of $g^{(h)}_\o(\xx)$ given by (\ref{2.38}), it
also follows that
\be \prod_{v\ {\rm not}\ {\rm e.p.}}
{1\over s_v!}\int \prod_{l\in T_v} d(\xx_l-\yy_l)\,
||g^{(h_v)}_{\o_l}(\xx_l-\yy_l)||\le c^n \prod_{v\ {\rm not}\ {\rm e.p.}}
\fra{1}{s_v!} \g^{-h_v(s_v-1)}\;.\label{2.55}\ee
The bound (\ref{2.17}) on the kernels produced by the ultraviolet integration
implies that
\be \int\prod_{l\in T^*\setminus\cup_v T_v}d(\xx_l-\yy_l)
\prod_{i=1}^n |K_{v_i^*}^{(h_i)}(\xx_{v_i^*})|\le \prod_{i=1}^n C^{p_i}
|U|^{\,\frac{p_i}2-1}\;,\label{2.56}\ee
where $p_i=|P_{v_i^*}|$. Combining the previous bounds, we find that
(\ref{2.53}) can be bounded above by
\be  \sum_{n\ge 1}\sum_{\t\in {\cal T}_{h,n}}
\sum_{\PP\in{\cal P}_\t\atop |P_{v_0}|=2l}\sum_{T\in{\bf T}}
C^n \Big[\prod_{v\ {\rm not}\ {\rm e.p.}} \fra{1}{s_v!}
\g^{{h_v}\left(\sum_{i=1}^{s_v}|P_{v_i}|-|P_v|-3(s_v-1)\right)}\Big]
\Big[\prod_{i=1}^n C^{p_i}
|U|^{\,\frac{p_i}2-1}\Big]
\label{2.57}\ee
Let us define $n(v)=\sum_{i: v_i^*>v}\,1$ as the number of endpoints following
$v$ on $\t$ and $v'$ as
the vertex immediately preceding $v$ on $\t$. Recalling that $|I_v|$
is the number of field labels associated to the endpoints following $v$
on $\t$ (note that $|I_v|\ge 4 n(v)$) and using that
\bea && \sum_{v\ {\rm not}\ {\rm e.p.}}\Big[\big(\sum_{i=1}^{s_v}
|P_{v_i}|\big)-|P_v|\Big]=|I_{v_0}|-|P_{v_0}|\;,\nn\\
&&
\sum_{v\ {\rm not}\ {\rm e.p.}}(s_v-1)=n-1\;,\label{2.58}\\
&& \sum_{v\ {\rm not}\ {\rm e.p.}}
(h_v-h)\Big[\big(\sum_{i=1}^{s_v}
|P_{v_i}|\big)-|P_v|\Big]=\sum_{v\ {\rm not}\ {\rm e.p.}}
(h_v-h_{v'})(|I_v|-|P_v|)\;,\nn\\
&&\sum_{v\ {\rm not}\ {\rm e.p.}}(h_v-h)(s_v-1)=
\sum_{v\ {\rm not}\ {\rm e.p.}}(h_v-h_{v'})(n(v)-1)\;,\nn\eea
we find that (\ref{2.57}) can be bounded above by
\bea &&\sum_{n\ge 1}\sum_{\t\in {\cal T}_{h,n}}
\sum_{\PP\in{\cal P}_\t\atop |P_{v_0}|=2l}\sum_{T\in{\bf T}}
C^n  \g^{h(3-|P_{v_0}|+|I_{v_0}|-3n)}\cdot\nn\\
&&\hskip1.truecm \cdot
\Big[\prod_{v\ {\rm not}\ {\rm e.p.}} \fra{1}{s_v!}
\g^{(h_v-h_{v'})(3-|P_v|+|I_v|-3n(v))}\Big]
\Big[\prod_{i=1}^n C^{p_i}
|U|^{\,\frac{p_i}2-1}\Big]\;.\label{2.59}\eea
Finally, let $\bar n(v)$ be the number of endpoints following
$v$ but not following any of the vertices $w>v$ and let $\bar p(v)$ be
the number of field labels associated to endpoints following
$v$ but not following any of the vertices $w>v$. Using the identities
\bea &&\g^{h  n}
\prod_{v\ {\rm not}\ {\rm e.p.}}
\g^{(h_v-h_{v'}) n(v)}=\prod_{v\ {\rm not}\ {\rm e.p.}}
\g^{h_v \bar n(v)}\;,\nn\\
&& \g^{h  |I_{v_0}|}
\prod_{v\ {\rm not}\ {\rm e.p.}}
\g^{(h_v-h_{v'}) |I_v|}=\prod_{v\ {\rm not}\ {\rm e.p.}}
\g^{h_v \bar p(v)}\;,\label{2.60}\eea
we obtain
\bea&& \frac1{\b|\L|}\int d\xx_1\cdots d\xx_{2l}|W^{(h)}_{2l,\ul\r,\ul\o}
(\xx_1,\ldots,\xx_{2l})|\le
\sum_{n\ge 1}\sum_{\t\in {\cal T}_{h,n}}
\sum_{\PP\in{\cal P}_\t\atop |P_{v_0}|=2l}\sum_{T\in{\bf T}}
C^n  \g^{h(3-|P_{v_0}|)}\cdot\nn\\
&&\hskip.2truecm \cdot \Big[\prod_{v\ {\rm not}\ {\rm e.p.}}
\fra{1}{s_v!} \g^{-(h_v-h_{v'})(|P_v|-3)}\Big]\Big[\prod_{v\ {\rm not}\ {\rm
e.p.}} \g^{h_v(\bar p(v)-3\bar n(v))} \Big] \Big[\prod_{i=1}^n
C^{p_i} |U|^{\,\frac{p_i}2-1}\Big]\;.\label{2.61} \eea
Note that, if $v$ is not an endpoint, $|P_v|-3\ge 1$ by the
definition of $\RR$. Moreover $\bar p(v)-3\bar n(v)\ge 0$ and
$\sum_{v\ {\rm not}\ {\rm e.p.}} (\bar p(v)-3\bar n(v))\ge n$; in particular,
this means that there exists at least one vertex $v^*$ that is not an endpoint,
such that $\bar p(v^*)-3\bar n(v^*)\ge 1$. Therefore, we get
\be \prod_{v\ {\rm not}\ {\rm
e.p.}} \g^{h_v(\bar p(v)-3\bar n(v))}\le \g^{h_*} \label{2.61a}\;,\ee
with $h_*$ the highest scale label of the tree. Now, note that
the number of terms in $\sum_{T\in {\bf T}}$ can be bounded by
$C^n\prod_{v\ {\rm not}\ {\rm
e.p.}} s_v!$. Using also that $|P_v|-3\ge 1$ and $|P_v|-3\ge|P_v|/4$,
we find that the l.h.s. of (\ref{2.61}) can be bounded as
\bea&& \frac1{\b|\L|}\int d\xx_1\cdots d\xx_{2l}|W^{(h)}_{2l,\ul\r,\ul\o}
(\xx_1,\ldots,\xx_{2l})|\le \g^{h(3-|P_{v_0}|)}
\sum_{n\ge 1}C^n\sum_{\t\in {\cal T}_{h,n}}\g^{h_{*}}\cdot\label{2.61b}\\
&&\cdot\big(
\prod_{v\ {\rm not}\ {\rm e.p.}}
\g^{-\th(h_v-h_{v'})}\g^{-(1-\th)(h_v-h_{v'})/2}\big)
\sum_{\PP\in{\cal P}_\t\atop |P_{v_0}|=2l}\big(\prod_{v\ {\rm not}\ {\rm e.p.}}
\g^{-(1-\th)|P_v|/8}\big)\prod_{i=1}^n
C^{p_i} |U|^{\,\frac{p_i}2-1}\;.\nn \eea
Now, the sum over $\PP$ can be bounded using the following combinatorial
inequality (see for instance \S A6.1 of \cite{GM}): let $\{p_v, v\in \t\}$, 
with $\t\in\TT_{h,n}$, a set of integers such that
$p_v\le \sum_{i=1}^{s_v} p_{v_i}$ for all $v\in\t$ which are not endpoints;
then, if $\a>0$,
$$\prod_{\rm v\;not\; e.p.} \sum_{p_v} \g^{-{\a p_v}}
\le C_\a^n\;.$$
This implies that
$$\sum_{\PP\in{\cal P}_\t\atop
|P_{v_0}|=2l}\big(\prod_{v\ {\rm not}\ {\rm e.p.}}
\g^{-(1-\th)|P_v|/8}\big)\prod_{i=1}^n
C^{p_i} |U|^{\,\frac{p_i}2-1}\le C_\th^n|U|^n\;.$$
Finally, using that $\g^{h_{*}}
\prod_{v\ {\rm not}\ {\rm e.p.}}
\g^{-\th(h_v-h_{v'})}\le \g^{\th h}$, and that, for $0<\th<1$,
$$\sum_{\t\in {\cal T}_{h,n}}
\prod_{v\ {\rm not}\ {\rm e.p.}}
\g^{-(1-\th)(h_v-h_{v'})/2}\le C^n\;,$$
as it follows by the fact that the number of non trivial vertices in $\t$ 
is smaller than $n-1$ and that the number of trees in ${\cal T}_{h,n}$ is 
bounded by ${\rm const}^n$, and collecting all the previous bounds, we obtain
\be \frac1{\b|\L|}\int d\xx_1\cdots d\xx_{2l}|W^{(h)}_{2l,\ul\r,\ul\o}
(\xx_1,\ldots,\xx_{2l})|\le \g^{h(3-|P_{v_0}|+\th)}
\sum_{n\ge 1}C^n |U|^n\;,\label{2.61e}\ee
which is the desired result.

It remains to prove the assumption \pref{2.51}. We proceed by induction.
The assumption is valid for $h=0$, as it follows by (\ref{2.17}) and by
the discussion in Appendix \ref{B}. Now,
assume that \pref{2.51} is valid for all $h\ge k+1$, and let us prove it
for $k-1$. The functions $-iz_k k_0+\s_k(\kk')$ and $\d_k(ik'_1-\o k'_2)+
\t_{k,\o}(\kk')$ admit a representation in terms of
$W^{(k)}_{2,\ul\r,(\o,\o)}(\xx,\yy)$. In particular,
\be \max\{|z_k|,|\d_k|\}\le \frac1{\b|\L|}\int d\xx_1d\xx_{2}
|\xx-\yy||W^{(k)}_{2,\ul\r,(\o,\o)}(\xx,\yy)|\;,\label{2.61c}\ee
and
\be\sup_{|\kk'| s.t.
\c_k(\kk')\not=0}\{||\partial_{\kk'}^n
\s_k(\kk')||,||\partial_{\kk'}^n \t_{k,\o}(\kk')||\}\le C\g^{2k}
\frac1{\b|\L|}\int d\xx_1d\xx_{2}
|\xx-\yy|^{n+2}|W^{(k)}_{2,\ul\r,(\o,\o)}(\xx,\yy)|\;.\label{2.61d}\ee
The same proof leading to (\ref{2.61e}) shows that the r.h.s. of (\ref{2.61c})
can be bounded by the r.h.s. of (\ref{2.61e}) times $\g^{-k}$ (that is
the dimensional estimate for $|\xx-\yy|$), and that
the r.h.s. of (\ref{2.61c})
can be bounded by the r.h.s. of (\ref{2.61e}) times $\g^{2k}\g^{-(n+2)k}$
(where $\g^{-k(n+2)}$ is
the dimensional estimate for $|\xx-\yy|^{n+2}$).
This concludes the proof of Theorem 2.\qed

\subsection{The two point Schwinger function}\label{IIId}

In this section we describe how to modify the expansion for the
free energy described in previous sections in order to compute the
Schwinger functions at distinct space-time points.
For simplicity, we shall restrict our attention to
the case of the two point Schwinger function. The general case can be worked
out along the same lines.

The Schwinger functions can be derived from the {\it generating
function} defined as
\be \WW(\phi)=\log\int P(d\Psi)e^{-\VV(\psi)+\int d\xx
\left[\phi^+_{\xx,\s,\r} \Psi^-_{\xx,\s,\r}+\Psi^+_{\xx,\s,\r}
\phi^-_{\xx,\s,\r}\right]}
\ee
where summation over repeated indices is understood and
the variables $\phi^\e_{\xx,\s,\r}$ are Grassmann variables,
anticommuting among themselves and with the variables
$\Psi^\e_{\xx,\s,\r}$.
The two--point Schwinger
function $S(\xx-\yy)_{\r,\r'}\defin
S_2(\xx,\s,-,\r;\yy,\s,+,\r')$ is given by
\be S(\xx-\yy)_{\r,\r'}={\dpr^2\over \dpr\phi^+_{\xx,\s,\r}
\dpr\phi^-_{\yy,\s,\r'}}\WW(\phi)\Big|_{\phi=0} \;.\label{3.0}\ee
We start by studying the generating function and, in analogy with
the procedure described before, we begin by decomposing the field
$\Psi$ in an ultraviolet and an infrared component:
$\Psi=\Psi^{(u.v.)}+\Psi^{(i.r.)}$, with $\Psi^{(i.r.)\pm}_{\xx,\s,\r}=
\sum_{\o=\pm}e^{i\vec p_F^\o\vec x}\Psi^{(\le 0)\pm}_{\xx,\s,\r,\o}$.
After the integration of the
$\Psi^{(u.v.)}$ variables, and after rewriting $\phi^{\pm}_{\xx,\s,\r}=
\sum_{\o=\pm}e^{i\vec p_F^\o\vec x}\phi^{\pm}_{\xx,\s,\r}$, we get:
\bea &&e^{{\cal W}(\phi)} = e^{-\b|\L| F_0+S^{(\ge 0)}(\phi)}\int
P_{\c_0,A_0}(d\Psi^{(\le 0)}) \cdot\label{3.1}\\
&&\cdot e^{-{\cal V}^{(0)}(\psi^{(\le 0)})-B^{(0)}(\Psi^{(\le 0)},\,\phi)
+\int d\xx\left[\phi^+_{\xx,\s,\r,\o} \Psi^{(\le 0)-}_{\xx,\s,\r,\o}+
\Psi^{(\le 0)+}_{\xx,
\s,\r,\o}\phi^-_{\xx,\s,\r,\o}\right]}\nn\eea
where $S^{(\ge 0)}(\phi)$ (chosen in such a way that $S^{(\ge 0)}(0)=0$)
collects the terms depending on $\phi$
but not on $\Psi^{(\le 0)}$ and $B^{(0)}(\Psi^{(\le 0)},\phi)$ the
terms depending both on $\phi$ and $\Psi^{(\le 0)}$ generated by the
ultraviolet integration.

Proceeding as in Section \ref{IIIc}, we inductively show (see below for
details) that, if $h\le 0$, $e^{\WW(\phi)}$ can be rewritten as:
\bea &&e^{{\cal W}(\phi)} = e^{-\b|\L| F_h+S^{(\ge h)}(\phi)} \int
P_{\c_h,A_h} (d\Psi^{(\le h)}) \cdot\label{3.1a}\\
&&\cdot e^{-{\cal V}^{(h)}(\Psi^{(\le h)})-B^{(h)}(\Psi^{(\le h)},\,\phi)
+\int d\kk'\big[\hat \phi^+_{\kk',\s,\r,\o} \hat Q^{(h+1)}_{\kk',\o,\r,\r'}
\hat \Psi^{(\le
h)-}_{\kk',\s,\r',\o}+\hat \Psi^{(\le
h)+}_{\kk',\s,\r,\o} \hat Q^{(h+1)}_{\kk',\o,\r,\r'}
\hat \phi^{-}_{\kk',\s,\r',\o}\big]}\nn
\eea
where: $\int d\kk'$ must be interpreted as equal to
$(\b|\L|)^{-1}\sum_{\kk\in\DD_{\b,L}^\o}$;
$B^{(h)}(\Psi^{(\le h)},\phi)$
can be written as $B^{(h)}_\phi(\Psi^{(\le h)})+W_R^{(h)}$,
with $W_R^{(h)}$ containing the terms of third or higher
order in $\phi$ and $B^{(h)}_\phi(\Psi^{(\le h)})$ of the form
\bea && \int d\xx\Big[
\phi^+_{\cdot,\s,\r_1,\o}*\,G^{(h+1)}_{\o,\r_1,\r_2}*\,\frac{\partial
{\cal V}^{(h)}(\Psi^{(\le h)})}{\partial\Psi^{(\le h)+}_{\cdot,\s,\r_2,\o}}+
\frac{\partial
{\cal V}^{(h)}(\Psi^{(\le h)})}{\partial\Psi^{(\le h)-}_{\cdot,\s,\r_1,\o}}
\,*G^{(h+1)}_{\o,\r_1,\r_2}*\phi^-_{\cdot,\s,\r_2,\o}+\label{3.2}\\
&&+\phi^+_{\cdot,\s_1,\r_1,\o_1}*G^{(h+1)}_{\o_1,\r_1,\r_2}*
\frac{\dpr^2}{\dpr\Psi^{(\le h)+}_{\cdot,\s_1,\r_2,\o_1} \dpr\Psi^{(\le
h)-}_{\cdot,\s_2,\r_3,\o_2}}\RR{\cal V}^{(h)}(\Psi^{(\le
h)})*G^{(h+1)}_{\o_2,\r_3,\r_4}*\phi^-_{\cdot,\s_2,\r_4,\o_2}\Big]
\;,\nn\eea
with
\be \hat G^{(h+1)}_{\o,\r,\r'}(\kk')=\sum_{k= h+1}^1
\hat g^{(k)}_{\o,\r,\r''}(\kk')\hat Q^{(k)}_{\kk',\o,\r'',\r'}
\label{3.3}\ee
and $\hat Q^{(h)}_{\kk',\o,\r,\r'}$ defined inductively by the
relations
\be \hat Q^{(h)}_{\kk',\o,\r,\r'}=\hat
Q^{(h+1)}_{\kk',\o,\r,\r'}-W_{2,\r,\r'',(\o,\o)}^{(h)}(\kk') \hat
G^{(h+1)}_{\o,\r'',\r'}(\kk')\;,
\quad\quad Q^{(1)}_{\kk',\o,\r,\r'}\=\d_{\r,\r'}\;,\label{3.4}\ee
where $W_{2,\ul\r,\ul\o}^{(h)}$ is the kernel of
${\cal L}{\cal V}^{(h)}$, as defined in
(\ref{2.30a}). In (\ref{3.3}), $\hat g^{(1)}_\o$ is defined as
$$\hat g^{(1)}_\o(\kk')=\hat g^{(u.v.)}(\kk'+\pp_F^\o)\Big[\openone
\big(||\kk'||<||\kk'+\pp_F^\o-\pp_F^{-\o}||\big)+\frac12\openone
\big(||\kk'||=||\kk'+\pp_F^\o-\pp_F^{-\o}||\big)\Big]\;,$$
where $\pp_F^\o\defin(0,\vec p_F^\o)$.
Note that, by the compact
support properties of $\hat g^{(h)}_\o(\kk')$, if $\hat g^{(h)}_\o(\kk')
\not=0$, $h<0$, then $\hat
g^{(j)}(\kk)=0$ for $|j-h|
>1$, so that
$$\hat Q^{(h)}_{\kk',\o,\r,\r'}=1 - \hat W_{2,\r,\r_1,(\o,\o)}^{(h)}(\kk')
\hat g^{(h+1)}_{\o,\r_1,\r_2}(\kk')\hat Q^{(h+1)}_{\kk',\o,\r_2,\r'}
\;, $$
and, therefore, proceeding by induction, we see that on the support of
$\hat g^{(h)}_\o(\kk')$ we have
\be ||\hat Q^{(h)}_{\kk',\o}-1||\le C|U|\g^{\th h}\;,\quad\quad
||\dpr_{\kk'}^n\hat Q^{(h)}_{\kk',\o}||\le C_n|U|\g^{(\th-n)h}\;.\label{3.5}\ee
In order to derive \pref{3.5}, we used Theorem 2 and
the decay bounds (\ref{2.38}).

Using \pref{3.5}, the definition \pref{3.3} and the decay bounds
(\ref{2.38}), we find that
\be \int d\xx\,|\xx|^j\,||G^{(h)}_\o(\xx)||\le C_j\g^{-(1+j)h}\;.\label{3.6}\ee
Let us now prove (\ref{3.1a}). We proceed by induction. For $h=0$
(\ref{3.1a}) is clearly true (it coincides with \pref{3.1}).
Assuming inductively that the representation (\ref{3.1a})
is valid up to a certain value of $h<0$,
we can show that the same representation is valid for $h-1$. In fact,
we can rewrite the term ${\cal V}^{(h)}$ in the exponent of (\ref{3.1a}) as
${\cal V}^{(h)}={\cal L}{\cal V}^{(h)}+{\cal R}{\cal V}^{(h)}$, as in
\pref{2.loc}, and we can
``absorb'' ${\cal L}{\cal V}^{(h)}$
in the fermionic integration, as explained in Section \ref{IIIc}, see
(\ref{2.32})--(\ref{2.34}). Similarly we rewrite
\be
\fra\dpr{\dpr\Psi^{(\le
h)\pm}_{\xx,\s,\r,\o}}{\cal V}^{(h)}(\Psi^{(\le h)})= \int d\yy\,
W_{2,(\r,\r'),(\o,\o)}^{(h)}(\xx,\yy)\Psi^{(\le
h)\mp}_{\yy,\s,\r',\o}+ \fra\dpr{\dpr\Psi^{(\le
h)\pm}_{\xx,\s,\r,\o}}\RR{\cal V}^{(h)}(\Psi^{(\le h)})\;,\label{3.7}\ee
This rewriting
induces a decomposition of the first line of \pref{3.2} into two pieces,
the first proportional to $W_2^{(h)}$, the second
identical to the first line of \pref{3.2} itself, with ${\cal V}^{(h)}$
replaced by $\RR{\cal V}^{(h)}$, that we will call
$\RR B_\phi^{(h)}(\Psi^{(\le h)})$.
We choose to ``absorb'' the term proportional to
$W_2^{(h)}$ into the definition of $Q^{(h)}$, and this gives the recursion
relation (\ref{3.4}). Moreover, note that combining
$\RR B_\phi^{(h)}(\Psi^{(\le h)})$ with $\RR\VV^{(h)}(\Psi^{(\le h)})$
we find:
\be \RR\VV^{(h)}(\Psi^{(\le h)})+\RR B_\phi^{(h)}(\Psi^{(\le h)})=
\RR\VV^{(h)}(\Psi^{(\le h)}+G^{(h+1)}*\phi)+W_{R,1}^{(h)}\;,\label{3.7a}\ee
with $W_{R,1}^{(h)}$ containing terms of third or higher
order in $\phi$. We define $\lis W_R^{(h)}=W_{R}^{(h)}+W_{R,1}^{(h)}$.

After these splittings and redefinitions, we can rewrite (\ref{3.1a}) as
\bea&& e^{{\cal W}(\phi)} = e^{-\b|\L| (F_h+e_h)+S^{(\ge h)}(\phi)} \int
P_{\c_{h-1},A_{h-1}} (d\Psi^{(\le h-1)})
\int
P_{f^{-1}_h,\lis A_{h-1}} (d\Psi^{(h)}) \cdot\label{3.7b}\\
&&\cdot e^{-\RR{\cal V}^{(h)}(\Psi^{(\le h)}+G^{(h+1)}*\phi)-
\lis W^{(h)}_R
+\int d\kk'\big[\hat \phi^+_{\kk'}
\hat Q^{(h)}_{\kk'}
\hat \Psi^{(\le h)-}_{\kk'}
+\hat \Psi^{(\le h)+}_{\kk'}
\hat Q^{(h)}_{\kk'}
\hat \phi^{-}_{\kk'}
\big]}\;.\nn\eea
Integrating the field $\Psi^{(h)}$, we get the analogue of (\ref{2.39}):
\bea&& \int P_{f^{-1}_h,\lis A_{h-1}}
(d\Psi^{(h)})\,
e^{-\RR{\cal V}^{(h)}(\Psi^{(\le h)}+G^{(h+1)}*\phi)-
\lis W^{(h)}_R
+\int d\kk'
\big[\hat \phi^+_{\kk'}\hat Q^{(h)}_{\kk'}\hat \Psi^{(\le h)-}_{\kk'}
+\hat \Psi^{(\le h)+}_{\kk'}\hat Q^{(h)}_{\kk'}\hat \phi^{-}_{\kk'}
\big]}
=\nn\\
&&=e^{-\lis e_h \b|\L|
-{\cal V}^{(h-1)}(\Psi^{(\le h-1)}+G^{(h)}*\phi)
+\int d\kk'
\hat \phi^+_{\kk'}\hat Q^{(h)}_{\kk'}\hat g^{(h)}(\kk')
\hat Q^{(h)}_{\kk'}\hat \phi^-_{\kk'}-W_{R,2}^{(h-1)}}\cdot\label{3.7c}\\
&&\hskip7.truecm\cdot e^{\int d\kk'\big[
\hat \phi^+_{\kk'}\hat Q^{(h)}_{\kk'}\hat \Psi^{(\le h-1)-}_{\kk'}
+\hat \Psi^{(\le h-1)+}_{\kk'}\hat Q^{(h)}_{\kk'}\hat \phi^{-}_{\kk'}
\big]}
\;,\nn\eea
with $G^{(h)}$ defined by the recursion relation (\ref{3.3}) and
$W_{R,2}^{(h-1)}$ a term of third or higher order in $\phi$.
Eq.(\ref{3.7c}) can be proved by making use of
a formal change of Grassmann variables $\hat \Psi'_{\kk'}=\hat \Psi_{\kk'}-
\hat g^{(h)}(\kk')Q^{(h)}_{\kk'}\hat \phi_{\kk'}$,
as described in Ch.4 of \cite{BG}. At this point it is straightforward to
check that the final expression for $e^{\cal W(\phi)}$
that we end up with is given by the r.h.s.
of (\ref{3.1a}), with $h$ replaced by $h-1$, and the inductive assumption
is proved.\\

From the definitions and the construction above, we get
\bea&& S_{\r,\r'}(\xx-\yy)=\sum_{\o=\pm}e^{-i\vec p_F^\o(\vec
x-\vec y)}S_{\o,\r,\r'}(\xx-\yy)\=
\sum_{\o=\pm}e^{-i\vec p_F^\o(\vec
x-\vec y)}\cdot\label{3.7d}\\
&&\cdot\sum_{h=-\io}^1
\Big[\big(Q^{(h)}_{\o,\r,\r_1}*g^{(h)}_{\o,\r_1,\r_2}*Q^{(h)}_{\o,
\r_2,\r'}\big)(\xx-\yy)-
\big(G^{(h)}_{\o,\r,\r_1}*W^{(h-1)}_{2,(\r_1,\r_2),(\o,\o)}*G^{(h)}_{\o,
\r_2,\r'}\big)(\xx-\yy)\Big]\;.\nn\eea
Analyticity of $S_{\r,\r'}(\xx-\yy)$ follows from this representation and
the results of Theorem 2. Concerning the representation (\ref{1.9}), let us
take the Fourier transform of $S_{\o,\r,\r'}(\xx-\yy)$. If we define
$h_\kk=\min\{ h: \hat g^{(h)}_\o(\kk')\not\=0\}$, we get,
for $\kk'$ inside the support of $\Psi^{(\le 0)}_{\kk',\s,\r,\o}$
\bea \hat S_{\o,\r,\r'}(\kk')&=& \sum_{j=h_\kk}^{h_\kk+1}
Q^{(j)}_{\kk',\o,\r,\r_1} g^{(j)}_{\o,\r_1,\r_2}(\kk')^{(j)}
Q^{(j)}_{\kk',\o,\r_2,\r'}-\nn\\
&-&  \sum_{j=h_\kk}^{h_\kk+1}
G^{(j)}_{\o,\r,\r_1}(\kk')
W^{(j-1)}_{2,(\r_1,\r_2),(\o,\o)}(\kk')G^{(j)}_{\o,\r_2,\r'}(\kk')
\;,\label{4.11} \eea
which readily implies (\ref{1.9}): in fact, using the explicit
expression of $g^{(h)}_\o$ and the inductive bounds on $Q^{(h)}$, see
(\ref{3.5}), it is easy to see that the term in the first line of (\ref{4.11})
can be written as in (\ref{1.9}) and that their only singularity is located
at $\kk'=\V0$.

The contributions from
the second line can be bounded using the bounds on $W_2^{(h)}$ proved in
Theorem 2, and we find that they can be bounded by
$C|U|\g^{h_{\kk'}(-1+\th)}$, which means that they only contribute
to the error term appearing in (\ref{1.9}). This also implies that no other
singularity, besides the one at the Fermi points, can be produced by such
terms.

Finally, if $\kk$ does not belong to the support of $\Psi^{(\le 0)}$,
we can write
\be \hat S_{\r,\r'}(\kk)=\hat S_{\r,\r'}^{(u.v.)}(\kk)=
g^{(u.v.)}(\xx-\yy)-
\big(g^{(u.v.)}_{\r,\r_1}*W_{2,(\r_1,\r_2)}*g^{(u.v.)}_{\r_2,\r'}\big)
(\xx-\yy)\;,\label{4.12}\ee
with $W_{2,\ul\r}$ defined by (\ref{2.16}). The bounds discussed in Section
\ref{IIIb} and Appendix \ref{B} imply that $S^{(u.v.)}_{\r,\r'}(\xx-\yy)$
decays faster than any power, so that no singularity can appear in its Fourier
transform.

A similar expansion can be obtained for higher order Schwinger functions,
but we will not belabor the details here. This concludes the proof of Theorem
1.\qed

\appendix
\section{The non-interacting theory}\lb{A}
\setcounter{equation}{0}
\renewcommand{\theequation}{\ref{A}.\arabic{equation}}

In this Appendix we give some details about the computation of the Schwinger
functions of the non interacting theory, i.e., of model (\ref{1.1}) with $U=0$.
In this case the Hamiltonian of interest reduces to
\be H_{0,\L}=
-\sum_{\vec x\in \L\atop i=1,2,3}\sum_{\s=\uparrow\downarrow} \Big(
a^+_{\vec x,\s} b_{\vec x+\vec \d_i,\s}^-+
b_{\vec x+\vec \d_i,\s}^+ a^-_{\vec x,\s} \Big)\;,\label{A.1}\ee
with $\L$, $a^{\pm}_{\vec x,\s}$, $b^{\pm}_{\vec x+\vec\d_i,\s}$ defined
as in items (1)--(4) after (\ref{1.1}).

First of all, let us remind that, being $H_{0,\L}$ quadratic, the $2n$-point
Schwinger functions satisfy the Wick rule, i.e.,
\bea&& \media{{\bf T}\{\Psi_{\xx_1,\s_1,\r_1}^-\cdots \Psi_{\xx_n,\s_n,\r_n}^-
\Psi_{\yy_1,\s_1',\r_1'}^+\cdots\Psi_{\yy_n,\s_n',\r_n'}^+\}}_{\b,\L}
=-\det G\;,\nn\\
&& G_{ij}=\d_{\s_i\s_j'}\media{{\bf T}\{\Psi^-_{\xx_i,\s_i,\r_i}
\Psi^+_{\yy_j,\s'_j,\r'_j}\}}_{\b,\L}
\;.
\label{A.2}\eea
Moreover, every $n$--point
Schwinger function $S_n^{\b,\L}(\xx_1,\e_1,\s_1,\r_1;\ldots;\xx_n,\e_n,\s_n,
\r_n)$ with $\sum_{i=1}^n\e_i\neq 0$
is identically zero. Therefore, in order to construct the whole set of
Schwinger functions of $H_{0,\L}$, it is enough to compute the $2$--point
function $S_0^{\b,\L}(\xx-\yy)=\media{{\bf T}\{\Psi^-_{\xx,\s,\r}
\Psi^+_{\yy,\s,\r'}\}}_{\b,\L}$, and in order to do this, it is convenient
to first diagonalize $H_{0,\L}$. Let us proceed as follows.

We identify $\L$ with the set of vectors in a fundamental cell, and
we write
\be \L= \{n_1 {\vec a_1}+n_2 {\vec a_2}\ :\ 0\le n_1,n_2\le L-1\} \;,
\label{A.2a}\ee
with $\vec a_{1}=\frac12(3,\sqrt3)$ and $\vec a_2=\frac12(3,-\sqrt3)$.
The reciprocal lattice $\L^*$ is the set of vectors such
that $e^{i\vec K \vec x}=1$, if $\vec x\in \L$.
A basis $\vec b_1,\vec b_2$ for $\L^*$ can be obtained by the inversion
formula:
\be\pmatrix{b_{11}&b_{12}\cr b_{21}&b_{22}\cr}=2\p\pmatrix{a_{11}&a_{21}\cr
a_{12}&a_{22}\cr}^{-1}\;,\label{A.3a}\ee
which gives
\be {\vec b_1}={2\pi\over 3}(1,\sqrt{3})\;,\qquad {\vec b_2}={2\pi\over
3}(1,-\sqrt{3})\;.\label{A.4a}\ee
We call $\DD_L$ the set of quasi-momenta $\vec k$ of the form
\be \vec k={m_1\over L}{\vec b_1}+{m_2\over L}{\vec b_2}\;,\qquad m_1,m_2
\in\ZZZ\;,\label{A.5a}\ee
identified modulo $\L^*$; this means that $\DD_L$ can be identified
with the vectors $\vec k$ of the form (\ref{1.2}) and restricted to the
{\it first Brillouin zone}:
\be \DD_L=\{\vec k={m_1\over L}{\vec b_1}+
{m_2\over L}{\vec b_2}\ :\ 0\le m_1,m_2\le L-1\}\;.
\label{A.6a}\ee
Given a periodic function $f:\L\to\RRR$, its Fourier transform is
defined as
\be  f(\vec x)={1\over |\L|}
\sum_{\vec k\in \DD_L} e^{i\vec k\vec x} \hat f(\vec k)\;,\label{A.7a}\ee
which can be inverted into
\be \hat f(\vec k)=\sum_{\vec x\in \L}
e^{-i\vec k\vec x} f(\vec x)\;,\qquad \vec k\in\DD_L\;,\label{A.8}\ee
where we used the identity
\be \sum_{\vec x\in\L} e^{i\vec k\vec x}= |\L| \d_{\vec k, \vec 0}\label{A.9}
\ee
and $\d$ is the periodic Kronecker delta function over $\L^*$.

We now associate to the set of creation/annihilation operators
$a^{\pm}_{\vec x,\s}$, $b^{\pm}_{\vec x+\vec\d_i,\s}$ the corresponding set
of operators in momentum space:
\be a^\pm_{\vec x,\s}={1\over |\L|} \sum_{\vec k\in \DD_L} e^{\pm
i\vec k\vec x} \hat a^\pm_{\vec k,\s}\;,\qquad\qquad b^\pm_{\vec x+\vec
\d_1,\s}={1\over |\L|} \sum_{\vec k\in \DD_L} e^{\pm i\vec k\vec x}
\hat b^\pm_{\vec k,\s} \;.\label{A.10}\ee
Note that, using (\ref{A.7a})--(\ref{A.9}), we find that
\be \hat a^\pm_{\vec k,\s}=\sum_{\vec x\in\L}e^{\mp i\vec k\vec x}
a^\pm_{\vec x,\s}\;,\qquad\qquad \hat b^\pm_{\vec k,\s}
=\sum_{\vec x\in\L}e^{\mp i\vec k\vec x}
b^\pm_{\vec x+\vec\d_1,\s}
\label{A.11}\ee
are fermionic creation/annihilation operators satisfying
\be \{a^\e_{\vec k,\s},a^{\e'}_{\vec k',\s'}\}=|\L|\d_{\vec k,\vec k'}
\d_{\e,-\e'}\d_{\s,\s'}\;,\qquad \qquad \{b^\e_{\vec k,\s},
b^{\e'}_{\vec k',\s'}\}=|\L|\d_{\vec k,\vec k'}
\d_{\e,-\e'}\d_{\s,\s'}\label{A.12}\ee
and $\{a^\e_{\vec k,\s},b^{\e'}_{\vec k',\s'}\}=0$.
With these definitions, we can rewrite
\bea &&H_{0,\L}=-\sum_{\vec x\in \L\atop
i=1,2,3}\sum_{\s=\uparrow\downarrow} (a^+_{\vec x,\s} b_{\vec
x+\vec \d_i,\s}^-+ b_{\vec
x+\vec \d_i,\s}^+a^-_{\vec x,\s} )=\label{A.13}\\
&=&-\frac1{|\L|^2}\sum_{\vec x\in \L\atop i=1,2,3}\sum_{\s=\uparrow\downarrow}
\sum_{\vec k,\vec k'\in \DD_L}\big(
e^{+i\vec k\vec x}e^{- i\vec k' (\vec
x+\vec \d_i-\vec \d_1)}  \hat a^+_{\vec k,\s}
\hat b^-_{\vec k',\s}+ e^{-i\vec k\vec x}e^{+i\vec k' (\vec
x+\vec \d_i-\vec \d_1)} \hat b^+_{\vec k',\s} \hat a^-_{\vec k,\s}
\big)=\nn\\
&=&-{1\over
|\L|}\sum_{\vec k\in \DD_L} \sum_{\s=\uparrow\downarrow}\big(
v^*_{\vec k} \hat a^+_{\vec k,\s} \hat b^-_{\vec
k,\s}+ v_{\vec k} \hat b^+_{\vec k,\s}\hat a^-_{\vec k,\s}\big)\;,\nn\eea
with
\be v_{\vec k}=\sum_{i=1}^3 e^{i(\vec \d_i-\vec \d_1) \vec k}=1+2
e^{-i \frac32 k_1}\cos{\sqrt{3}\over 2}k_2 \;.\label{A.14}\ee
The Hamiltonian $H_{0,\L}$ can be diagonalized by introducing
the fermionic operators
\be \hat \a_{\vec k,\s}={\hat a_{\vec k,\s}\over \sqrt2}+{v^*_{\vec k}\over
\sqrt{2}
|v_{\vec k}|}\hat b_{\vec k,\s}\;,\quad
\quad \hat \b_{\vec k,\s}={\hat a_{\vec k,\s}\over \sqrt2}-{v^*_{\vec k}\over
{\sqrt{2}}
|v_{\vec k}|}\hat b_{\vec k,\s}\;,\label{A.15}\ee
in terms of which we can re-write
\be H_{0,\L}=\frac1{|\L|}\sum_{\vec k\in\DD_L}\sum_{\s=\uparrow\downarrow}
\big(-|v_{\vec k}|\hat \a^+_{\vec k,\s}
\hat \a_{\vec k,\s}+|v_{\vec k}|\hat \b^+_{\vec k,\s}\hat \b_{\vec k,\s}
\big)\;,\label{A.16}\ee
with
\be |v_{\vec k}|=\sqrt{\big(1+2\cos(3k_1/2)\cos(\sqrt{3} k_2/2)\big)^2+
4\sin^2(3 k_1/2)\cos^2(\sqrt{3} k_2/2)}\;,\label{A.17}\ee
which is vanishing iff $\vec k=\vec p_F^\o$, $\o=\pm$, with
\be \vec p_{F}^{\ \o}=({2\pi\over 3},\o{2\pi\over 3\sqrt{3}})\;.\label{A.18}\ee
Now, for $\vec x\in\L$, we define $\a_{\vec x,\s}^\pm=
|\L|^{-1}\sum_{\vec k\in\DD_L} e^{\pm i\vec k\vec x}\hat \a_{\vec
k,\s}$ and $\b_{\vec x,\s}^\pm= |\L|^{-1}\sum_{\vec k\in\DD_L}
e^{\pm i\vec k\vec x}\hat \a_{\vec k,\s}$; moreover, if
$\xx=(x_0,\vec x)$ we define
$\a_{\xx,\s}^\pm=e^{H_{0,\L}x_0}\a_{\vec x,\s}^\pm
e^{-H_{0,\L}x_0}$ and $\b_{\xx,\s}^\pm=e^{H_{0,\L}x_0}\b_{\vec
x,\s}^\pm e^{-H_{0,\L}x_0}$.  A straightforward computation, see,
e.g., Appendix 1 of \cite{BG}, shows that, if $-\b<x_0-y_0\le \b$,
\bea &&\media{{\bf T}\{\a_{\xx,\s}^-\a_{\yy,\s'}^+\}}_{\b,\L}=\\
&&={\d_{\s,\s'}\over|\L|}\sum_{\vec k\in\DD_L}e^{-i\vec k( \vec x-\vec
y)}\Big[\openone\big(x_0-y_0>0\big)\frac{e^{(x_0-y_0) |v_{\vec
k}|}} {1+e^{\b|v_{\vec k}|}}-\openone\big(x_0-y_0\le
0\big)\frac{e^{(x_0-y_0+\b) |v_{\vec k}|}} {1+e^{\b|v_{\vec k}|}}
\Big]\label{A.19aa}\nn\eea
\bea &&\media{{\bf T}\{\b_{\xx,\s}^-\b_{\yy,\s'}^+\}}_{\b,\L}=\\
&&={\d_{\s,\s'}\over|\L|}\sum_{\vec k\in\DD_L}e^{-i\vec k( \vec x-\vec
y)}\Big[\openone\big(x_0-y_0>0\big)\frac{e^{-(x_0-y_0) |v_{\vec
k}|}} {1+e^{-\b|v_{\vec k}|}}-\openone\big(x_0-y_0\le
0\big)\frac{e^{-(x_0-y_0+\b) |v_{\vec k}|}} {1+e^{-\b|v_{\vec
k}|}} \Big] \label{A.19}\nn\eea
and $\media{{\bf T}\{\a_{\xx,\s}^-\b_{\yy,\s'}^+\}}_{\b,\L}=
\media{{\bf T}\{\b_{\xx,\s}^-\a_{\yy,\s'}^+\}}_{\b,\L}=0$. A priori 
Eq.(\ref{A.19aa}) and (\ref{A.19}) are defined only for
$-\b<x_0-y_0\le \b$, but we can extend them periodically over the 
whole real axis; the periodic extension of the propagator is 
continuous in the time variable for $x_0-y_0\not\in\b \ZZZ$, and it has
jump discontinuities at the points $x_0-y_0\in\b\ZZZ$. 
Note that at $x_0-y_0=\b n$, the difference between the right and left 
limits is equal to
$(-1)^n\d_{\vec x,\vec y}$, so that the propagator is discontinuous only
at $\xx-\yy=\b\ZZZ\times \vec 0$. For $\xx-\yy\not\in\b\ZZZ\times \vec 0$, 
we can write
\bea &&\media{{\bf T}\{\a_{\xx,\s}^-\a_{\yy,\s'}^+\}}_{\b,\L}=
\lim_{M\to\io}\frac{\d_{\s,\s'}}
{\b|\L|}\sum_{\kk\in\DD_{\b,L}}e^{-i\kk(\xx-\yy)}
\frac1{-ik_0-
|v_{\vec k}|}\;, \label{A.19a1}\\
&&\media{{\bf
T}\{\b_{\xx,\s}^-\b_{\yy,\s'}^+\}}_{\b,\L}=\lim_{M\to\io}\frac{\d_{\s,\s'}}
{\b|\L|}\sum_{\kk\in\DD_{\b,L}}e^{-i\kk(\xx-\yy)}
\frac1{-ik_0+ |v_{\vec k}|}\;. \label{A.19a2}\eea
Note indeed that for $x_0- y_0\not\in\b\ZZZ$ the sums over $k_0$ in
(\ref{A.19a1}) are convergent, uniformly in $M$;
if $x_0-y_0=\b n$ and $\vec x\neq\vec y$, the r.h.s. of \pref{A.19a1} is equal 
to 
\bea && \frac12\Big(\lim_{x_0-y_0\to(\b n)^+} \media{{\bf
T}\{\a_{\xx,\s}^-\a_{\yy,\s'}^+\}}_{\b,\L}+\lim_{x_0-y_0\to (\b n)^-}
\media{{\bf T}\{\a_{\xx,\s}^-\a_{\yy,\s'}^+\}}_{\b,\L}\Big)=\label{A.19b}\\
&&\hskip9.truecm=\media{{\bf
T}\{\a_{\xx,\s}^-\a_{\yy,\s'}^+\}}_{\b,\L}
\Big|_{x_0-y_0=\b n}
\;.\nn\eea
A similar remark is valid for $\media{{\bf
T}\{\b_{\xx,\s}^-\b_{\yy,\s'}^+\}}_{\b,\L}$. If we now re-express
$\a_{\xx,\s}^\pm$ and $\b^\pm_{\xx,\s}$ in terms of
$a^\pm_{\xx,\s}$ and $b^\pm_{\xx+\dd_1,\s}$, using (\ref{A.15}),
we get (\ref{1.5}).
Note finally that if $\xx=\yy$
\be \lim_{M\to\io}\fra{1}{\b|\L|}\sum_{\kk\in\DD_{\b,L}} \hat g_{\kk}=
\sum_{\kk\in\DD_{\b,L}} \frac{1}{k_0^2+|v(\vec k)|^2}
\pmatrix{0&-v^*(\vec k)\cr -v(\vec k)&0}\;,\label{A.20}\ee
so that the diagonal part is vanishing; on the contrary,
using (\ref{A.15}) and the fact that $\media{\hat \a^+_{\vec k,\s}
\hat \b_{\vec k',\s'}}_{\b,\L}=\media{\hat \b^+_{\vec k,\s}
\hat \a_{\vec k',\s'}}_{\b,\L}=0$, we get
\bea S_0(0^-,\vec 0)_{1,1}&=&S_0(0^-,\vec 0)_{2,2}=
-\frac12\big(\media{\a^+_{\vec x,\s}\a_{\vec x,\s}}_{\b,\L}+
\media{\b^+_{\vec x,\s}\b_{\vec x,\s}}_{\b,\L}\big)=\nn\\
&=&-\frac1{2|\L|}\sum_{\vec k\in\DD_L}
\Big(\frac{e^{\b |v_{\vec k}|}} {1+e^{\b|v_{\vec k}|}}+
\frac{e^{-\b |v_{\vec k}|}} {1+e^{-\b|v_{\vec k}|}}\Big)
=-\frac12\;,\label{A.21}\eea
and this explains why there
are no quadratic terms in $V(\Psi)$, see (\ref{2.6xyz}).

\section{The ultraviolet integration}\lb{B} \setcounter{equation}{0}
\renewcommand{\theequation}{\ref{B}.\arabic{equation}}

In order to prove Eq.(\ref{2.16})--(\ref{2.17}), a simple application
of (\ref{2.46}) and determinant bounds is not enough, because
$g^{(u.v.)}(\xx)$ does not admit a Gram representation, which is a
key property needed for the implementation of standard fermionic
cluster expansion methods. As mentioned in Section \ref{IIIb}, a
way out of this problem is to decompose the ultraviolet propagator
into a sum of propagators, each admitting a Gram representation,
and performing a simple multiscale analysis of the ultraviolet
problem, in analogy with the standard strategy for ultraviolet
problems in fermionic Quantum Field Theories \cite{GK,L}. This
multiscale analysis is very similar to (but much simpler than) the
one describe in Section \ref{IIIc}; it has been performed in
several previous papers \cite{BGPS94,BM1,BGM} and it is reported
here for completeness.

Let $M$ be the integer introduced after (\ref{1.5}), and let us
write
\be g^{(u.v.)}(\xx)=\sum_{h=1}^{h_M} g^{(h)}(\xx)\;,\label{B.1}\ee
where
\be
g^{(h)}(\xx)={1\over \b|\L|}\sum_{\kk\in\DD_{\b,L}} f_{u.v.}(\kk)
H_h(k_0)e^{-i\kk\xx}\hat g_\kk\;,\label{B.2}\ee
with $H_1(k_0)=\c_0(|k_0|)$, $H_h(k_0)=\c_0(\g^{-h+1}|k_0|)-
\c_0(\g^{-h+2}|k_0|)$ and $h_M$ is the smallest integer such that
$f_{u.v.}(\kk) H_j(k_0)\=0$ for all $j>h_M$ (note that
$h_M\simeq\log(M/\b)$). Note that $g^{(h)}({\bf 0})=0$ and, for
any integer $K\ge 0$, $g^{(h)}(\xx)$ satisfies the bound
\be |g^{(h)}(\xx)|\le {C_K\over 1+ (\g^h |x_0|_\b + |\vec x|_\L)^K}
\;,\label{B.3}\ee
where $|\cdot|_\b$ is the distance from the origin on
the one dimensional torus of size $\b$, while $|\cdot|_\L$ is the
distance on $\L$. Moreover, $g^{(h)}(\xx)$ admits a Gram representation:
$g^{(h)}(\xx-\yy)=\int d\zz\, A_h^*(\xx-\zz)\cdot B_h(\yy-\zz)$, with
\bea A_h(\xx)&=&{1\over \b|\L|}\sum_{\kk\in\DD_{\b,L}} \sqrt{f_{u.v.}(\kk)
H_h(k_0)}\frac{e^{-i\kk\xx}}{k_0^2+|v(\vec k)|^2}\pmatrix{1&0\cr 0& 1}\;,
\nn\\
B_h(\xx)&=&{1\over \b|\L|}\sum_{\kk\in\DD_{\b,L}} \sqrt{f_{u.v.}(\kk)
H_h(k_0)}\,e^{-i\kk\xx}\pmatrix{i k_0 & -v^*(\vec k)\cr
-v(\vec k)& i k_0}\label{B.4}\eea
and
\be ||A_h||^2=\int d\zz |A_h(\zz)|^2\le C\g^{-3h}\;,\quad\quad
||B_h||^2\le C \g^{3h}\;,\label{B.5}\ee
for a suitable constant $C$.

Our goal is to compute
\bea
e^{-\b |\L|F_0-\VV(\Psi^{(i.r)})}&=&\lim_{M\to\io}\int P(d\Psi^{[1,h_M]})
e^{V(\Psi^{(i.r.)}+\Psi^{[1,h_M]})}\;,
\label{B.6}\eea
where $P(d\Psi^{[1,h_M]})$ is the fermionic ``Gaussian integration''
associated with the propagator $\sum_{h=1}^{h_M}\hat g^{(h)}(\kk)$
(i.e., it is the same as $P(d\Psi^{(u.v.)})$).
We perform the integration of (\ref{B.6}) in an iterative fashion,
analogous to the procedure described in Section \ref{IIIc} for the
infrared integration. We can inductively prove the
analogue of (\ref{2.28}), i.e.,
\be e^{-\b |\L|F_0-\VV(\Psi^{(i.r)})}=e^{-\b|\L|F_h}\int P(d\Psi^{[1,h]})
e^{\VV^{(h)}(\Psi^{(i.r.)}+\Psi^{[1,h]})}\label{B.7}\ee
where $P(d\Psi^{[1,h]})$  is the fermionic ``Gaussian integration''
associated with the propagator $\sum_{h=1}^{h}\hat g^{(h)}(\kk)$
and
\be {\cal V}^{(h)}(\Psi^{[1,h]})
=\sum_{n=1}^\io \sum_{\ul\r,\ul\s}\int\,d\xx_1\cdots d\xx_{2n}
\Big[\prod_{j=1}^n \Psi^{[1,h]+}_{\xx_{2j-1},\s_j,\r_{2j-1}}
\Psi^{[1,h]-}_{\xx_{2j},\s_j,\r_{2j}}\Big]
W_{2n,\ul\r}^{(h)}(\xx_1,\ldots,
\xx_{2n})\;.\label{B.8}\ee
In order to inductively prove (\ref{B.7})-(\ref{B.8})
we simply use the addition principle to rewrite
\be \int P(d\Psi^{[1,h]})
e^{\VV^{(h)}(\Psi^{(i.r.)}+\Psi^{[1,h]})}=
\int P(d\Psi^{[1,h-1]})\int P(d\Psi^{(h)})
e^{\VV^{(h)}(\Psi^{(i.r.)}+\Psi^{[1,h-1]}+\Psi^{(h)})}\;,\label{B.9}\ee
where $P(d\Psi^{(h)})$ is the fermionic Gaussian integration with
propagator $\hat g^{(h)}(\kk)$. After the integration of $\Psi^{(h)})$
we define
\be e^{\VV^{(h-1)}(\Psi^{(i.r.)}+\Psi^{[1,h-1]})
-\b|\L|\lis e_h}=\int P(d\Psi^{(h)})
e^{\VV^{(h)}(\Psi^{(i.r.)}+\Psi^{[1,h-1]}+\Psi^{(h)})}\;,\label{B.10}\ee
which proves (\ref{B.7}). In analogy with (\ref{2.40}) we have
\be \lis e_h+{\cal V}^{(h-1)}(\Psi)=
\sum_{n\ge 1}{1\over n!}(-1)^{n+1}\EE^T_h({\cal V}^{(h)}\big(
\Psi+\Psi^{(h)}\big);n)
\;.\label{B.11}\ee
As described in Section \ref{IIIc}, the iterative action of $\EE^T_{h_i}$
can be conveniently represented in terms of trees
$\t\in \TT_{M;h,n}$, where $\TT_{M;h,n}$ is a set of labelled trees,
completely analogous to the set $\TT_{h,n}$ described before Eq.(\ref{2.41}),
unless for the following modifications:
\begin{enumerate}
\item a tree $\t\in\TT_{M;h,n}$ has vertices $v$ associated with scale
labels $h+1\le h_v\le h_M+1$, while the root $r$ has scale $h$;
\item with each end-point $v$ we associate $V(\Psi^{[1,h_M]})$, with $V(\Psi)$
defined in (\ref{2.6xyz}).
\end{enumerate}

In terms of these trees, the effective potential ${\cal V}^{(h)}$,
$0\le h\le h_M$ (with $\VV^{(0)}(\Psi^{(i.r.)})$ identified with
$\VV(\Psi^{(i.r.)})$), can be written as
\be {\cal V}^{(h)}(\Psi^{[1,h]}) + \b|\L| \lis e_{h+1}=
\sum_{n=1}^\io\sum_{\t\in\TT_{M;h,n}}
{\cal V}^{(h)}(\t,\Psi^{[1,h]})\;,\label{B.12}\ee
where, if $v_0$ is the first vertex of $\t$ and $\t_1,\ldots,\t_s$
($s=s_{v_0}$) are the subtrees of $\t$ with root $v_0$,
${\cal V}^{(h)}(\t,\Psi^{[1,h]})$
is defined inductively as follows:\begin{enumerate}
\item[i)] if $s>1$,
then
\be {\cal V}^{(h)}(\t,\Psi^{[1,h]})={(-1)^{s+1}\over s!} \EE^T_{h+1}
\big[\bar{\cal V}^{(h+1)}(\t_1,\Psi^{[1, h+1]});\ldots; \bar{\cal V}^{(h+1)}
(\t_{s},\Psi^{[1,h+1]})\big]\;,\label{B.13}\ee
where $\bar{\cal V}^{(h+1)}(\t_i,\Psi^{[1, h+1]})$ is equal to ${\cal
V}^{(h+1)}(\t_i,\Psi^{[1, h+1]})$ if the subtree $\t_i$ contains
more than one end-point, or if it contains one end-point but it is
not a trivial subtree;
it is equal to $V(\Psi^{[1, h+1]})$
if $\t_i$ is a trivial subtree;
\item[ii)] if $s=1$, then ${\cal V}^{(h)}(\t,\Psi^{(\le h)})$ is equal to
$\EE^T_{h+1}\big[{\cal V}^{(h+1)}(\t_1,\Psi^{[1, h+1]})\big]$
if $\t_1$ is not a trivial
subtree; it is equal to $\EE^T_{h+1}\big[V(\Psi^{[1, h+1]})- V(\Psi^{[1,h]})
\big]$
if $\t_1$ is a trivial subtree.
\end{enumerate}

Note that, with $V(\Psi)$ defined as in (\ref{2.6xyz})
and with the present choice of the ultraviolet cutoff
(such that $g^{(h)}(\V0)=0$), we get
$\EE^T_{h+1}\big[V(\Psi^{[1, h+1]})- V(\Psi^{[1,h]})
\big]=0$. This implies
that, if $v$ is not an endpoint and $n(v)$ is the number of endpoints 
following $v$ on $\t$,
and if $\t$ has a vertex $v$ with $n(v)=1$, then its value vanishes:
therefore, in the sum over the trees, we can freely impose the constraint that
$n(v)>1$ for all vertices $v\in\t$. From now on we shall assume that
the trees in $\TT_{M;h,n}$ satisfy this constraint.

\begin{figure}[ht]
\hspace{-5.5truecm}
\includegraphics[height=8.truecm]{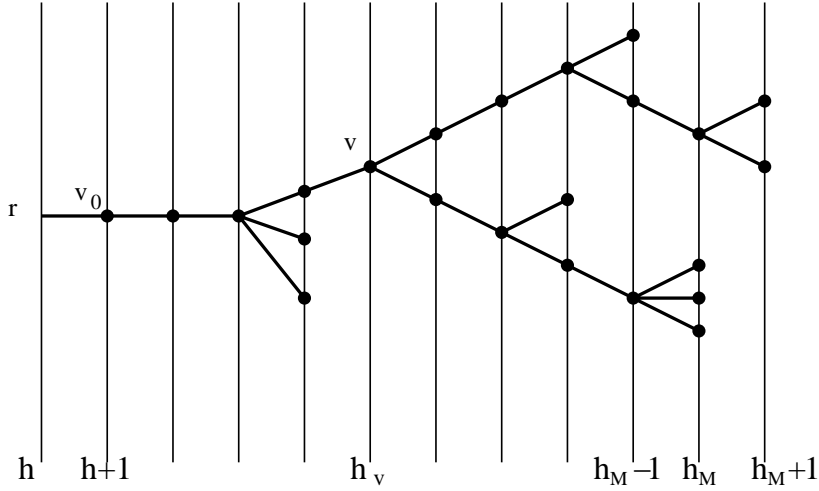}
\vspace{-0.7truecm}
\caption{A tree $\t\in\TT_{M;h,n}$ with its scale labels.}
\end{figure}

Repeating step by step the discussion leading to (\ref{2.43}), (\ref{2.49})
and (\ref{2.50}), and using analogous definitions, we find that
\be {\cal V}^{(h)}(\t,\PP) = \sum_{T\in {\bf T}} \int d\xx_{v_0}
\widetilde\Psi^{(\le h)}(P_{v_0}) W_{\t,\PP,T}^{(h)}(\xx_{v_0})
\= \sum_{T\in {\bf T}}
{\cal V}^{(h)}(\t,\PP,T)\;,\label{B.14}\ee
where
\be \widetilde\Psi^{(\le h)}
(P_{v})=\prod_{f\in P_v}\Psi^{ (\le
h)\e(f)}_{\xx(f),\s(f),\r(f)}\label{B.15}\ee
and
\be
W_{\t,\PP, T}(\xx_{v_0}) =U^n 
\Bigg\{\prod_{v\,\atop\hbox{\ottorm not e.p.}}{1\over s_v!} \int
dP_{T_v}({\bf t}_v)\;{\rm det}\, G^{h_v,T_v}({\bf t}_v)\Biggl[
\prod_{l\in T_v} \d_{\s^-_l,\s^+_l}\,
\big[g^{(h)}(\xx_l-\yy_l)\big]_{\r^-_l,\r^+_l}\,\Biggr]
\Bigg\}\;.\label{B.16}\ee
Moreover, $G^{h_v,T_v}({\bf t}_v)$ is a matrix, analogous to (\ref{2.48}),
with $\d_{\o_l^+,\o_l^-}$ replaced by $1$ and $g^{(h)}_{\o_l}$ replaced by
$g^{(h)}$.

As in the proof of Theorem 2, we get the bound
\bea&& \frac1{\b|\L|}\int d\xx_1\cdots d\xx_{2l} |W^{(h)}_{2l,\ul\r}(\xx_1,
\ldots,\xx_{2l})|\le
\sum_{n\ge 1}|U|^n\sum_{\t\in {\cal T}_{M;h,n}}
\sum_{\PP\in{\cal P}_\t\atop |P_{v_0}|=2l}\sum_{T\in{\bf T}}
\int\prod_{l\in T^*}
d(\xx_l-\yy_l) \cdot
\nn\\
&&\hskip1.truecm\cdot\left[\prod_{i=1}^n|K_{v_i^*}|\right]
\cdot\Bigg[\prod_{v\ {\rm not}\ {\rm e.p.}}{1\over s_v!}
\max_{{\bf t}_v}\big|{\rm det}\, G^{h_v,T_v}({\bf t}_v)\big|
\prod_{l\in T_v}
\big|\big|g^{(h)}(\xx_l-\yy_l)\big|\big|\Bigg]\label{B.17}\eea
and, using the analogues of the estimates (\ref{2.54}), (\ref{2.55})
and (\ref{2.56}), taking into account the new scaling of the propagator,
we find that (\ref{B.17}) can be bounded above by
\be \sum_{n\ge 1}\sum_{\t\in {\cal T}_{M;h,n}}
\sum_{\PP\in{\cal P}_\t\atop |P_{v_0}|=2l}\sum_{T\in{\bf T}}
C^n|U|^n \Big[\prod_{v\ {\rm not}\ {\rm e.p.}} \fra{1}{s_v!}
\g^{-h_v(s_v-1)}\Big]\;.
\label{B.18}\ee
Using (\ref{2.58}) we find that the latter expression can be rewritten as
\be \sum_{n\ge 1}\sum_{\t\in {\cal T}_{M;h,n}}
\sum_{\PP\in{\cal P}_\t\atop |P_{v_0}|=2l}\sum_{T\in{\bf T}}
C^n|U|^n \g^{-h(n-1)}\Big[\prod_{v\ {\rm not}\ {\rm e.p.}} \fra{1}{s_v!}
\g^{-(h_v-h_{v'})(n(v)-1)}\Big]\;,
\label{B.18a}\ee
where we remind the reader that $n(v)>1$ for any $\t\in\TT_{M;h,n}$.
Performing the sums over $T, \PP$ and $\t$ as in the proof of Theorem 2,
we finally find
\be \frac1{\b|\L|}\int d\xx_1\cdots d\xx_{2l} |W^{(h)}_{2l,\ul\r}(\xx_1,
\ldots,\xx_{2l})|\le C|U|^{\max\{1,n-1\}}\;,\label{B.19}\ee
which is a special case of (\ref{2.17}). The proof of the general case is
completely analogous.

\section{Graphene as asymptotic infrared massive $QED_{2+1}$
}\lb{C} \setcounter{equation}{0}
\renewcommand{\theequation}{\ref{C}.\arabic{equation}}

In this Appendix we describe the relation between 2D graphene and
a regularized version of euclidean $QED_{2+1}$ with a {\it
massive} photon, massless dirac fermions and an ultraviolet
cut-off. Let us first introduce the model of regularized $QED_{2+1}$
and let us next describe
its connections with the graphene model described in this paper.

We consider the following
{\it generating function} for euclidean $QED_{2+1}$:
\be e^{\WW_{L,a}(J,\,\phi)}= \int P(d\psi) P(d A)\,e^{\int d\xx
(e_0 A_{\m,\xx}\bar\psi_\xx \g_\m \psi_\xx+
J_{\m,\xx}\bar\psi_\xx\g_\m\psi_\xx+ \phi_\xx\bar\psi_\xx+
\bar\phi_\xx\psi_\xx)}\;,\label{C.1} \ee
where:\begin{enumerate}
\item if $c$ is the speed of light, $\int d\xx$ is a shorthand for
$a^3c^{-1}\sum_{\xx\in\L_a}$,
$a$ is the lattice spacing and
$\L_{a}$ is a periodic lattice of side $Lc^{-1}$ in the time direction, of
side $L$ in the two spatial directions, and with sites labelled by
$x_0=n_0 a c^{-1}$, $\vec x=\vec n a$, with $La^{-1}$ integer and
$n_\m=0,\ldots, La^{-1}-1$, $\m=0,1,2$;
\item summation over repeated indices $\m=0,1,2$ is understood;
\item $e_0$ is a constant,
$J_\m$ and $\phi$ are the external fields, and
$\g_\m$ are euclidean gamma matrices, satisfying $\{\g_\m,\g_\n\}=
-2\d_{\m\n}$, and defined as
\be \g_0=-i\pmatrix{0 & \s_0\cr \cr \s_0 &0}\;,\quad\quad
\g_1=\pmatrix{0 & \s_2\cr \cr -\s_2 &
0}\;,\quad\quad\g_2=\pmatrix{0 & \s_1\cr \cr -\s_1 & 0}\;,\label{C.2}\ee
with $\s_\m$, $\m=0,1,2$, the Pauli matrices:
\be \s_0=\pmatrix{1 & 0\cr \cr 0 & 1}\;,\quad\quad \s_1=\pmatrix{0
& 1\cr \cr 1& 0}\;,\quad\quad\s_2=\pmatrix{0 & -i\cr \cr i & 0}\;;
\label{C.3}\ee
\item $\psi_\xx$ is a 4-components
{\it Grassmann spinor} of components $\psi_{\xx,i}$,
$i=1,\ldots,4$; moreover, $\bar\psi_\xx=\psi^+_\xx\g_0$, with $\psi^+_\xx$
a Grassmann spinor of components $\psi^+_{\xx,i}$;
\item
let $\DD$ be the set of space-time
momenta $\kk$ with $k_0=2\p cL^{-1}(m_0+{1\over 2})$,
$\vec k=2\p L^{-1}\vec m$, with $m_\m=0,1,\ldots,La^{-1}-1$,
$\m=0,1,2$; if we define
$\hat \psi^\pm_{\kk,i}=a^3c^{-1}\sum_{\xx\in\L_a}e^{\mp i \kk\xx}\psi_{\xx,i}$,
the {\it fermionic integration} can be written as
\be P(d\psi) = \frac1{\cal N}\big[\prod_{\kk\in\DD}\prod_{i=1}^4
d\hat\psi^+_{\kk,i}
d\hat\psi_{\kk,i}\big]\exp \Big\{-\frac{Z}{L^3c^{-1}}\sum_{\kk\in\DD}
\chi_0^{-1}(|\kk|)\bar\psi_\kk
i\!\not\!\kk\,\psi_\kk\}\;,\label{C.4}\ee
where $\not\!\kk=\g_\m k_\m$, $c$ is the speed of light,
$Z$ is the wave function renormalization, $\NN$ is a normalization constant
and $\chi_0$ is the cut-off
function introduced in Sec.\ref{IIIb};
\item
$A_{\m,\xx}$ is a euclidean gaussian boson
field associated to the gaussian measure $P(dA)$ with covariance
\be v_{\m,\n}(\xx-\yy)=\d_{\m\n} v(\xx-\yy)\=\d_{\m\n}\frac{c}{L^3}
\sum_{\pp\in\DD} e^{-ip_0(\xx-\yy)}\frac{\chi_0(|\pp|)}{\pp^2+M^2}\;,
\label{C.5}\ee
with $M>1$ the ``photon mass''.
\end{enumerate}

Integrating out the gaussian boson field, we can rewrite:
\be e^{\WW_{L,a}(J,\phi)}= \int P(d\psi) e^{-\VV(\psi)+\int d\xx
J_{\m,\xx}\bar\psi_\xx\g_\m\psi_\xx+\int d\xx
(\phi_\xx\bar\psi_\xx+ \bar\phi_\xx\psi_\xx )}\;,\label{C.6}\ee
where
\be \VV(\psi)=-\frac{e_0^2}2\int d\xx d\yy \,
(\bar\psi_\xx\g_\m\psi_\xx)v(\xx-\yy)(\bar\psi_\yy\g_\m\psi_\yy)\;.
\label{C.7}\ee
The four dimensional version of the above model was studied in \cite{M1}
by RG methods; the
analysis (that can be repeated for the three dimensional model considered
here without any relevant difference)
is essentially identical to the one described in this paper for the 2D
Hubbard model. Note in particular that, identifying
the spinor $\hat \psi_\kk$ with
$\big(\hat\Psi_{\kk,\s,1,+},\hat\Psi_{\kk,\s,2,+},\hat\Psi_{\kk,\s,2,-},
\hat\Psi_{\kk,\s,1,-}\big)$, both the fermionic integration $P(d\psi)$
and the effective interaction $\VV(\psi)$ are invariant under a number of
symmetries, analogous to
(4)--(8) of Lemma 1, i.e.,\\
\\
\noindent(4') $\hat \Psi_{(k_0,k_1,k_2),\s,1,\o}^\pm\to
\hat \Psi_{(k_0,k_2,k_1),\s,2,\o}^\pm$, $\hat \Psi_{(k_0,k_2,k_1),\s,2,\o}^\pm
\to(\mp i\o)\hat \Psi_{(k_0,k_1,k_2),\s,1,\o}^\pm$;\\
\noindent(5') $\hat \Psi^{\pm}_{\kk,\s,\r,\o}
\to \hat\Psi^\pm_{-\kk,\s,\r,-\o}$, $c\to c^*$, where $c$ is a generic
constant appearing in $P(d\Psi)$ and/or in $\VV(\psi)$;\\
\noindent(6'.a) $\hat\Psi^{\pm}_{(k_0,k_1,k_2),
\s,1,\o}
\otto \hat\Psi^\pm_{(k_0,-k_1,k_2),\s,2,\o}$;\\
\noindent(6'.b) $\hat\Psi^{\pm}_{(k_0,k_1,k_2),\s,\r,\o}
\to \hat\Psi^\pm_{(k_0,k_1,-k_2),\s,\r,-\o}$;\\
\noindent(7') $\hat\Psi^{\pm}_{(k_0,\vec k),\s,\r,\o}\to
i\hat\Psi^{\mp}_{(k_0,-\vec k),\s,\r,-\o}$;\\
\noindent(8') $\hat\Psi^{\pm}_{(k_0,\vec k),\s,\r}\to
i(-1)^\r\hat\Psi^{\pm}_{(-k_0,\vec k),\s,\r}$.\\
\\
It is important to note that, in addition to the symmetries (4')--(8')
above, $QED_{2+1}$ also admits extra symmetries, related to its relativistic
invariance, which have no counterpart in the Hubbard model, e.g.,\\
\\
\noindent(9') $\psi_\kk\to e^{\frac{\th}4[\g_0,\g_1]}
\psi_{R_{\th}^{-1}\kk}$, $\bar\psi_\kk\to\bar\psi_{R_\th^{-1}\kk}
e^{-\frac{\th}4[\g_0,\g_1]}$,
where $R_\th\kk=
(k_0\cos\th -ck_1\sin\th,k_1\cos\th+
c^{-1}k_0\sin\th,k_2)$. Note that in the limit $L,a^{-1}\to\io$, there is
no constraint on the choice of $\th$, while for finite $L$ and $a$ we are
forced to choose $\th=\p/2$. The proof of the invariance of the model under the
symmetry (9') is a simple consequence of the remark that
\be e^{-\frac{\th}4[\g_0,\g_1]}(\g_0,\g_1,\g_2)e^{\frac{\th}4[\g_0,\g_1]}=
(\g_0\cos\th-\g_1\sin\th\;,
\g_1\cos\th+\g_0\sin\th\;,\g_2)\;,\label{C.8}
\ee
which implies that $\sum_{\kk\in\DD}\bar\psi_\kk \not\!\kk\psi_\kk$ is
invariant under (9'). In particular, if $\th=\p/2$, in terms of the
components $\hat \Psi_{\kk,\r,\s,\o}$ of the spinor,
(9') reads as follows:
\bea&& \hat\Psi_{(k_0,k_1,k_2),\s,\r,\o}\to\frac1{\sqrt2}(\s_0+i\s_2)_{\r,\r'}
\hat\Psi_{(ck_1,-c^{-1}k_0,k_2),\s,\r',\o}\;,\nn\\
&&\hat\Psi_{(k_0,k_1,k_2),\s,\r,\o}^+\to
\frac1{\sqrt2}\hat\Psi_{(ck_1,-c^{-1}k_0,
k_2),\s,\r',\o}^+(\s_0+i\s_2)_{\r',\r}\;.\label{C.9}\eea
This symmetry also implies that the kernels of the quadratic part of the
effective potentials have a special structure. In fact, repeating the proof
of Lemma 2, using symmetries (4')--(8'),
and if $\hat W_{2,(\r_1,\r_2),\o}^{(h)}(\kk)$ is the kernel of the
quadratic part of the effective action at scale $h$, we find the analogue
of (\ref{2.31}):
\be  \kk'\partial_{\kk'}
\hat W^{(h)}_{2,(\r_1,\r_2),\o}({\bf 0})=
\pmatrix{ -iz_hk_0 & \d_h(ik_1-\o k_2) \cr
\d_h(-ik_1-\o k_2) & -iz_hk_0}_{\r_1,\r_2}\;.\label{C.10}\ee
On the other hand, for $QED_{2+1}$ we also know that
\be \sum_{\kk\in\DD}\hat \Psi_{\kk,\s,\cdot,\o}^+\pmatrix{-iz_hk_0 &
\d_h(ik_1-\o k_2) \cr
\d_h(-ik_1-\o k_2) & -iz_hk_0}\hat \Psi_{\kk,\s,\cdot,\o}\ee
must be invariant under (\ref{C.9}), which implies $c z_h=\d_h$, i.e.,
{\it the speed of light is not renormalized}.

The same proof shows that if,
in relativistic notation, $\sum_{\kk}\bar\psi_\kk\kk_\m W_\m\psi_\kk$ is
invariant under (4')--(9'), then $W_\m=C\g_\m$, for some constant $C$.
This is precisely the same as in four dimensional euclidean QED. Therefore,
we can repeat step by step the construction in \cite{M1} and, in particular,
we find that the following Ward Identity (WI) is valid:
\be i Z e_0\pp_\m \la j_{\m,\pp};\psi_\kk\bar\psi_{\kk-\pp}\ra= e
\big[\la  \psi_{\kk-\pp}\bar\psi_{\kk-\pp}\ra-\la
\psi_\kk\bar\psi_\kk\ra\big] (1+H_{0}(\kk,\pp))\;,\label{C.11}
\ee
where:\begin{enumerate}
\item $\la j_{\m,\pp};\psi_\kk\bar\psi_{\kk-\pp}\ra=\int d\xx\int d\zz
e^{-i\pp(\zz-\yy)}e^{i\kk(\xx-\yy)}\media{j_{\m,\zz};\psi_\xx\bar\psi_{\yy}}$,
with
\be \media{j_{\m,\zz};\psi_\xx\bar\psi_{\yy}}=
\lim_{L,a^{-1}\to\io}\frac{\partial^3 {\cal W}_{L,a}(J,\phi)}{\dpr\bar\phi_\xx
\dpr\phi_\yy\dpr J_{\m,\zz}}\Big|_{J=\phi=\bar\phi=0}\;;\label{C.12a}\ee
similarly,
\be
\media{\psi_\xx\bar\psi_{\yy}}=\frac{c}{L^3}\sum_{\kk\in\DD}e^{-i\kk(\xx-\yy)}
\la\psi_\kk\bar\psi_{\kk}\ra=
\lim_{L,a^{-1}\to\io}\frac{\partial^2 {\cal W}_{L,a}(J,\phi)}{\dpr\bar\phi_\xx
\dpr\phi_\yy}\Big|_{J=\phi=\bar\phi=0}\;;\label{C.12}\ee
\item $e=e_0-c_+e_0^3+O(e_0^5)$, with $c_+$ a suitable constant;
\item the correction $H_0(\kk,\pp)$ is such that, for
momenta $\kk,\pp,\kk-\pp$ all on the same scale $h$ (i.e,
all belonging to the support of $f_h$, for some finite $h\le 0$)
\be |H_{0}(\kk,\pp)|\le C |e| \g^{\th h}\;,\label{C.13}\ee
for some $0<\th<1$.
\end{enumerate}
Note that the above WI differs from the {\it formal} WI
obtained by neglecting the ultraviolet cut-off, because of
the presence of the {\it renormalized charge} $e=e_0- c_+ e_0^3+O(e_0^5)$
and of the correction $H_{0}(\kk,\pp)$.\\

There is a strong connection between the above model and the
Hubbard model. Indeed from (\ref{4.12}) we know that
\be S(\xx-\yy)=S^{(1)}(\xx-\yy)+
\sum_{\o=\pm} e^{-i \vec p_F^\pm (\vec x-\vec y)}
S_{\o}^{(\le 0)} (\xx-\yy)\;, \label{C.14} \ee
where $S_{\o}^{(\le 0)} (\xx-\yy)$ is given by the sum in the second line
of (\ref{3.7d}) restricted to $h\le 0$, and $|S^{(1)}(\xx-\yy)|
\le C |\xx-\yy|^{-2-\th}$ for $|\xx-\yy|\ge 1$;
this means that, for large distances, $S^{(1)}$ is asymptotically
negligible with respect to $S_\o^{(\le 0)}(\xx-\yy)$.

By (\ref{1.9}) and the construction in Sections \ref{IIIc} and \ref{IIId},
we expect that the Grassmann spinor
$\big(\hat\Psi^{(\le 0)}_{\kk,\s,1,+},\hat\Psi^{(\le 0)}_{\kk,\s,2,+},
\hat\Psi^{(\le 0)}_{\kk,\s,2,-},
\hat\Psi^{(\le 0)}_{\kk,\s,1,-}\big)$ plays the same role as the spinor
$\psi_\kk$ in the $QED_{2+1}$ model.
In order to make this intuition precise, it is convenient to combine
$S_{\pm}^{(\le 0)}(\xx-\yy)$ in the following matrix
\be \GG(\xx-\yy) =\pmatrix{0 & S_{+}^{(\le 0)}(\xx-\yy)\cr
S^{(\le 0)T}_{-}(\xx-\yy) & 0}\;, \label{C.14a}\ee
where $S^{(\le0)T}_\o$ is the transpose of $S^{(\le 0)}_\o$.
$\GG(\xx-\yy)$ will play the same
role as the correlation $\media{\psi_\xx\bar\psi_\yy} $
defined in (\ref{C.13}), in a sense to be made precise below.
Similarly, the role of $\media{j_{\m,\zz};\psi_\xx\bar\psi_\yy}$ will be
played by the correlation $S_{2,1;\m}(\zz;\xx,\yy)$, $\m=0,1,2$,
defined as
\be S_{2,1;\m}(\zz;\xx,\yy)_{\r,\r'}=\media{\bT\{
\Psi^{-}_{\xx,\s,\r}\Psi^{+}_{\yy,\s,\r'}\Psi^+_{\zz,\s,\cdot}\s_\m
\Psi^-_{\zz,\s,\cdot}\}}-\media{\bT\{
\Psi^{-}_{\xx,\s,\r}\Psi^{+}_{\yy,\s,\r'}\}}\cdot
\media{\Psi^+_{\zz,\s,\cdot}\s_\m
\Psi^-_{\zz,\s,\cdot}}\;. \label{C.15}\ee
By an analysis similar to the one in Section \ref{IIId}, we get
\be S_{2,1;\m}(\zz;\xx,\yy)=S^{(1)}_{2,1,\m}(\zz;\xx,\yy)+
S_{2,1;\m}^{+}(\zz;\xx,\yy)+S_{2,1;\m}^{-}(\zz;\xx,\yy)\;,  \label{C.16}\ee
where the first term is asymptotically negligible with respect to the
last two for large distances.
The terms $S_{2,1;\m}^{+}(\zz;\xx,\yy)$ and $S_{2,1;\m}^{-}(\zz;\xx,\yy)$
correspond to contributions to the correlation function coming from the
infrared integration, whose computation requires, as in Sections \ref{IIIc}
and \ref{IIId}, the decomposition of the infrared field into the sum
of {\it quasi-particle} fields indexed by $\o=\pm$ and
supported, in momentum space, around the two different Fermi points
$\vec p_F^\o$. By the compact support properties of the infrared fields,
in the terms contributing to $S_{2,1;\m}^\pm(\zz;\xx,\yy)$,
the quasi-particle indeces
corresponding to the fields located at $\zz$ are the same, and will be denoted
by $\o_\zz$; similarly, the quasi-particle indeces corresponding to the
fields located at $\xx$ and $\yy$ are the same, and will be denoted by
$\o_{\xx\yy}$. Finally,
$S_{2,1;\m}^+(\zz;\xx,\yy)$ is defined as the sum
of all the contributions such that $\o_\zz=\o_{\xx\yy}$, while
$S_{2,1;\m}^-(\zz;\xx,\yy)$ corresponds to the terms with $\o_\zz=
-\o_{\xx\yy}$.
By construction, $S_{2,1;\m}^{+}(\zz;\xx,\yy)$
can be written as a sum over two terms:
\be S_{2,1;\m}^{+}(\zz;\xx,\yy)= \sum_{\o=\pm}
e^{-i \vec p_F^{\o} (\vec x-\vec y)}
S_{2,1;\m,\o}^{+}(\zz;\xx,\yy) \label{C.17}\ee
and we can combine such terms in a single matrix
\be \G_{\m}(\zz;\xx,\yy) =\frac1{(\b|\L|)^2}\sum_{\kk,\pp}
e^{i\pp\zz}e^{-i\kk\xx}e^{i(\kk-\pp)\yy}\hat\G_\m(\pp,\kk)=
\pmatrix{0 & \hat S_{2,1;\m,+}^+(\zz;\xx,\yy)\cr
\cr \hat S_{2,1;\m,-}^{+,T}(\zz;\xx,\yy) & 0}
\;. \ee
It is clear from the multiscale construction of these correlation
functions that, with a proper choice of the parameters, such
matrices are asymptotically close to the Schwinger function of the
$QED_{2+1}$ model seen above, as explained by the following
theorem, which is, in fact, a corollary of the analysis
in the previous sections and of a finite dimensional fixed point argument.\\

{\bf Theorem 3.} {\it Let $U$ and $e_0$ be small enough.
It is possible to choose
$Z$ and $c$ in \pref{C.1}--\pref{C.5}
as functions of $U,e_0,M$ and $v(\V0)$, so that, if $\kk,\pp,\kk-\pp$
are all on the same scale $h$
(i.e., if $a_0\g^{h-1}\le |\kk|,|\pp|,|\kk-\pp|\le a_0\g^{h+1}$, $h\le 0$),
\bea
&&\GG(\kk)=\la\psi_\kk\bar \psi_\kk\ra(1+ O(\g^{\th h}))\;,\label{df1}\\
&&\G_{\m}(\kk,\pp)=Z_\m\la j^\m_\pp;\psi_\kk\bar
\psi_{\kk+\pp}\ra(1+ O(\g^{\th h}))\;,\label{df2}\eea
where $Z_0,Z_1,Z_2$ in (\ref{df2}) are suitable constants,
depending on $U,e_0,M$ and $v(\V0)$ and $0<\th<1$.}
\\

Theorem 3 says that, by choosing the wave function renormalization
and the velocity of light in the $QED$ model as suitable functions
of $U$, $e_0$, $M$ and $v(\V0)$, its two point Schwinger functions
coincide with the ones of the Hubbard model, up to corrections
which are negligible at small momenta. With this choice of $Z$ and
$c$, the vertex functions of $QED_{2+1}$ are asymptotically
proportional to those of the Hubbard model, provided that the
renormalizations $Z_\m$ are properly chosen. Note that, while 
in a relativistic QFT $Z_\m$ is $\m$-independent, here it is not 
\cite{GMprl}, the symmetry (9') being broken by the underlying
lattice; however, one can check, by arguments
similar to the ones used in the proof of Lemma 2, that the lattice 
symmetries imply that the renormalizations $Z_\m$ are still diagonal in $\mu$:
note, in fact, that in principle the r.h.s. of (\ref{df2}) could be 
of the form
$$\sum_{\n}Z_{\m,\n}\la j^{\n}_\pp;\psi_\kk\bar
\psi_{\kk+\pp}\ra(1+ O(\g^{\th h}))\;,$$
but, remarkably, $Z_{\m,\n}$ turns out to be diagonal.

Theorem 3 implies that the Schwinger functions of the 2D Hubbard model on
the honeycomb lattice obey to a Ward Identity
analogous to (\ref{C.11}), as it follows by
combining (\ref{C.11}) with (\ref{df1})--(\ref{df2}), see \cite{GMprl}. 
This is true
not only in the free case $U=0$ (in which case the WI can be verified by a
simple explicit computation) but also, remarkably, in the interacting case.
Note that, with respect to the WI for QED, the WI for the Hubbard model
is modified by the presence of some proportionality constants,
which take into account both
the relativistic renormalization of the charge and the fact
that the Hubbard model breaks some relativistic symmetries.

Let us conclude by remarking that, while here the WI can be proved
{\it a posteriori} of the construction of the correlation
functions, in the presence of Coulomb interactions the validity of
an analogous WI is believed to play a crucial role in the
construction of the model itself, as in one dimension \cite{M2}:
in fact, in that case, the interparticle interaction becomes
marginal in a RG sense \cite{GGV1} and the presence of WIs is a
key ingredient in the control of the flow of the beta function
equation, as in QED or in the Luttinger model.

\end{document}